\documentclass[preprint]{aastex}
\usepackage{graphicx}
\usepackage{subfigure}

\begin{document}

\title{The Detection and Statistics of Giant Arcs Behind CLASH Clusters}

\author{Bingxiao Xu\altaffilmark{1}, Marc Postman\altaffilmark{2}, Massimo 
Meneghetti\altaffilmark{3}, Stella Seitz\altaffilmark{4}, Adi Zitrin\altaffilmark{5,6}, Julian Merten\altaffilmark{7}, Dani Maoz\altaffilmark{8}, Brenda Frye\altaffilmark{9}, Keiichi Umetsu\altaffilmark{10}, Wei Zheng\altaffilmark{1}, 
Larry Bradley\altaffilmark{2},  Jesus Vega\altaffilmark{11}, Anton Koekemoer\altaffilmark{2}}
\email{bxu6@jhu.edu}
\altaffiltext{1}{Department of Physics and Astronomy, The Johns Hopkins 
University, 3400 North Charles Street, Baltimore, MD 21218, USA}
\altaffiltext{2}{Space Telescope Science Institute, 3700 San Martin Drive, 
Baltimore, MD 21208, USA}
\altaffiltext{3}{INAF, Osservatorio Astronomico di Bologna, \& INFN, Sezione di Bologna; Via Ranzani 1, I-40127 Bologna, Italy}
\altaffiltext{4}{Universitaets-Sternwarte, Fakultaet fuer Physik, Ludwig-Maximilians Universitaet Muenchen, Scheinerstr. 1, D-81679 Muenchen, Germany}
\altaffiltext{5}{Hubble Fellow}
\altaffiltext{6}{California Institute of Technology, MC 249-17, Pasadena, CA 91125, USA}
\altaffiltext{7}{University of Oxford, Department of Physics, Denys Wilkinson Building, Keble Road, Oxford, OX1 3RH, UK}
\altaffiltext{8}{School of Physics and Astronomy, Tel Aviv University, Tel-Aviv 69978, Israel}
\altaffiltext{9}{Steward Observatory/Department of Astronomy, University of Arizona, 933 N Cherry Ave, Tucson, AZ 85721, USA}
\altaffiltext{10}{Institute of Astronomy and Astrophysics, Academia Sinica, P.O. Box 23-141, Taipei 10617, Taiwan}
\altaffiltext{11}{Universidad Autonoma de Madrid, Ciudad Universitaria de Cantoblanco, 28049 Madrid, Spain}
\slugcomment{Accepted for publication in the {\it Astrophysical Journal}}

\begin{abstract}
We developed an algorithm to find and characterize gravitationally lensed 
galaxies (arcs) to perform a comparison of the observed and simulated arc 
abundance. Observations are from the Cluster Lensing And Supernova survey with 
Hubble (CLASH). Simulated CLASH images are created using the MOKA package and 
also clusters selected from the high resolution, hydrodynamical simulations, 
MUSIC, over the same mass and redshift range as the CLASH sample. The algorithm'
s arc elongation accuracy, completeness and false positive rate are determined 
and used to compute an estimate of the true arc abundance.  We derive a lensing 
efficiency of $4 \pm 1$ arcs (with length $\ge 6''$ and length-to-width ratio 
$\ge 7$) per cluster for the X-ray selected CLASH sample, $4 \pm 1$ arcs per 
cluster for the MOKA simulated sample and $3 \pm 1$ arcs per cluster for the 
MUSIC simulated sample. The observed and simulated arc statistics are in full 
agreement. We measure the photometric redshifts of all detected arcs and find a 
median redshift $z_s = 1.9$ with 33\% of the detected arcs having $z_s > 3$. We 
find that the arc abundance does not depend strongly on the source redshift 
distribution but is sensitive to the mass distribution of the dark matter halos 
(e.g. the $c-M$ relation). Our results show that consistency between the 
observed and simulated distributions of lensed arc sizes and axial ratios can 
be achieved by using cluster-lensing simulations that are carefully matched to 
the selection criteria used in the observations.
\end{abstract}


\section{Introduction}
The occurrence frequency of giant gravitationally lensed arcs -- those most 
elongated, highly non-linear lensing features -- is sensitive to the matter 
distribution within the cores of galaxy clusters. The statistics of giant arcs 
can thus provide useful tests of the structure formation. Cosmological models 
can be tested by comparing the observed giant arc abundance with the 
expected abundance from ray-tracing cosmological simulations. In an early study
of arc statistics, \cite{bar98} (hereafter B98) first suggested that the 
predicted arc abundance by $\Lambda CDM$ is lower than the observed abundance 
by approximately an order of magnitude. This ``order of magnitude" puzzle has 
stimulated a significant amount of research towards understanding the most 
important arc-producing effects. The proposed effects include the triaxiality of
cluster mass profiles \citep{ogu03,dal04,hen07,men10}, the amount of intervening
large scale structure \citep{wam05,hil07,puc09}, the rapid increase in the 
lensing cross section during the major mergers \citep{tor04,fed06,hen07,zit13}, 
the background galaxy redshift distribution \citep{wam04}, the cosmological 
parameters \citep{bar03,men05,fed08,jul10,dal11b,bol15}, the cluster selection 
criteria \citep{hor10,hor11}, the baryonic mass distribution, primarily in the 
form of brightest central galaxies (BCG) and substructures \citep{flo00, men00, 
men03, hen07, men10}, and the baryon-dragging effects due to cooling 
\citep{puc05, roz08}. Inclusion of these effects have gone a long way towards 
explaining the ``arc statistics problem." However, the tension between the 
observed arc abundance and the predicted number remained at the level of factor 
3, at least for clusters at low redshifts ($z < 0.3$) \citep{hor11}. Moreover, 
there has not yet been a consensus reached on which of these physical processes 
are the dominant ones.  

To see if the remaining discrepancies can be resolved, efforts need to be made 
on both theoretical and observational fronts. On the theoretical side, all 
effects which impact arc abundance should be included in the simulations to make
them more realistic. A straightforward example is to compare the lensing cross 
section in simulations with dark and baryonic matter against dark matter only 
simulations \citep{men03,puc05,kil12}. On the observational side, larger, 
carefully selected cluster samples with ample redshift information are needed. 
All analyses will also benefit from the utilization of automated procedures for selecting giant arcs as rigorous comparisons must be 
done using an identical arc selection process for both the actual data and 
simulated data. In this respect, visual inspection, by which early arc 
statistics studies were conducted, is not an optimal approach as unquantifiable 
biases can potentially be introduced when classification is done by eye. Several
groups have devised tools to search for arcs in an automated manner 
\citep{len04, hor05, ala06,sei07}. Most recently, \citet{hor10,hor11} measured 
the observed abundance of arcs in a sample of 100 clusters observed with the 
Hubble Space Telescope (HST), using an automated and objective arc finder. The 
observed statistics  were compared to those from a simulated dataset of cluster 
images. The simulated images were produced by ray-tracing through a large sample
of clusters produced in N-body simulations, realistically simulating the 
observational effects, and then searching for arcs in these simulated clusters 
using the same arc-finding algorithm. \citet{hor11} found excellent agreement 
between the observed and simulated arc statistics, particularly for their main 
sub-sample of X-ray selected clusters at redhsifts $0.3 < z < 0.6$. However, 
tension between the observations and simulations remained at other redshifts 
ranges, particularly for the subsample at $z < 0.3$. Moreover, none of the above
groups has quantified the performance of their arcfinders, such as the arc 
detection completeness or the false positive rate. Without that information, 
the arcfinders' ability to predict the ``true" arc abundance is limited.

In this paper, we measure the observed abundance of giant arcs from the CLASH 
(Cluster Lensing And Supernova survey with Hubble) sample \citep{pos12}. Giant 
arcs are found in the CLASH images, and in simulated images that mimic the CLASH
data, using an efficient automated arc-finding algorithm whose selection 
function has been carefully quantified. CLASH is a 524-orbit multi-cycle 
treasury program that targeted 25 massive clusters with 0.18  $< z <$  0.90. 
Twenty of the CLASH clusters are selected based on their X-ray characteristics.
The X-ray selected CLASH sample contains clusters with $T_{x} \ge $ 5 keV and 
with X-ray surface brightness profiles that have low asymmetry. The five 
remaining clusters were selected based on their expected lensing strength (large
Einstein radii, typically $\theta_{Ein} > 30''$ for $z_s = 2$ or high 
magnification areas). Although the cluster sample is smaller than the one 
analyzed by \citet{hor10,hor11}, the CLASH observations are deeper, and photometric redshift information is available for all arcs brighter than about 26 AB mag
(all the magnitudes hereafter are AB mag). In addition, our arcfinder is 
capable of detecting fainter arcs than previous studies. As a result, the total 
number of arcs that we find is comparable to that in the Horesh et al. studies.
We simulate artificial clusters with the same mass and redshift range as the 
CLASH sample by using the N-body simulation-calibrated semi-analytic tool -- 
MOKA \citep{gio12}, and directly from the high resolution, hydrodynamical 
simulations, MUSIC \citep{men14}, and perform ray-tracing simulation to prepare
large sets of realizations for the simulated cluster images. We correct the raw 
arc counts in both the observations and simulations for incompleteness, false 
positive detections and arc elongation measurement bias. This allows us to 
conduct a direct comparison between the data and the simulations under different
theoretical scenarios.

This paper is organized as follows: we describe the arcfinder algorithm and its implementation in section~\ref{s2} and in the appendices; we 
demonstrate the arcfinder detection efficiency and overall performance in 
Section~\ref{s3}; we present the arc abundance results for the CLASH observations  
in Section~\ref{s4}; we describe the cluster simulation and ray-tracing 
calculations in Section~\ref{s5}; we compare the observed and simulated arc abundance results in Section~\ref{s6},
including specifically testing the dependence of the abundance on the source redshift distribution and $c-M$ 
relation in Section~\ref{s7}; a discussion and summary are given in 
Section~\ref{s8} and \ref{s9}, respectively. Throughout the paper, we adopt a 
$\Lambda CDM$ cosmology with parameters $\Omega_{m} = 0.3$, $\Omega_{\Lambda} = 
0.7$, $\sigma_{8} = 0.83$, $H_{0} = 100h \quad \rm km s^{-1} Mpc^{-1}$, and 
$h = 0.7$ \citep{pla14}.

\section{Conceptual Development of the Arcfinder}
\label{s2}

In early works on arc statistics, arc detection was performed by visual 
inspection due, in part, to the complex shapes of arcs and the crowded 
environments in which they are found. An automated arc finding algorithm has 
three key advantages over visual search methods. First, the detection process 
is reproducible and can be implemented by anyone who learns how to run the code. 
Second, it can be applied to a large number of real and 
simulated images. Finally, the detection efficiency and false positive rate can 
be accurately quantified using artificial objects implanted in real data or using simulated
images created by ray-tracing sources through lens models.  The biggest challenge to developing such 
an algorithm is creating a definition of an arc for the purpose of detection 
that can be implemented in a robust manner using parameters that can be easily 
quantified from astronomical images.

An ideal arc finder should have the following characteristics:
\begin{enumerate}
  \item The arc finder should be able to suppress image noise to enhance the 
contrast of real, low surface brightness arcs without significantly altering 
the intrinsic shape characteristics of these faint objects.
  \item The selection of pixels belonging to arcs should, 
if possible, not be based on a global fixed intensity threshold as the
 intensity can vary significantly across a lensed image.
  \item The arc finder must employ rules to reject spurious detections such as 
diffraction spikes from bright stars or edge-on disk galaxies.
  \item The arc finder must be able to process many images in a reasonable 
amount of time.
\end{enumerate}

\begin{figure}
\centering
\includegraphics[width = 1.0\linewidth]{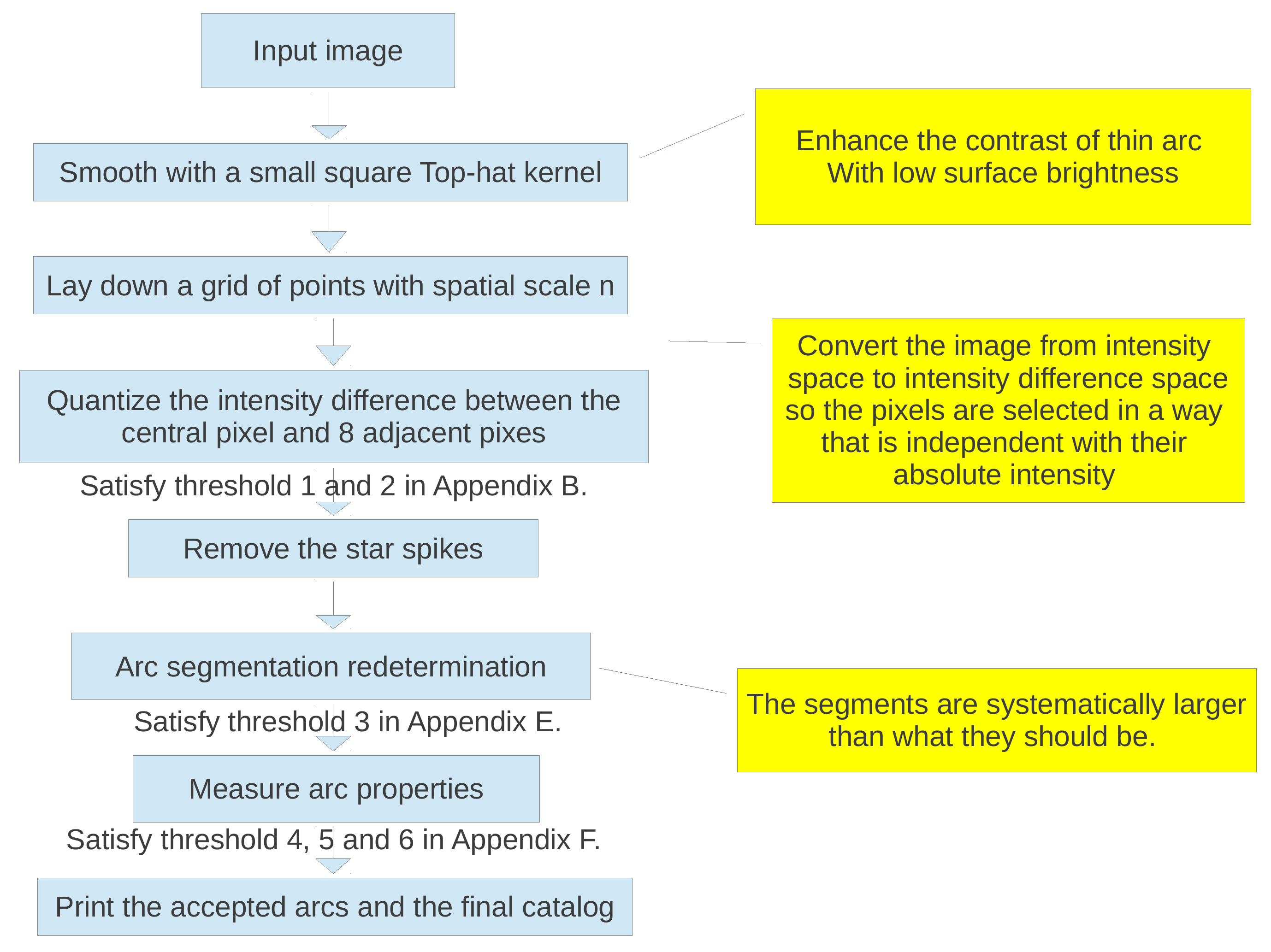}
\caption{Flowchart of the arcfinding algorithm.}
\label{fig0}
\end{figure}

Here we describe an algorithm for identifying giant arcs - the arcs we are most interested in 
analyzing in this work. The algorithm was designed to reasonably comply with 
the above criteria. The parameters that define what we consider to be a
giant arc, such as the minimum length and length-to-width ratio, are presented in Section~\ref{s3.2}. 
 Figure~\ref{fig0} shows a flowchart of the steps 
involved in the algorithm and summarizes its key components. The detailed 
descriptions of the various steps that comprise the algorithm can be found in appendices \ref{A} through \ref{F}.

\section{Arcfinder Testing and Performance}
\label{s3}

\subsection{Simulating Arcs}
\label{s3.1}

To compute the true arc abundance from the detected one, we must 
quantify the incompleteness and false positive rate of the arcfinder using a combination of actual
and simulated datasets. The most robust way to simulate arcs is via ray-tracing in which light rays from 
objects in a source plane are shot towards the observer, deflected by the lens plane, and projected onto the 
image plane (the ``sky" as seen by the observer). To quantify 
the incompleteness and false positive rate, one needs to trace large number of 
simulated arcs which is often very CPU intensive. Moreover, we need full 
control of all the input parameters of simulated arcs to perform the tests 
efficiently, and this becomes difficult to do solely by ray-tracing objects that
are placed randomly on the source plane. 

\citet{fur13} use a different approach to simulate the arcs. Their basic idea is
to represent an arc as a curved ellipse with its main axis being a segment of a circle. The model arc
is then superposed directly on an image at various locations. The arc's shape is
set by various parameters (e.g., length, width, curvature and orientation) chosen to
mimic the shapes of real lensed galaxies. The surface brightness distribution is set using a Sersic law profile. 
The intensity parameters include the Sersic index and the intensity at the center of the arc,
which allows one to assign any magnitude or the total flux to the simulated arcs.
However, this simple analytic prescription does not precisely reproduce the 
properties of real arcs. For example, the ``painted-on" arcs tend to have a 
deficit of surface brightness at their long ends, which can result in shape measurement biases, especially for faint arcs. 
For a robust comparison between the real 
data and simulations, the ``painted" arc method falls short of the fidelity 
that is required. 

\begin{figure}
\centering
\subfigure[$l/w = 6$]{
 \label{fig5a:subfig:a}
 \includegraphics[width = 0.4\linewidth]{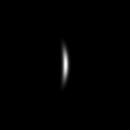}
}
\centering
\subfigure[$l/w = 7$]{
 \label{fig5a:subfig:b}
 \includegraphics[width = 0.4\linewidth]{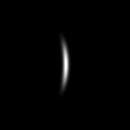}
}
\centering
\subfigure[$l/w = 8$]{
 \label{fig5a:subfig:c}
 \includegraphics[width = 0.4\linewidth]{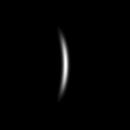}
}
\centering
\subfigure[$l/w = 9$]{
 \label{fig5a:subfig:d}
 \includegraphics[width = 0.4\linewidth]{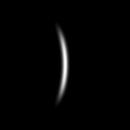}
}

\caption{Four arc templates with integer $l/w$ ratios of 6, 7, 8, and 9, 
respectively, produced by the ray-tracing simulations.}
\label{fig5a:subfig}
\end{figure}

We adopt a hybrid approach to simulate the arcs: simulate a representative set of arcs via 
ray-tracing with a range of $l/w$ ratios and surface brightnesses and then 
``paint" these template arcs onto the background images. This approach keeps the
advantages of both methods: realistic arc rendering and fast performance. First, we perform ray-tracing by 
 using simulated cluster lens with a NFW profile and a simulated background source with a 
Sersic profile. Second, we fine tune the distance from the source to the 
caustic line of the lens and carefully measure the $l/w$ ratio of the formed 
arcs. We keep those arcs with $l/w$ ratio that are closest to integer values as 
templates, as shown in Figure~\ref{fig5a:subfig}. We then create many additional
simulated arcs by arbitrarily rotating the template images and by adjusting the 
total flux as desired.  These arcs are then inserted into both simulated and actual CLASH images
for our arcfinder performance testing. A detailed discussion on the general detectability of arcs
as a function of source properties can be found in \citet{men08}.

%\clearpage

\begin{figure}
\centering
\subfigure[Simulated arcs]{
 \label{fig5b:subfig:a}
 \includegraphics[width = 0.4\linewidth]{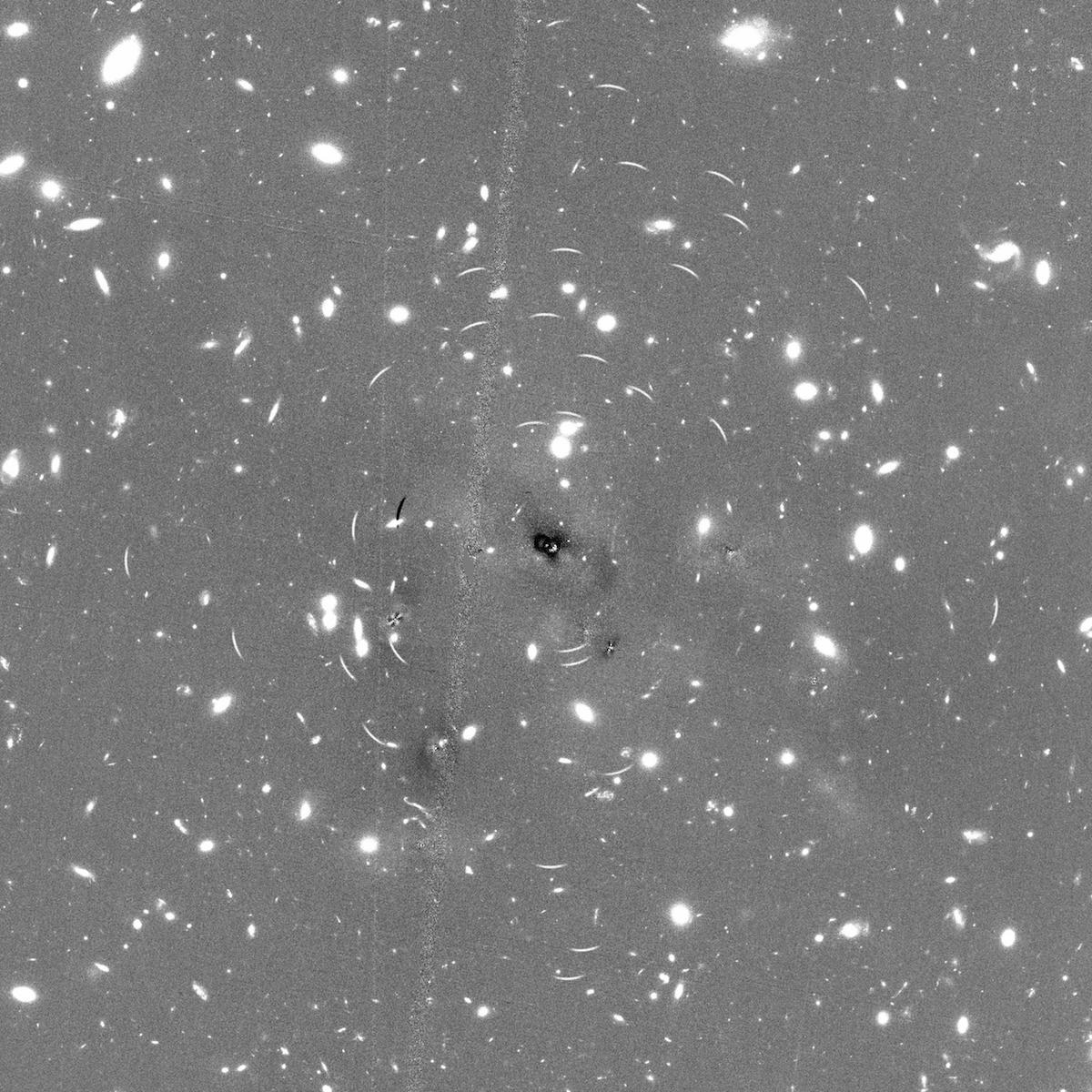}
}
\centering
\subfigure[Detected arcs]{
 \label{fig5b:subfig:b}
 \includegraphics[width = 0.4\linewidth]{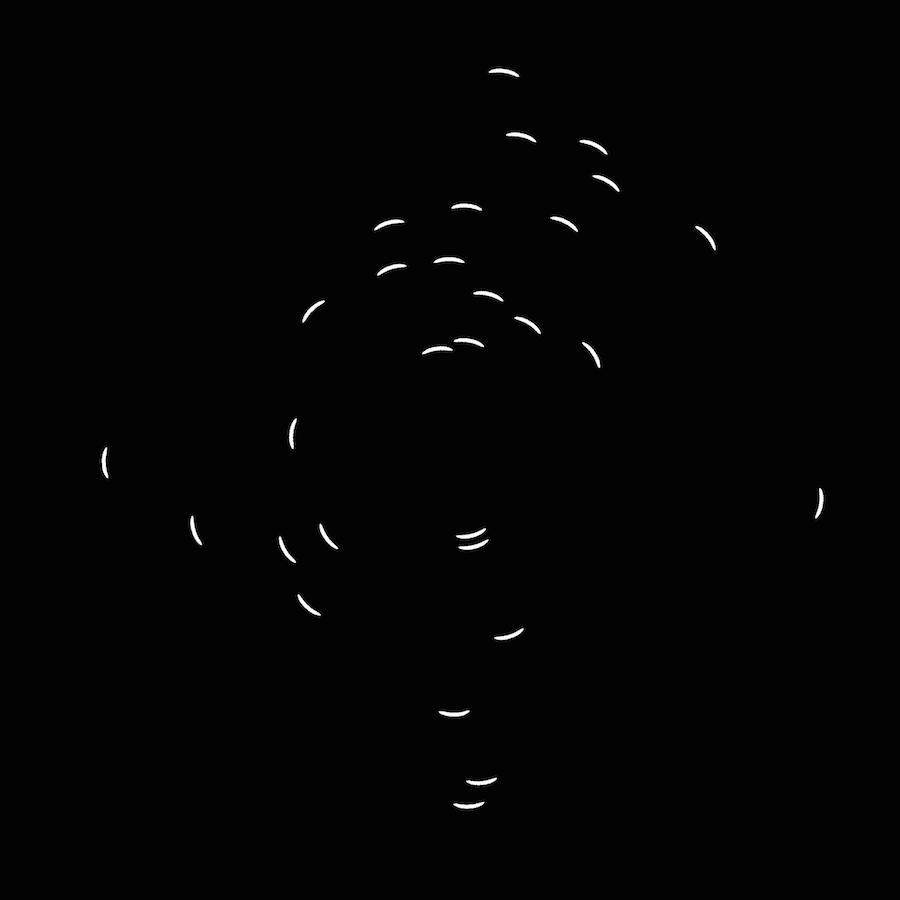}
}
\caption{(a) CLASH detection image for Abell 1423 shown with 30 ``painted" arcs with $l/w = 7$. The brightest cluster galaxy and a handful of its satellites are first subtracted off before the arcfinder is run. (b) The arcs that are detected are shown. The FOV of both images is $2.7' 
\times 2.7'$.} 
\label{fig5b:subfig}
\end{figure}

In order to accurately determine the arc detection completeness, we must 
account for the effects such as light contamination from 
cluster galaxies, variable sky background and instrumental noise. We accomplish this by inserting
the arc templates into actual CLASH 
detection images (a weighted co-addition of all of the ACS and WFC3-IR images for
a given cluster). An example of a CLASH detection image, with the brightest cluster galaxy subtracted out, is shown in 
Figure~\ref{fig5b:subfig}. We simulate a total of 14700 arcs spread over 7 
different $l/w$ values and 7 different total flux values. For the purpose of computing a measure of the algorithm's detection
completeness, we only look at the fraction of simulated arcs that are detected, even though we are inserting the simulated arcs
into real cluster data with real arcs. The completeness is then just the ratio of the number of the simulated arcs detected to
the total number of arcs simulated. 
The inverse of this ratio, $f_{incom} = N_{sim}/N_{det, sim}$, is then the 
multiplicative factor that we will apply to any raw arc count to correct for 
incompleteness.

We also utilize the F814W CANDELS \citep{gro11,koe11} 
images in the false positive rate test because there are no strong lensed 
sources in the CANDELS fields. We select images from the CANDELS ``Wide 
programs'' (e.g. the UDS and COSMOS fields), which have the similar total exposure times
as the CLASH detection images ($\sim50000s$) and split these mosaics into smaller images that 
match the angular size of the CLASH co-added images. We run the arcfinder on the CANDELS
data and compute the surface density of detections as functions of both the $l/w$ threshold and the
total arc length, $l$. This comprises the basis for our
false positive correction function.

\begin{figure}
\centering
\subfigure[]{
 \includegraphics[width = 0.45\linewidth]{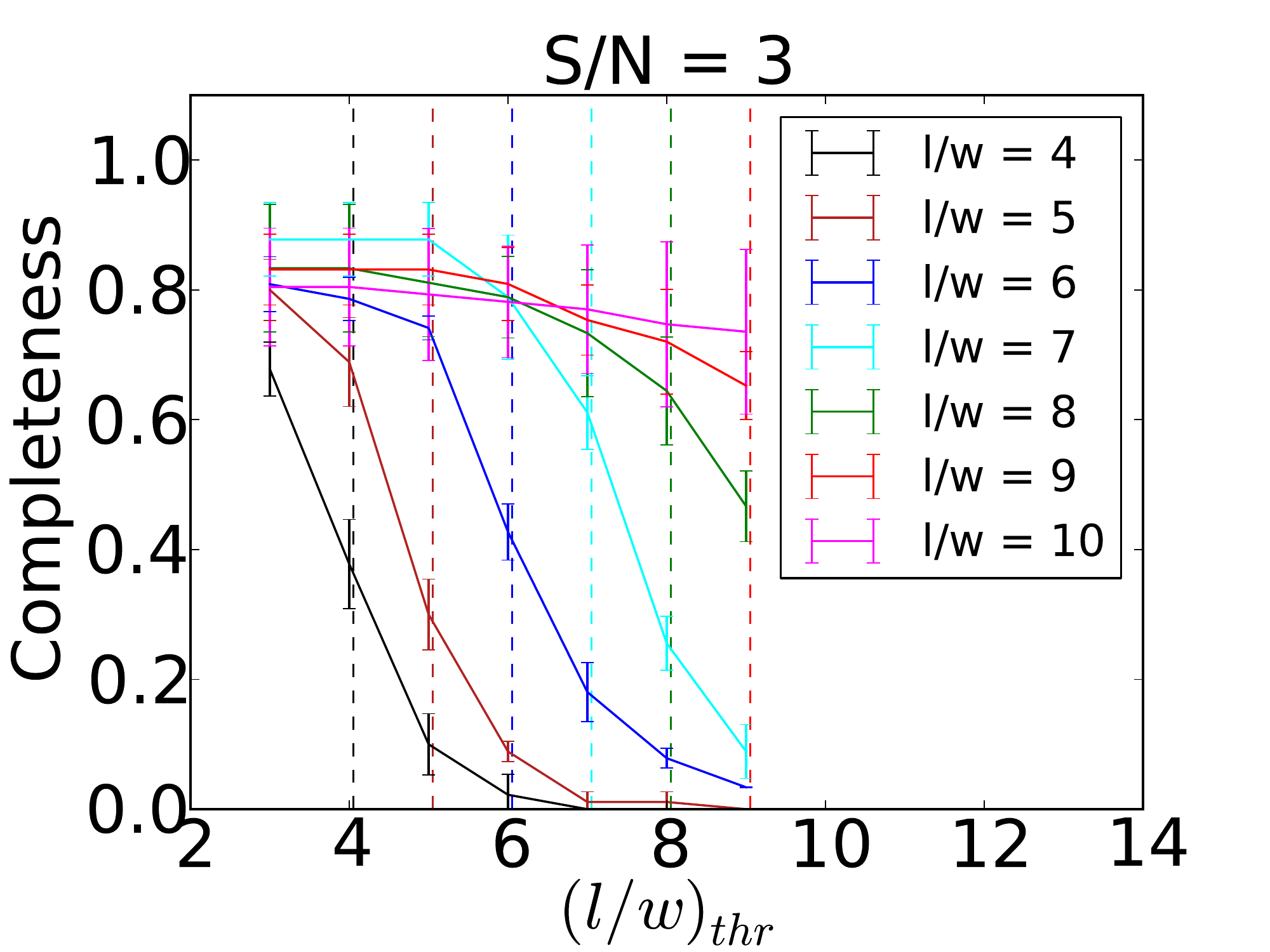}
}
\centering
\subfigure[]{
 \includegraphics[width = 0.45\linewidth]{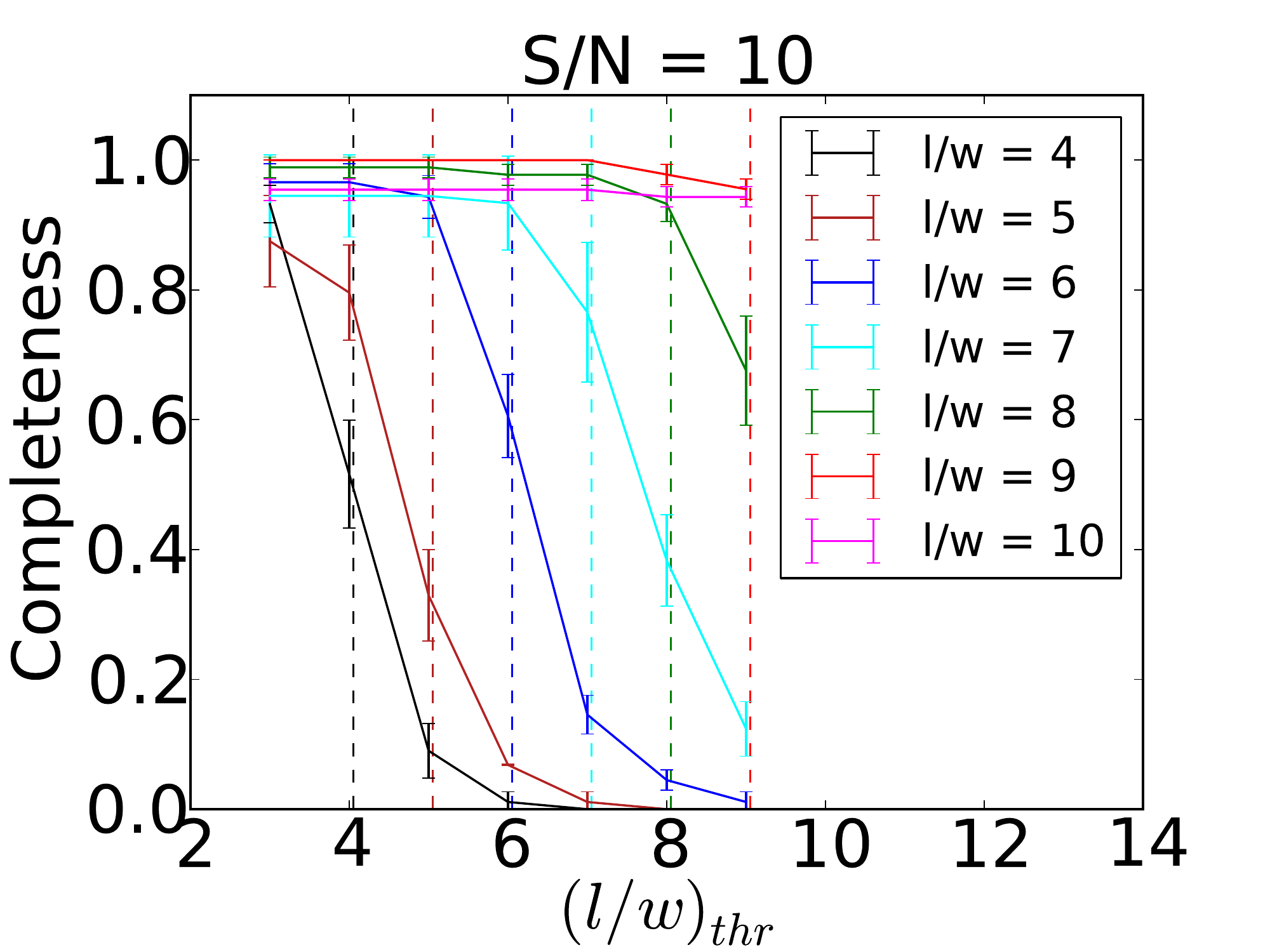}
}

\caption{Figure (a), (b) show the completeness as a function of the $l/w$
threshold for 7 different true $l/w$ ratios at two different $S/N$ levels. The 
dashed lines indicate the $l/w = (l/w)_{thr}$ and the errorbars denote the 
$1\sigma$ $rms$ error.}
\label{fig5c}
\end{figure}

\begin{figure}
\centering
\subfigure[]{
 \label{fig5d:subfig:a}
 \includegraphics[width = 0.45\linewidth]{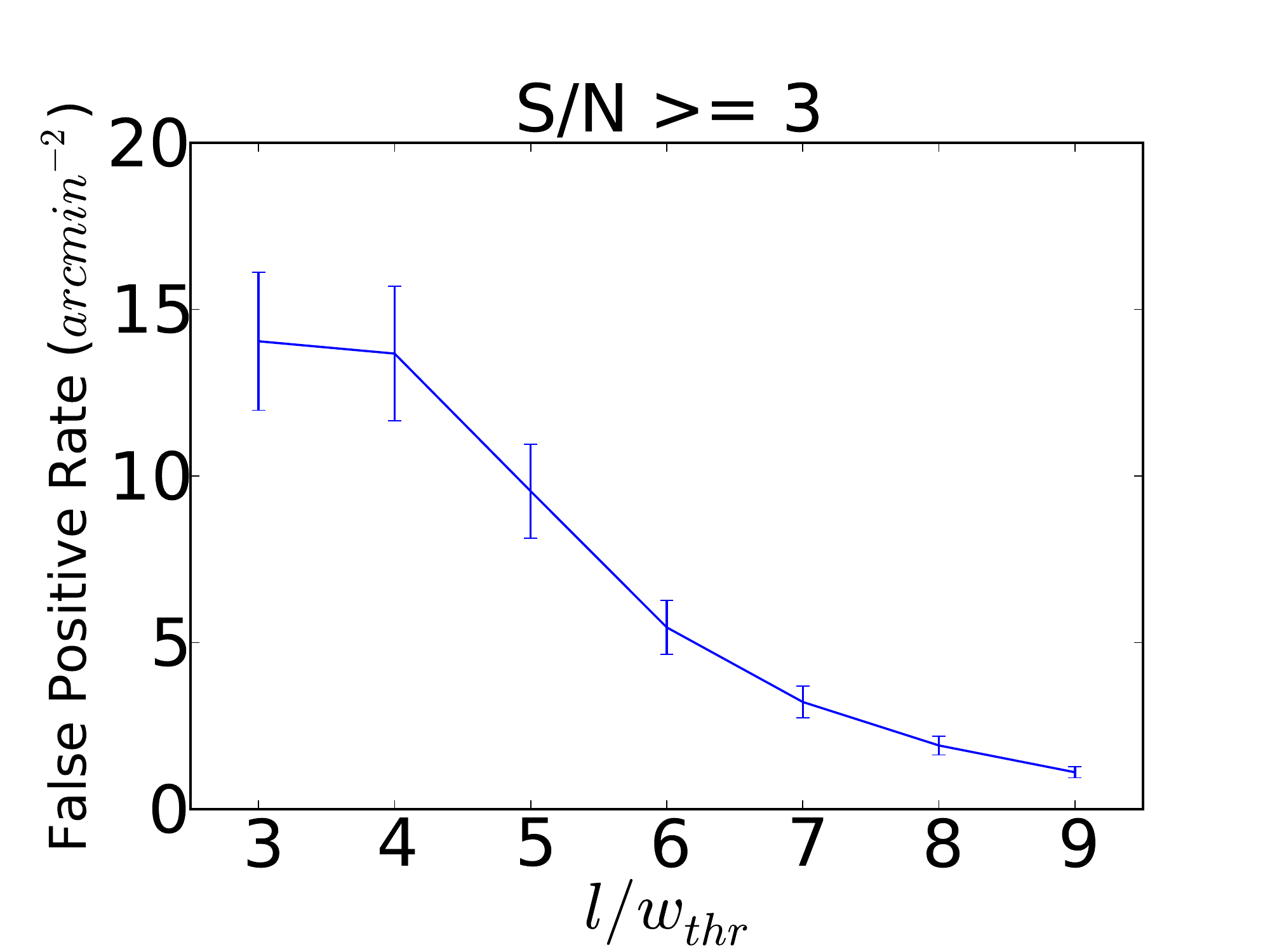}
}
\centering
\subfigure[]{
 \label{fig5d:subfig:b}
 \includegraphics[width = 0.45\linewidth]{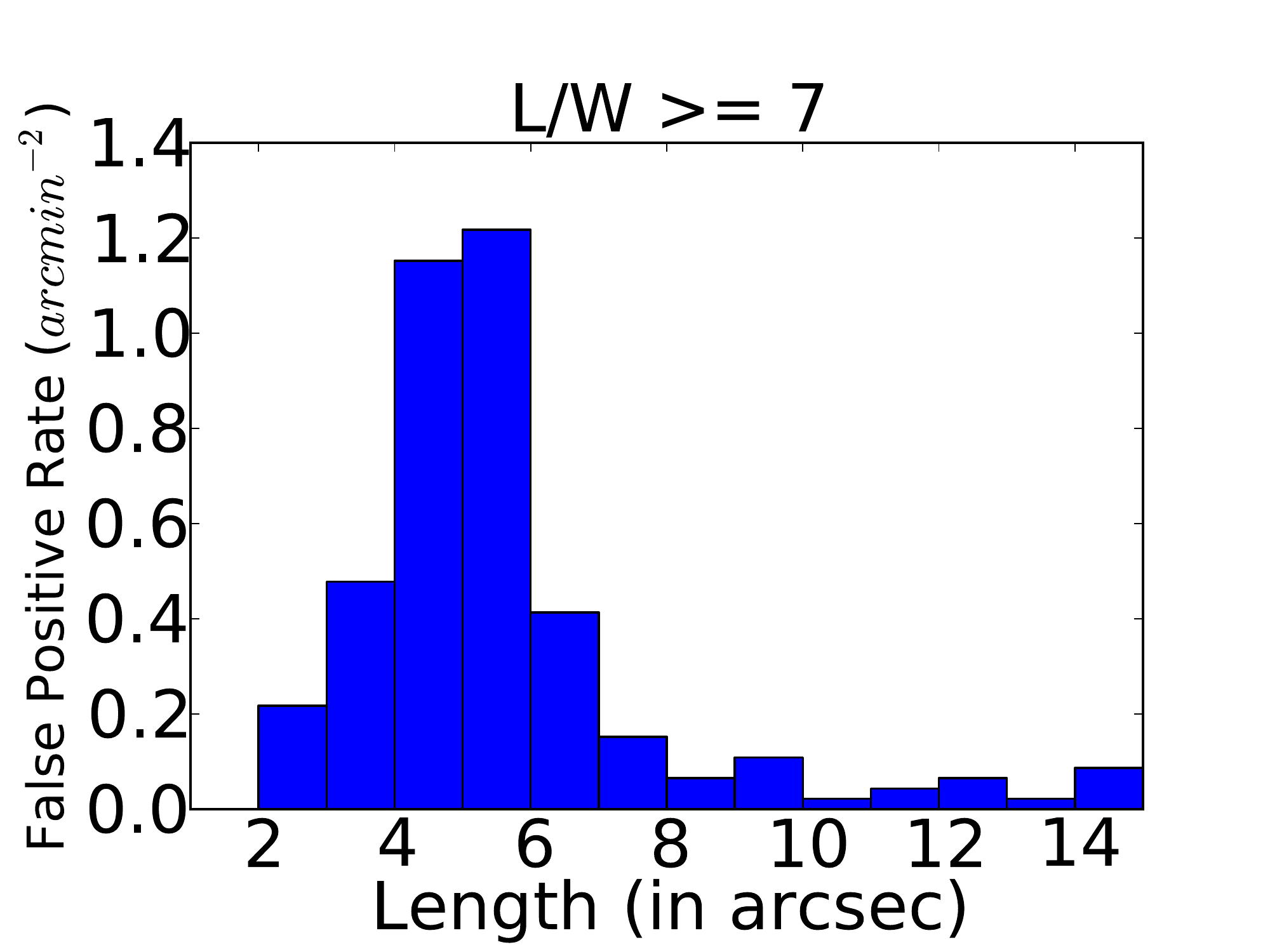}
}
\caption{(a) The false positive rate per unit area as a function of the 
detected $l/w$ threshold for arcs with SNR $\ge$ 3. (b) The false positive rate per unit area as a function of 
the arc length for arcs with $l/w \ge 7$. Results based on running the arcfinder
 on CANDELS data. The errorbars denote the $1\sigma$ $rms$ error.}
\label{fig5d:subfig}
\end{figure}

We use the CLASH data with simulated arcs to measure our arcfinder completeness as a function
of arc length, $l/w$ ratio, and arc signal-to-noise ratio (SNR). Here we define 
the SNR of arc as: $\sum I_i / \sqrt{ \sum (\sigma^{2}_{i,bn} + I_i) }$, where 
$I_i$ is the intensity from the source at pixel $i$ and $\sigma^{2}_{i,bn}$ is 
the  combined variance due to the sky background and all sources of detector 
noise at pixel $i$. In our completeness test,  the total flux of each drawn arc
is adjusted to match the assigned SNR value. 
We use the CANDELS images to 
assess the arcfinder false positive rate. Figure~\ref{fig5c} shows the 
completeness versus the $l/w$ detection threshold, $(l/w)_{thr}$, at $S/N = 3,10$. The 
completeness remains at a high level ($> 80 \%$ for $S/N = 10$) when the 
$l/w \ge (l/w)_{thr}$; Figure~\ref{fig5d:subfig:a} shows the false positive rate in the CANDELS data 
as a function the $l/w$ detection threshold when the minimum arc length is set 
to $2''$. The detected number of false positives is slightly above 10 $\rm 
arcmin^{-2}$ at low $(l/w)_{thr}$, while it decrease rapidly as $(l/w)_{thr}$ 
increases. Figure~\ref{fig5d:subfig:b} shows the false positive rate as a 
function of the length of the objects when the $l/w$ threshold is set to 7. 
The number of the false positive detections peaks in the length bin 
$5'' \le l < 6''$. The spurious detections can, thus, be suppressed if we adopt a minumum 
length threshold of  $l\ \ge 6''$. 

We have not applied this minimum length threshold to our completeness test because the
identification of the arcs does not depend on the length (only depends on $l/w$ and $S/N$).
Moreover, the intensity gradient along the ridge line of the arc should be smaller than that 
in the perpendicular direction, which mean that the length measurement should be more immune
to the noise effects. To test that, we measure the ratio of the detected length to the true 
length of the simulated arcs. Figure~\ref{fig5e:subfig:a} shows the distribution of the ratio 
at three different $S/N$ levels. The dashed lines indicate the median value of the ratio. We can 
see that both the distribution and the median value remain statistically similar at different $S/N$ levels.

\subsection{Determination of the Optimal $l/w$ Detection Threshold}
\label{s3.2}

\begin{figure}
\centering
\subfigure[]{
 \label{fig5e:subfig:a}
 \includegraphics[width = 0.45\linewidth]{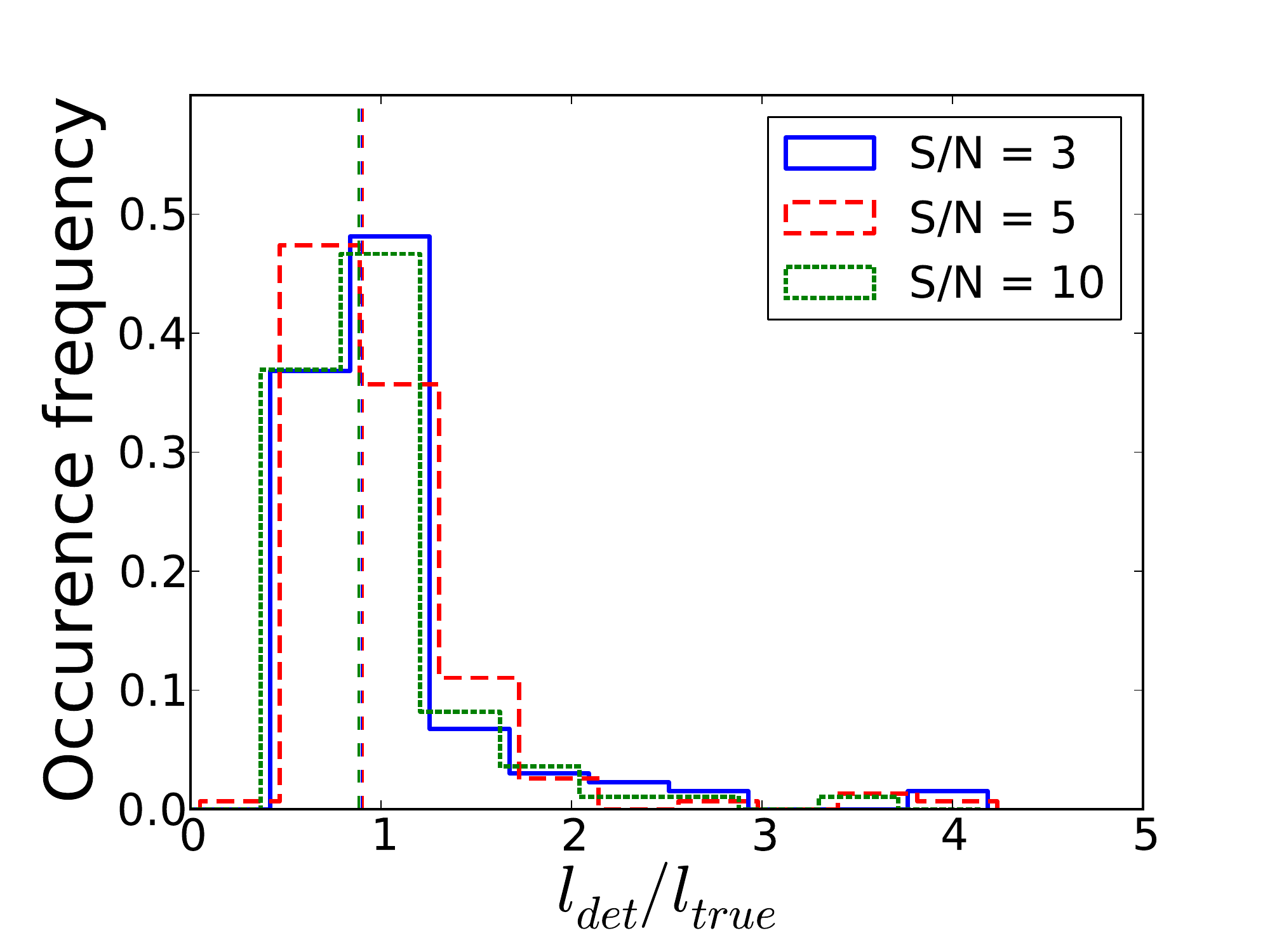}
}
\centering
\subfigure[]{
 \label{fig5e:subfig:b}
 \includegraphics[width = 0.45\linewidth]{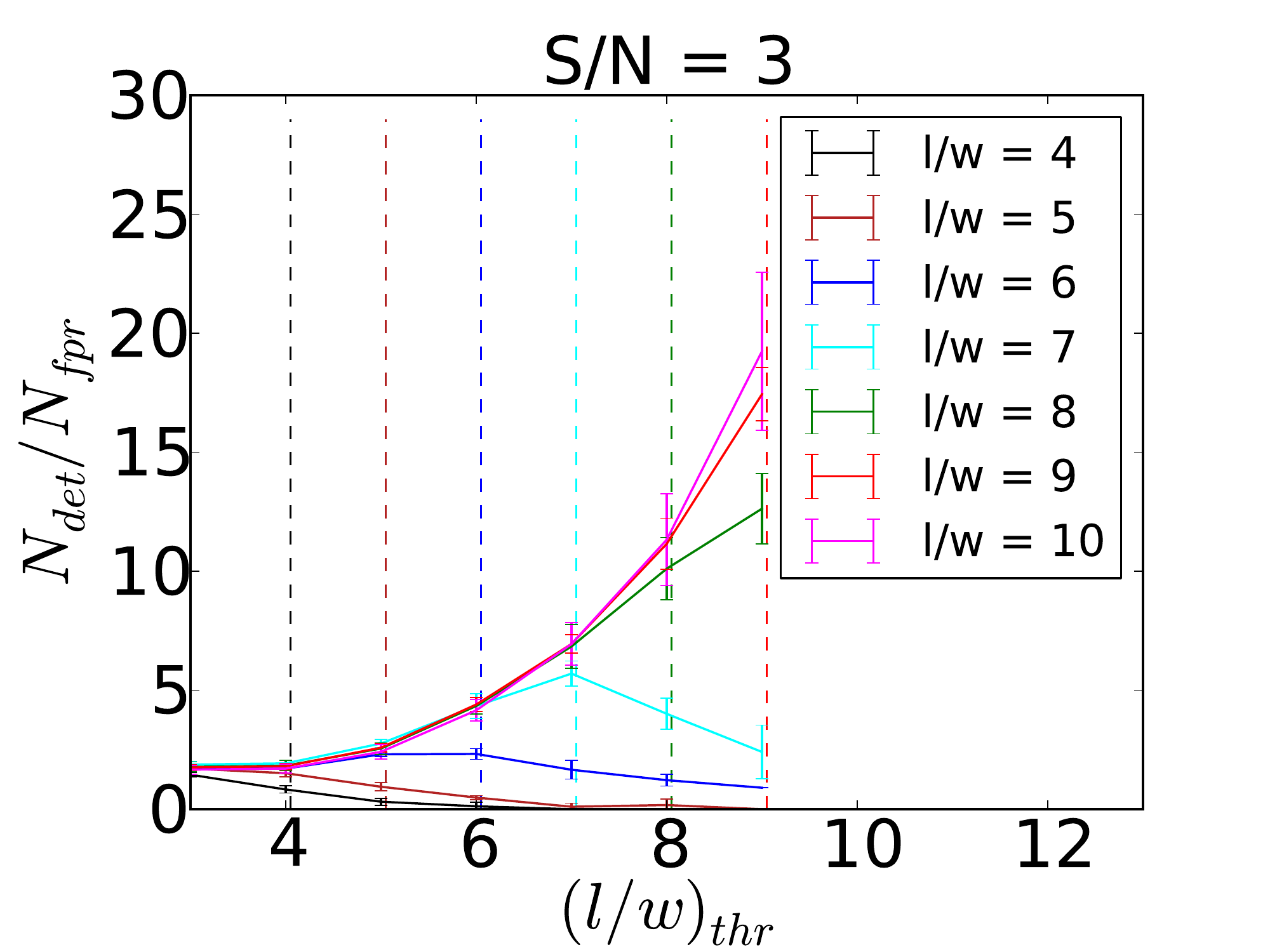}
}

\caption{Figure(a) shows the distribution of the ratio of the detected length 
to the true length at three different $S/N$ levels. The colored dashed lines
denote the median values of the distribution; figure(b) shows the ratio of the
number of detections per unit area to the number of false positive rate per unit
area as a function of the detected $l/w$ threshold. The dashed lines indicate
the $l/w = (l/w)_{thr}$ and the errorbars denote the $1\sigma$ $rms$ error.} 
\label{fig5e:subfig}
\end{figure}

In previous studies, the $l/w$ detection threshold is typically set to 7.5, 8 or
10. Generally, the reason to set a high $l/w$ threshold is to avoid the 
inclusion of highly elliptical and edge-on spiral galaxies into the arc sample. In general, 
the lower $l/w$ threshold one uses, the more contamination one gets. Hence it is
desirable to find a $l/w$ threshold that maximizes the completeness level and 
minimizes the false positive rate. We now use our measured estimates of the completeness and false positive rate 
as a function of the minimal $l/w$ to identify the optimal $l/w$ threshold to use in
the construction of our final arc catalog. We do this by identifying the smallest $l/w$
threshold at which the surface density of detected simulated arcs, $N_{det}$, exceeds the surface
density of false positive detections, $N_{fpr}$, by a factor of 5 or more. The results of this
test are shown in Figure~\ref{fig5e:subfig:b}. We find that the ratio $N_{det}(\ge (l/w)_{thr}) / 
N_{fpr}(\ge (l/w)_{thr})$ is always larger than 5 when the $l/w$
detection threshold is larger than 7. We thus adopt the $l/w$ detection threshold of 7  
in our analysis of the arc abundance. The  
false positive rate for $l/w \ge 7$ and an arc length threshold $l \ge 6''$ is $1.5 \pm 0.4$ arcmin$^{-2}$. 
We use this false positive rate to
correct our corresponding raw arc counts.

\subsection{$l/w$ Elongation Bias, Incompleteness and False Positive Rate 
Correction}
\label{s3.3}

\begin{figure}
\centering
 \includegraphics[width = 0.7\linewidth]{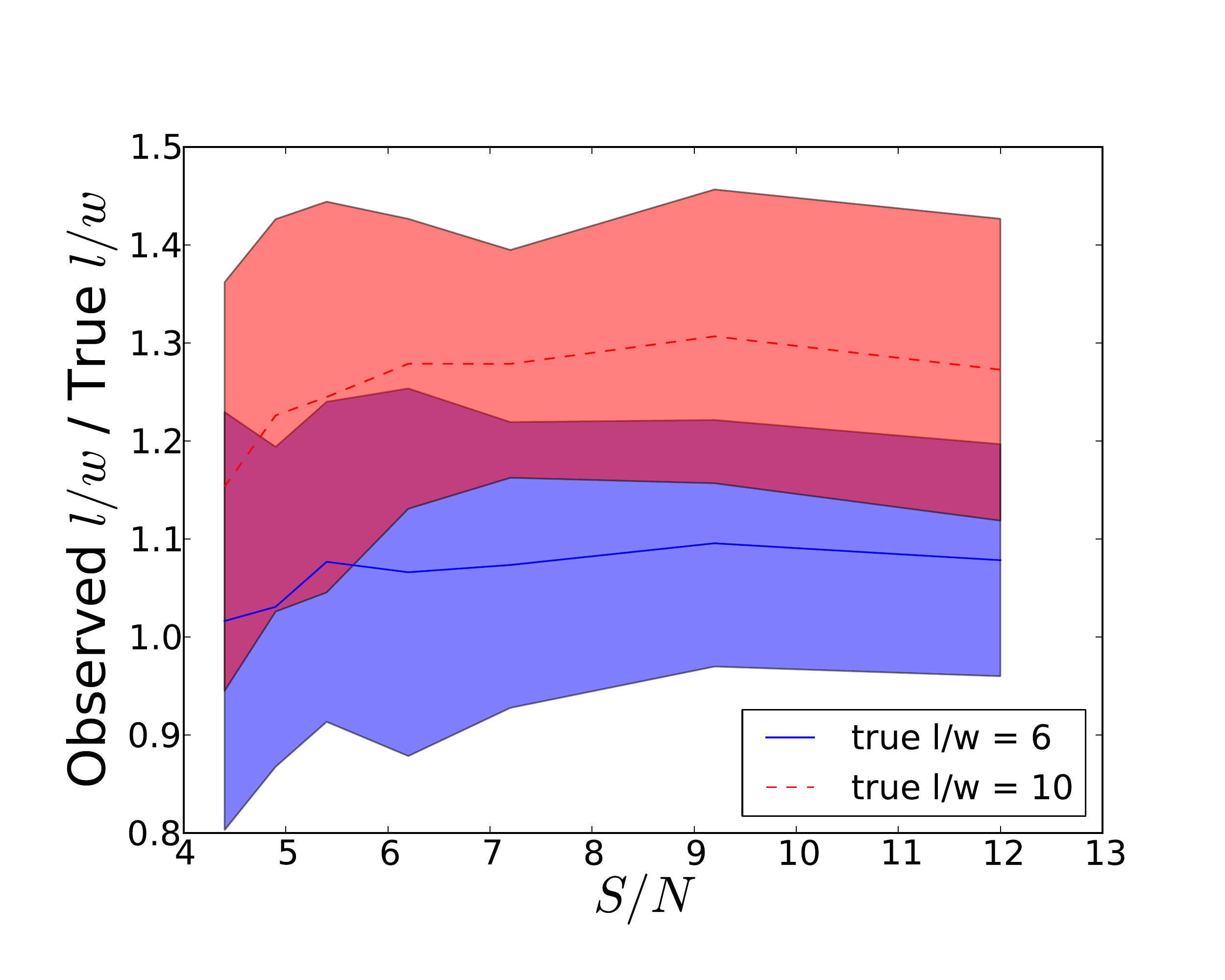}

\caption{The ratio of the observed $l/w$ to true $l/w$ as a function of the arc signal-to-noise ratio, $S/N$. The height of the shaded 
regions denotes the $1\sigma$ $rms$ errors on the ratio. }
\label{fig5f}
\end{figure}

There are three statistical corrections we need to apply to the raw counts of 
the giant arcs. First, the detected $l/w$ is not equal to the true 
$l/w$. The background noise and/or the segmentation boundaries of a detected 
object may systematically affect the determination of the $l/w$ ratio. We 
need to determine how the detected $l/w$ ratio deviates from the true $l/w$ 
ratio at different $S/N$ levels, and correct for this elongation bias in a 
statistical sense. For example, as shown in Figure~\ref{fig5f}, the detected 
$l/w$ ratio of arcs can be biased high by image noise, as the noise tends to 
make arcs appear thinner than they actually are. Second, we need to apply the incompleteness correction (presented above)
as there will always be some real arcs that are missed by our detection algorithm. Third, we need to apply a
false positive correction as there are always some objects misidentified by the
arcfinding algorithm. We apply all these three corrections in deriving the final observed and simulated arc 
abundances.

To compute the $l/w$ elongation bias correction, we collect all the detected 
arcs with measured $l/w \ge$ 6.5 \footnote{In practice, we set the $l/w$ 
threshold to be 6.5 instead of 7. The bias correction is then done by comparing
the arcs with detected ratios in the range 6.5 $\le l/w<$ 7.5 to the number with 
true $l/w = $ 7.}, and assign them to one of the three bins: 6.5 $\le l/w <$ 7.5,
7.5 $\le l/w<$ 8.5 and $l/w\ge $ 8.5. We also assign their corresponding true $l/w$ 
ratios into one of the three bins: $l/w =$ 7, $l/w =$ 8 and $l/w \ge $ 9. We 
further split each bin into three sub-bins by their $S/N$ ratios: $S/N<$ 5, 
5 $\le S/N<$ 10 and $S/N \ge $ 10. We then calculate the mean value and standard 
deviation of the correction factor for the elongation bias 
$f_{bias} = N_{true} / N_{det}$, where $N_{true}$ and $N_{det}$ are the number 
of simulated arcs and detected arcs in each bin, respectively. 

\begin{figure}
\centering
\subfigure[Elongation bias correction]{
 \label{fig5g:subfig:a}
 \includegraphics[width = 0.45\linewidth]{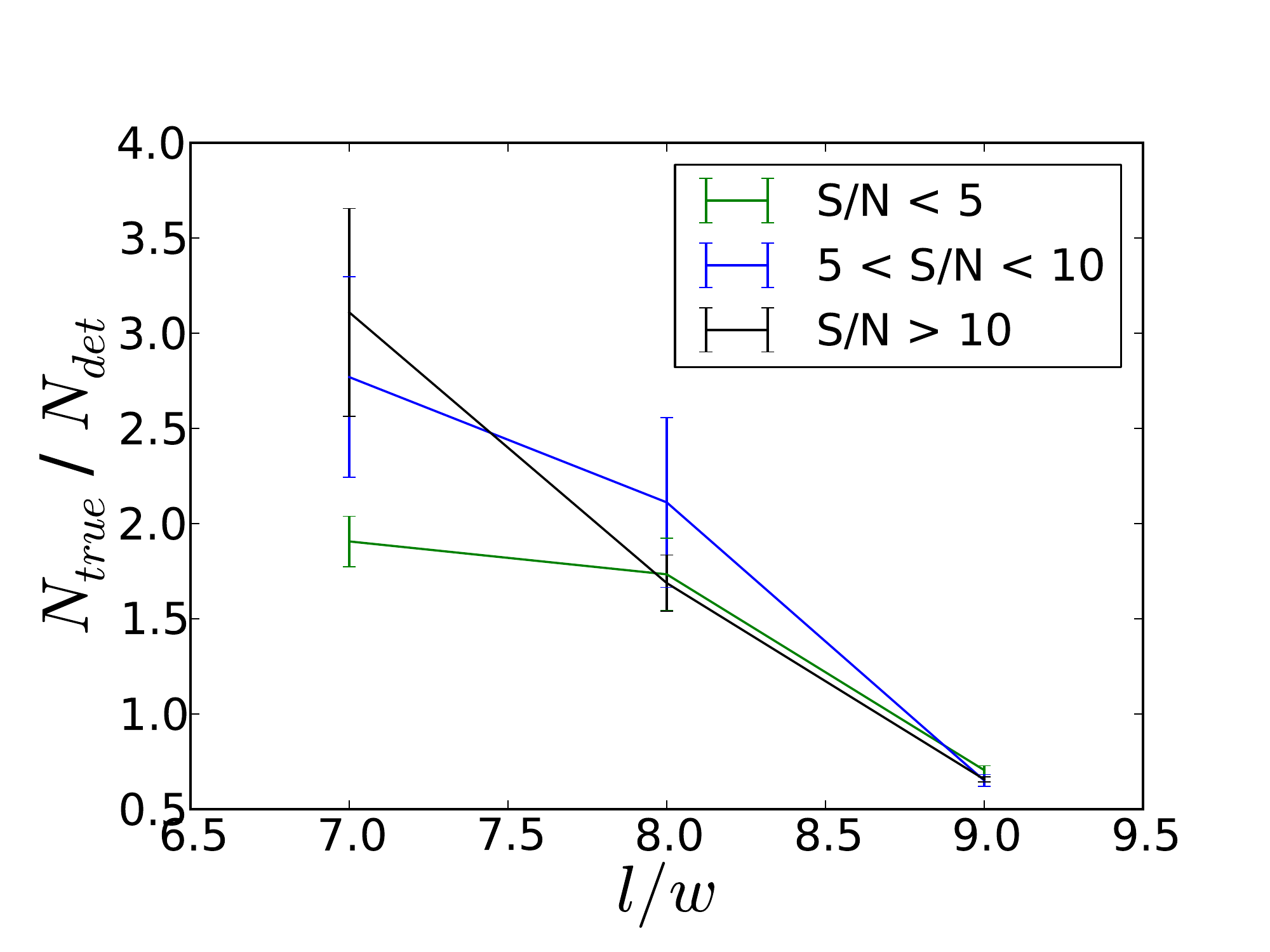}
}
\centering
\subfigure[Incompleteness correction]{
 \label{fig5g:subfig:b}
 \includegraphics[width = 0.45\linewidth]{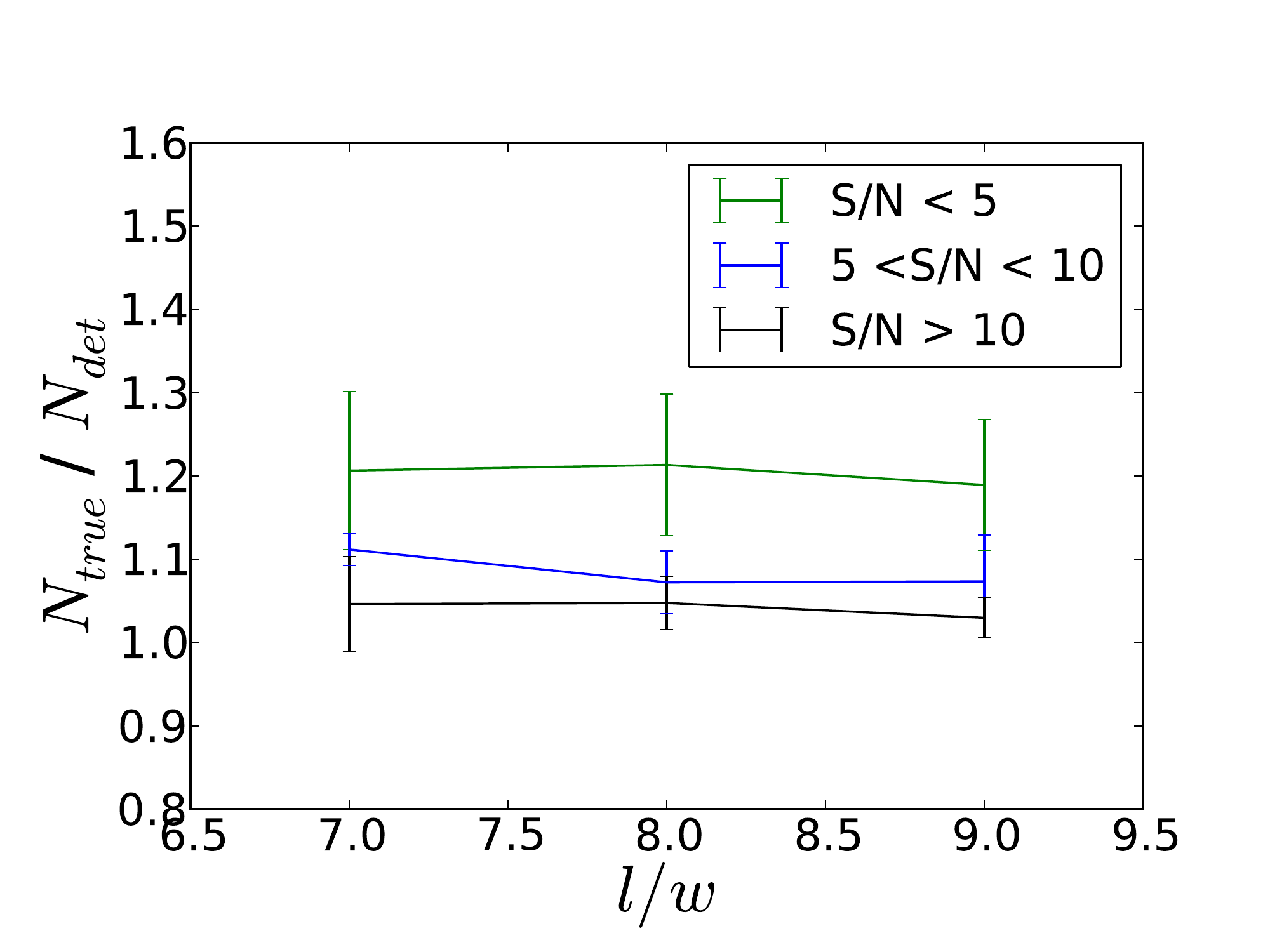}
}
\caption{All the detected arcs with $l/w \ge 6.5$ are assigned into three 
$l/w$ bins (horizontal axis). (a) shows the elongation bias correction factors 
in three $l/w$ bins; (b) shows the incompleteness correction factors in three 
$l/w$ bins. The errorbars denotes the $1\sigma$ $rms$ error.}
\label{fig5g:subfig}
\end{figure}

The true arc count, $N_{tru}$, is then computed as follows:

\begin{eqnarray}
&& N_{true} = \sum_{i} N_{det,i} \times f_{bias,i}\times f_{incom,i} - N_{false} \nonumber\\
&& \sigma_{true} = \sqrt{ N_{true}\times \sum_{i} 
[(\frac{\sigma_{bias,i}}{f_{bias,i}})^2 + (\frac{\sigma_{incom,i}}{f_{incom,i}} )^2] + \sigma_{false}^2}
\end{eqnarray}

\noindent where $N_{det,i}$ is the observed number of arcs in each bin and $i$ 
goes over all the bins. As shown in Figure~\ref{fig5g:subfig:a} and 
Figure~\ref{fig5g:subfig:b}, most of the measured $l/w$ are biased high, 
especially for the arcs with intrinsically low $l/w$.  The completeness remains above 80\% for all the cases. 
Here, biased high means that arcs with ``true" low $l/w$ have their $l/w$ values systematically overestimated. The 
mean ratio of the observed $l/w$ to the ``true" $l/w$ also appears to be dependent on
the true $l/w$ ratio as shown in Figure~\ref{fig5f}.

\subsection{Comparison of Arcfinder's Performance with a Previous Code}
\label{s3.4}

We compare the arc detection efficiency of our arcfinder to
that of the only publicly available arc-finding code from \citet{hor05}. 
We simulate a large amount of arc with different $l/w$ ratios and draw gaussian
random noise onto the arcs to produce simulated arc images with 7 different 
$S/N$ levels. We run both arcfinding algorithms on these simulated data sets.

\begin{figure}
\centering
 \includegraphics[width = 0.7\linewidth]{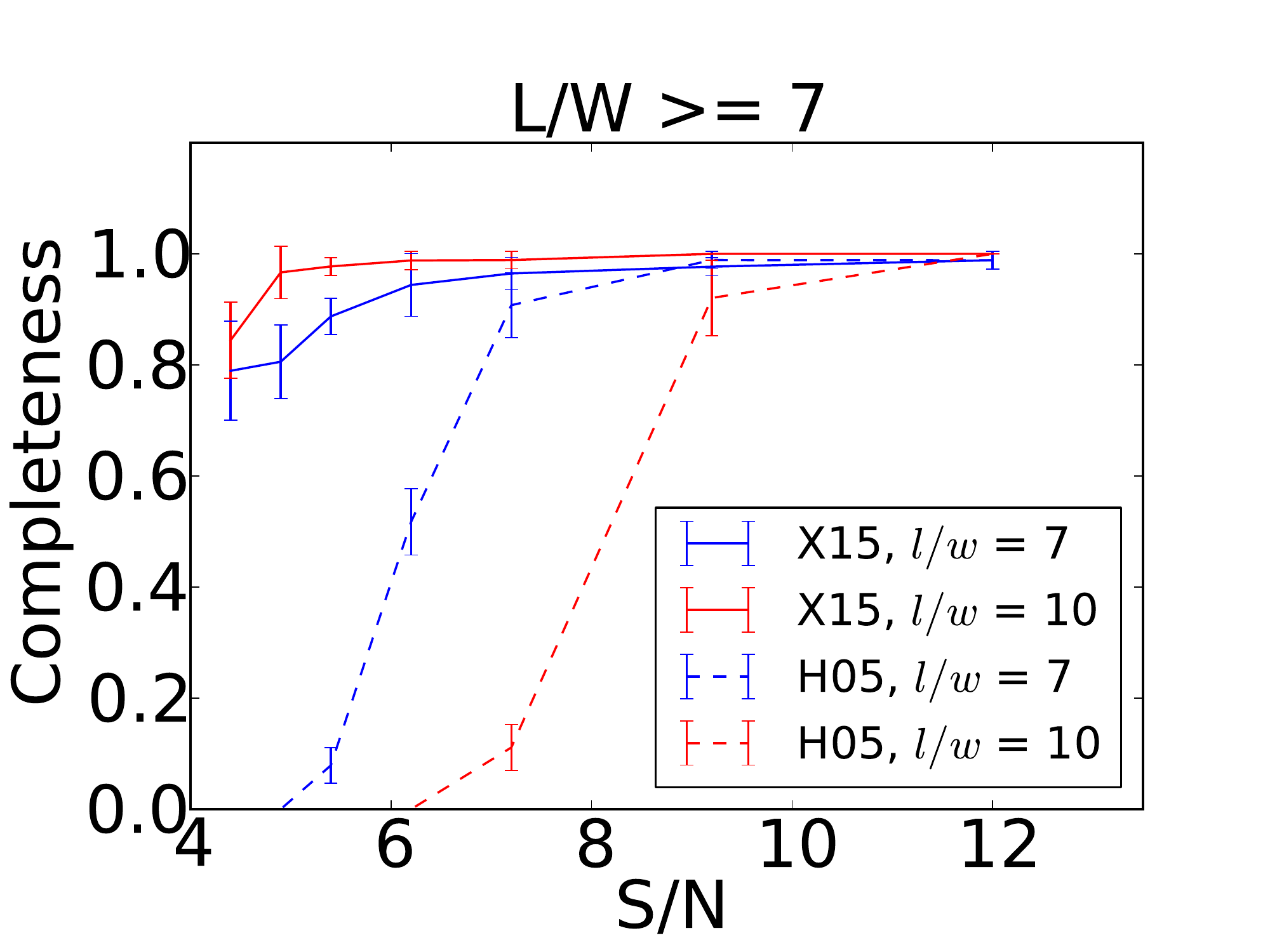}

\caption{Comparison of the arc detection completeness between our arcfinder 
(solid curves) and the \citet{hor05} arcfinder (dashed curves) for arcs with 
$l/w = $ 7, 10.  The errorbars denote the scatter. For bright arcs with 
$S/N > 10$ both arcfinders maintain a high detection rate ($> 90\%$); while for 
faint arcs with lower $S/N$ levels, our arcfinder exhibits considerably higher 
detection efficiency. The errorbar denotes the $1\sigma$ $rms$ error.}
\label{fig5h}
\end{figure}

Figure~\ref{fig5h} 
shows the detection rate versus the arc $S/N$ ratio level for arcs with true 
$l/w =$ 7, 10, at a detection threshold $l/w \ge$ 7. 
We have computed the $S/N$ ratio for detected arcs found using each of the 
algorithms using the definition 
given in \S\ref{s3.1}. For the bright arcs ($S/N 
>$ 10), the detection rates for both arcfinders remains high ($> 90\%$); for 
faint arcs (5 $\le  S/N <$ 10), the \citet{hor05} arcfinder's detection rate drops
rapidly, while our algorithm's detection efficiency remains higher than 90\%; 
for very faint arcs ($S/N <$ 5), our detection rate drops to about 80\%. The 
advantage of our intensity-gradient based arc-finding algorithm is nicely 
demonstrated in Figure~\ref{fig5h}, especially for the detection of large arcs 
with low-surface brightness. 

\section{Analyses of the CLASH Data}
\label{s4}

\subsection{Arc Statistics for the CLASH Sample}

\begin{figure}
\centering
 \includegraphics[width = 5.5in]{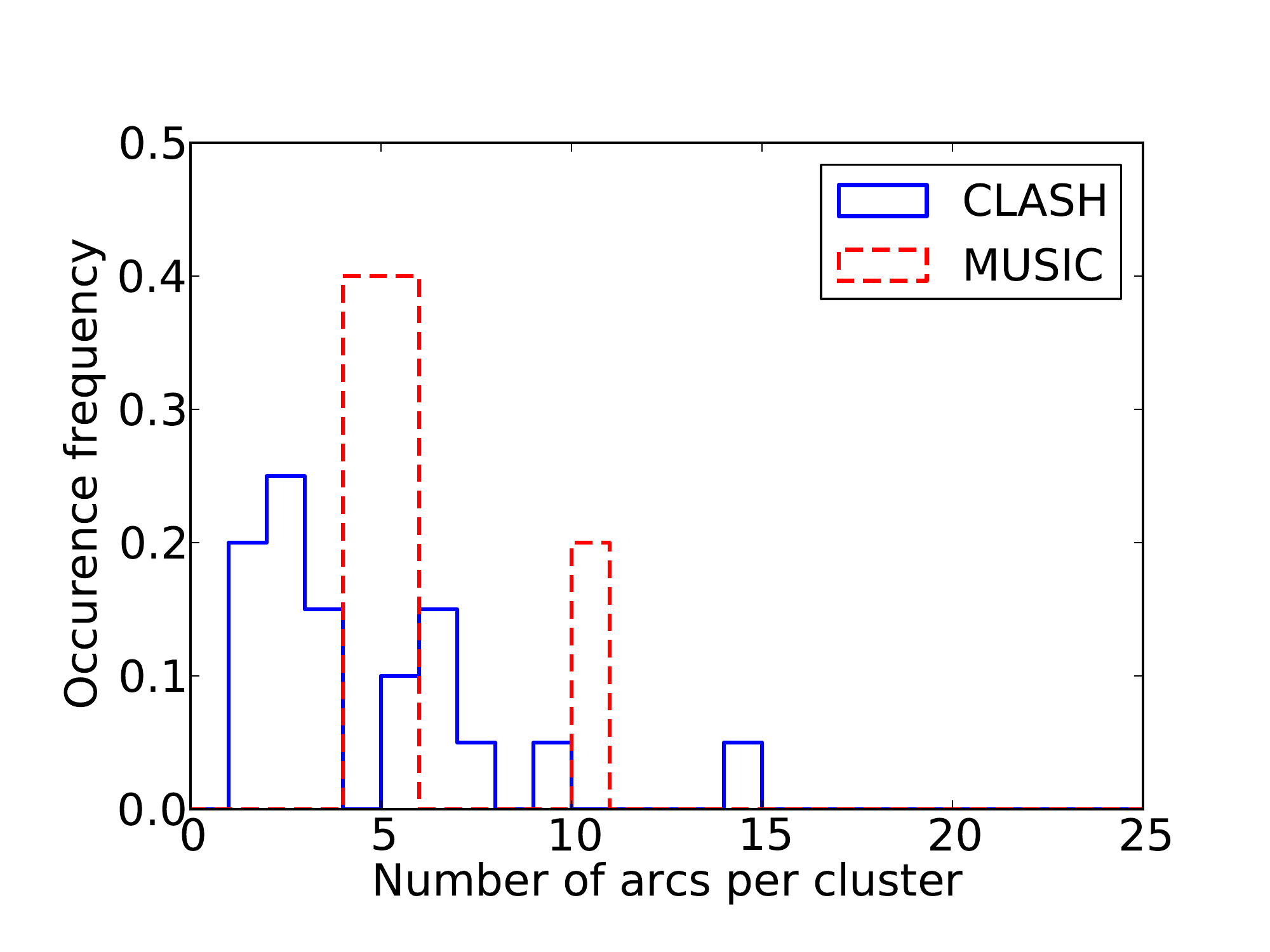}
\caption{The occurrence frequency of arcs per cluster for 20 X-ray selected 
CLASH clusters in blue and for 5 high lens magnification subsample of CLASH clusters in red. }
\label{fig6_1}
\end{figure}

The CLASH observations for each cluster consist of 16 broadband images 
(spanning the range $0.23\mu\ - 1.6\mu$) using the $WFC3/UVIS$, $WFC3/IR$, and
$ACS/WFC$ instruments onboard HST. The cluster properties are listed in Table~\ref{tab1}. We run 
our arcfinder on the detection ($ACS+WFC3/IR$) image created for each cluster. 
We detect a raw total of 187 arcs with $l/w \ge 6.5$ in 20 X-ray selected CLASH 
clusters. 
After applying our minimum arc length criterion $l \ge 6''$, the arc count drops to 81 giant arcs selected from the 
20 X-ray selected CLASH clusters. Correcting for the elongation bias and 
incompleteness brings the total number of detected arcs in 20 X-ray selected clusters 
is $104 \pm 12$. After further correcting for the false positive rate, we find 
a lensing efficiency  of $4 \pm 1$ arcs per X-ray selected 
cluster. Throughout this paper, the lensing efficiency 
denotes the number of arcs per cluster. 
There are 28 arcs with $l/w \ge 6.5$ and $l \ge 6''$ detected in the five 
high-magnification CLASH clusters, corresponding to a mean value $5 \pm 1$ arcs 
per cluster after all corrections are applied.  
Figure~\ref{fig6_1} shows the distributions of number of
arcs per cluster for the X-ray selected cluster sample and the high 
magnification cluster sample. 
Figure~\ref{fig6_2} shows the comparison of the detection images with the 
raw output of the arcfinder with $l/w > 7$ for five CLASH clusters.

\begin{figure*}
 \begin{minipage}[t]{0.5\linewidth}
  \centering
  \includegraphics[width = 0.9\linewidth]{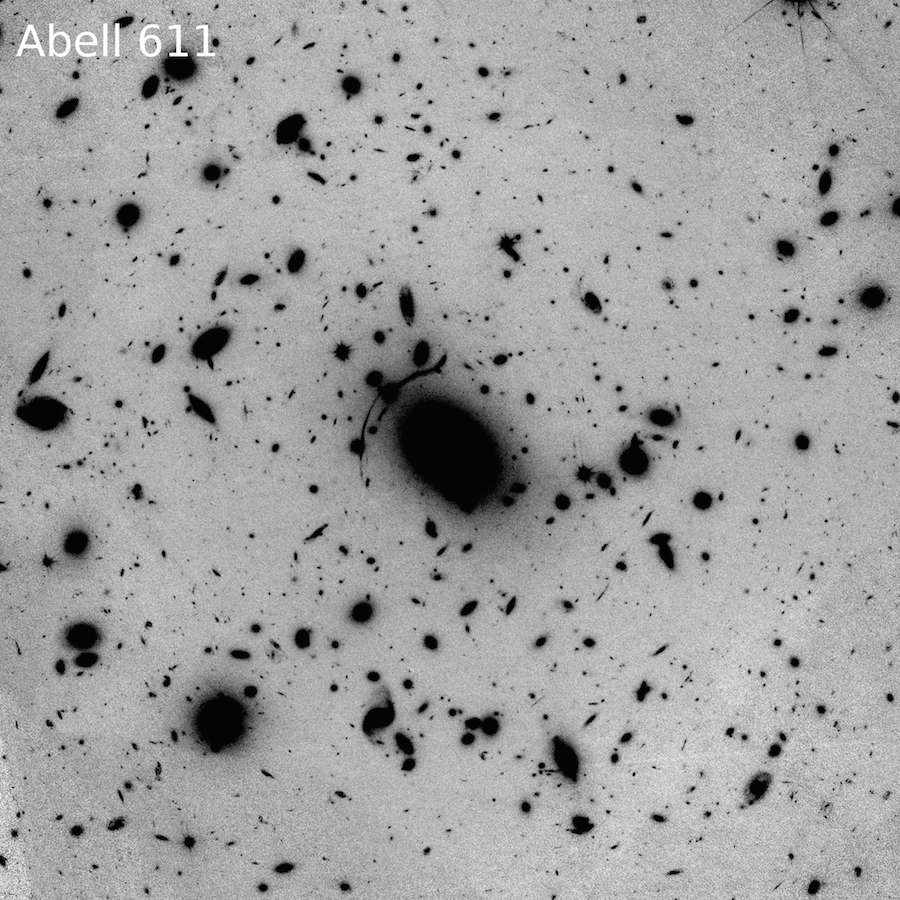}
 \end{minipage}%
 \begin{minipage}[t]{0.5\linewidth}
  \centering
  \includegraphics[width = 0.9\linewidth]{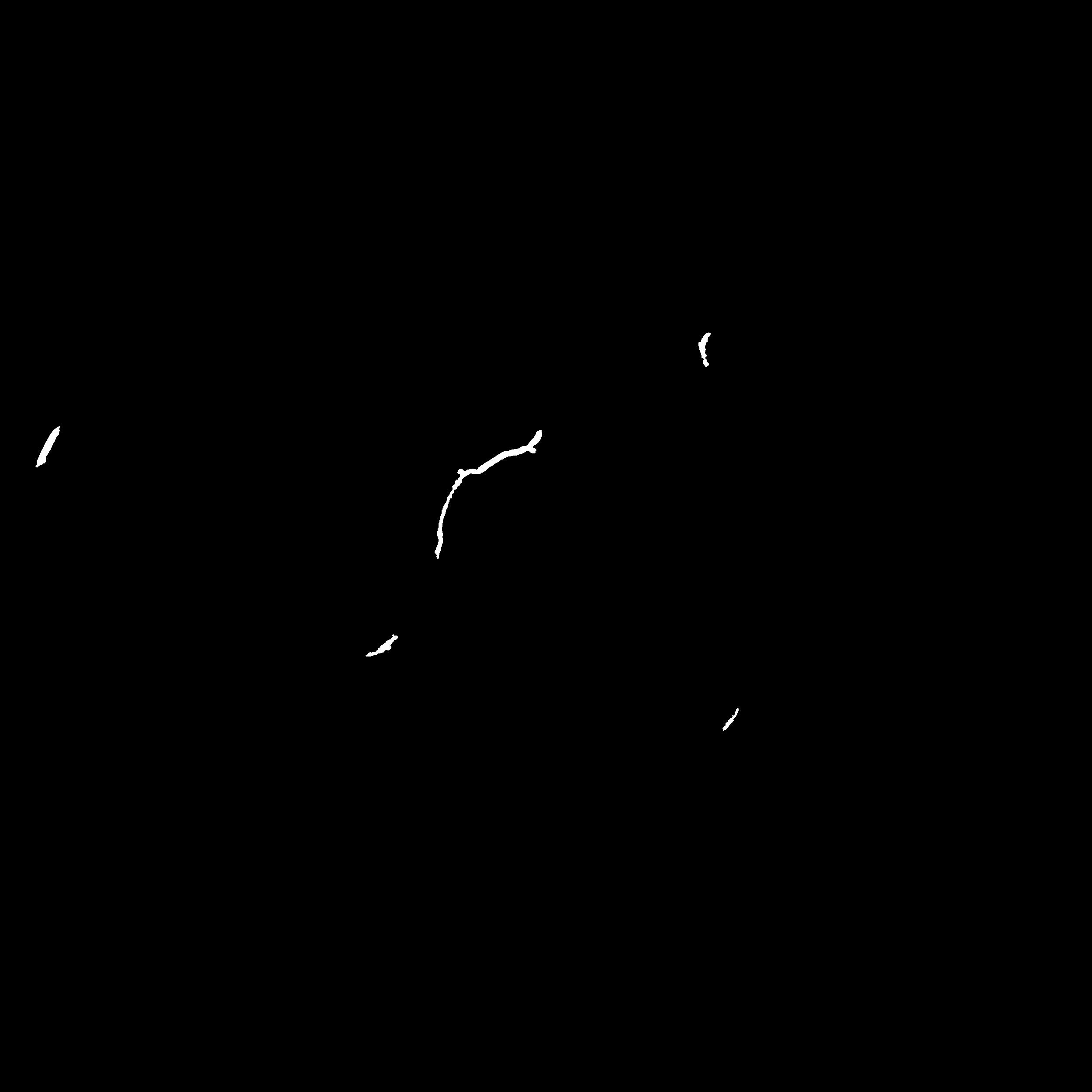}
 \end{minipage}

 \begin{minipage}[t]{0.5\linewidth}
  \centering
  \includegraphics[width = 0.9\linewidth]{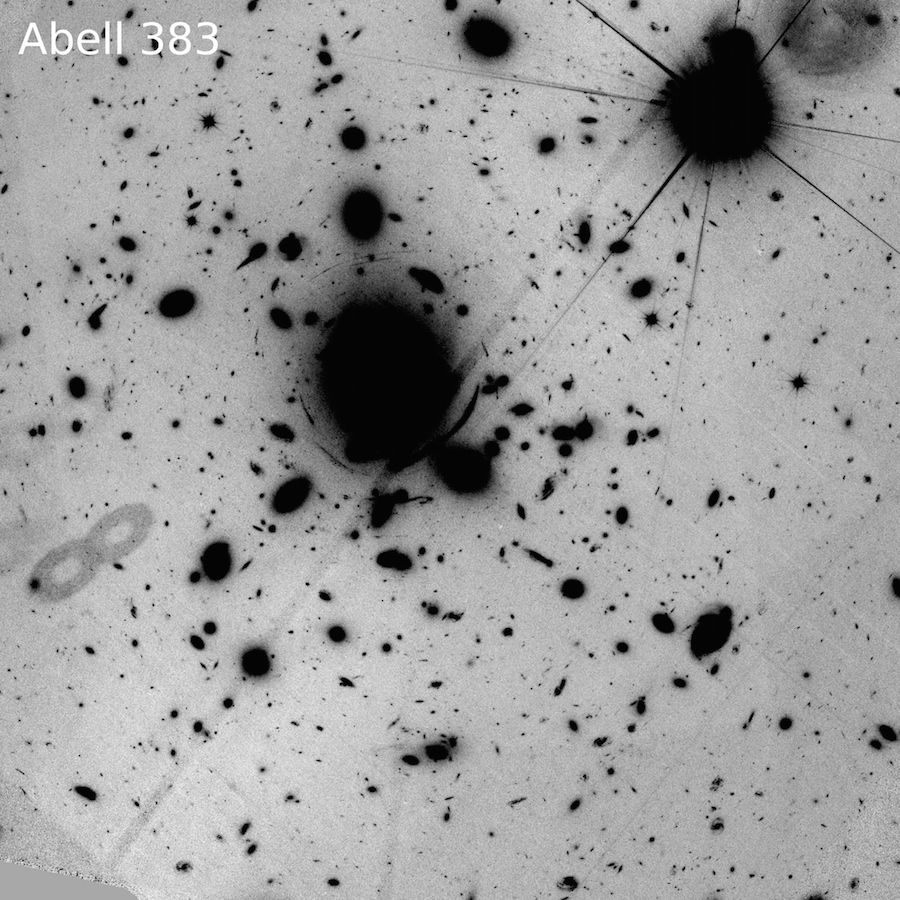}
 \end{minipage}%
 \begin{minipage}[t]{0.5\linewidth}
  \centering
  \includegraphics[width = 0.9\linewidth]{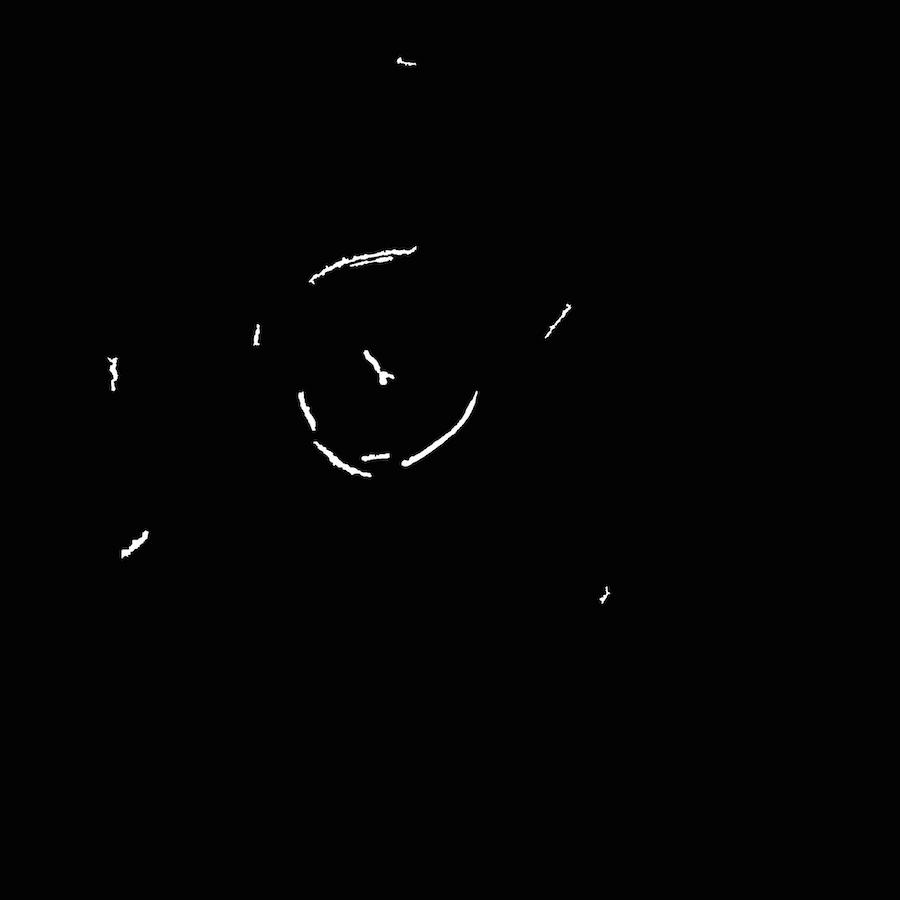}
 \end{minipage}

\begin{minipage}[t]{0.5\linewidth}
  \centering
  \includegraphics[width = 0.9\linewidth]{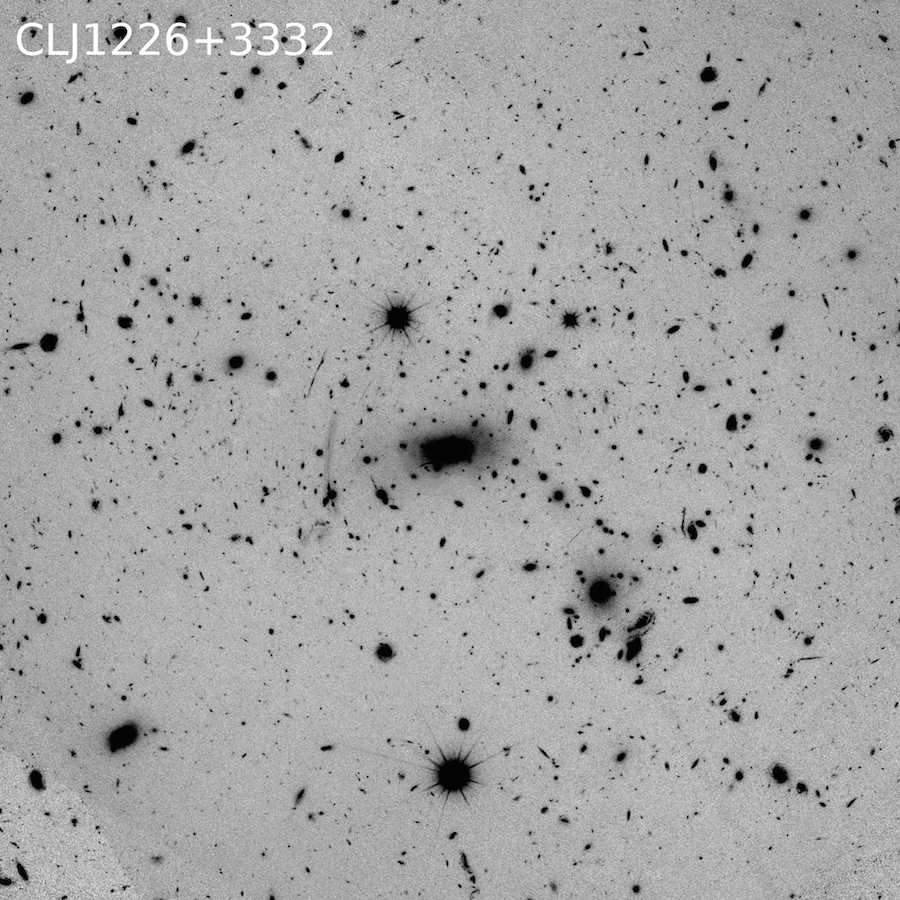}
 \end{minipage}%
 \begin{minipage}[t]{0.5\linewidth}
  \centering
  \includegraphics[width = 0.9\linewidth]{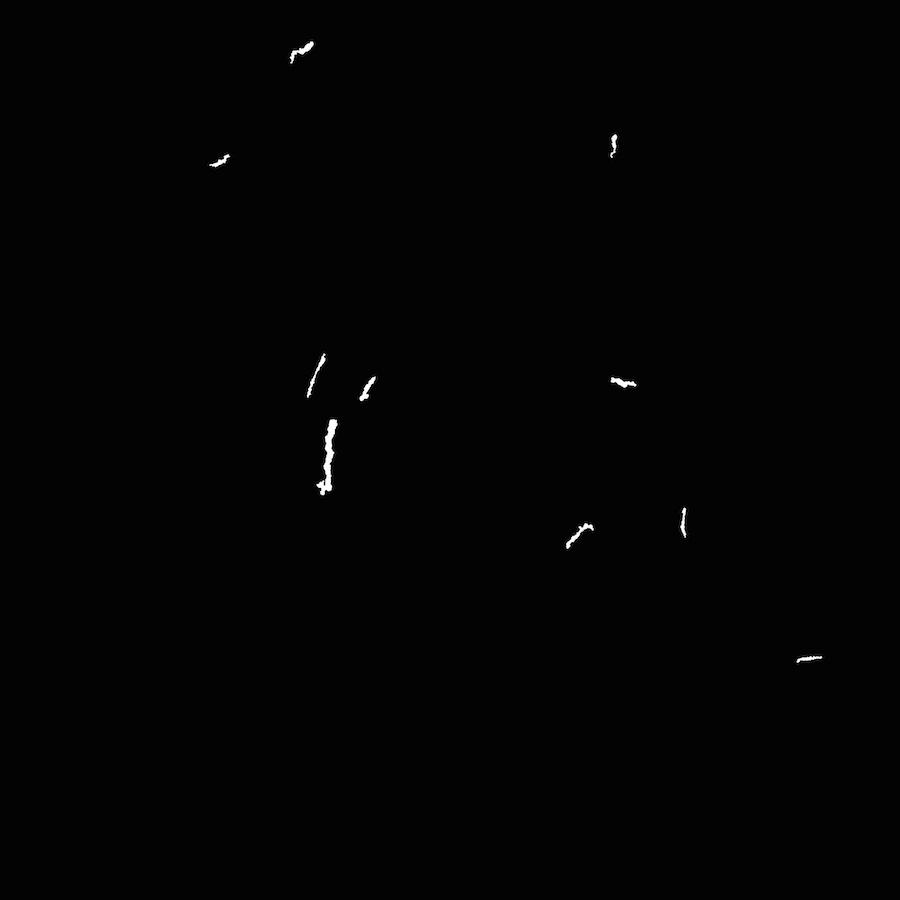}
 \end{minipage}

\caption{To be continued  }
\label{fig6_2}
\end{figure*}

\addtocounter{figure}{-1}

\begin{figure*}
 \begin{minipage}[t]{0.5\linewidth}
  \centering
  \includegraphics[width = 0.9\linewidth]{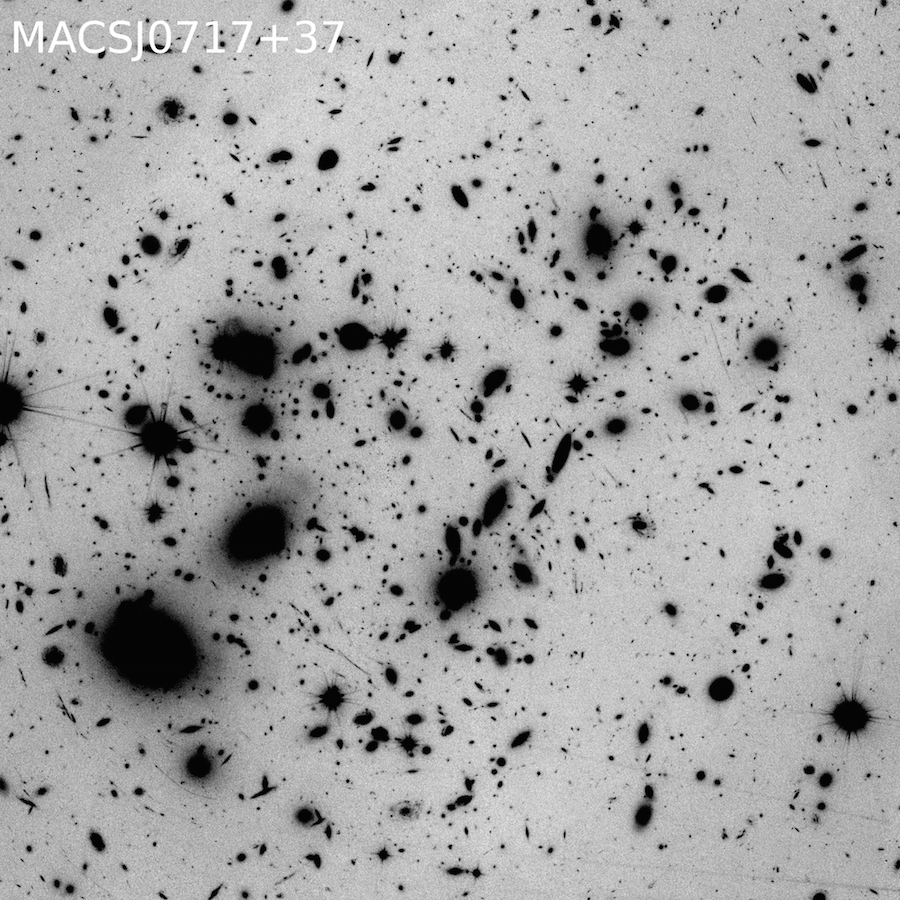}
 \end{minipage}%
 \begin{minipage}[t]{0.5\linewidth}
  \centering
  \includegraphics[width = 0.9\linewidth]{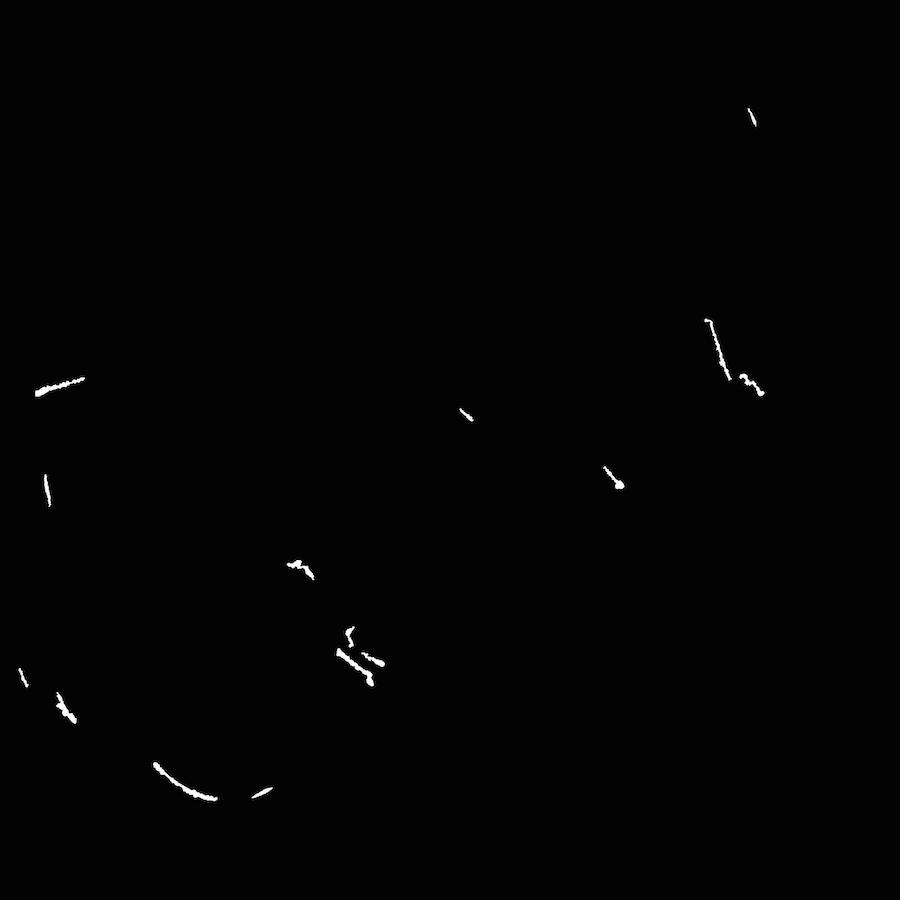}
 \end{minipage}

\begin{minipage}[t]{0.5\linewidth}
  \centering
  \includegraphics[width = 0.9\linewidth]{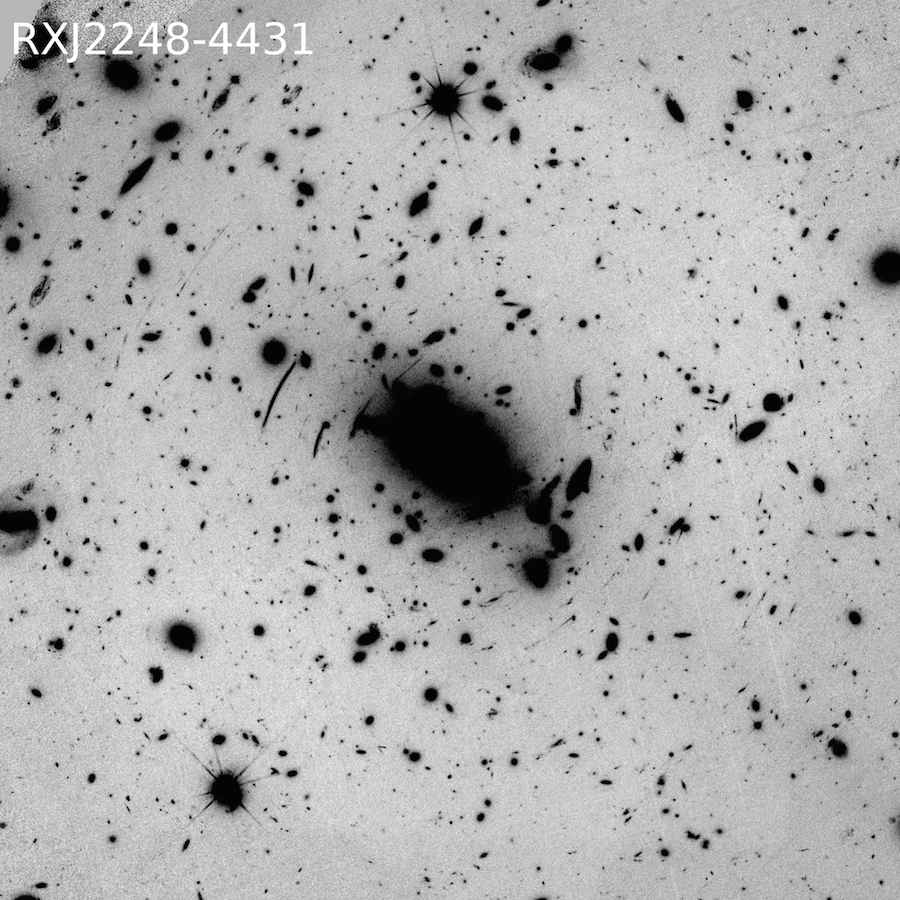}
 \end{minipage}%
 \begin{minipage}[t]{0.5\linewidth}
  \centering
  \includegraphics[width = 0.9\linewidth]{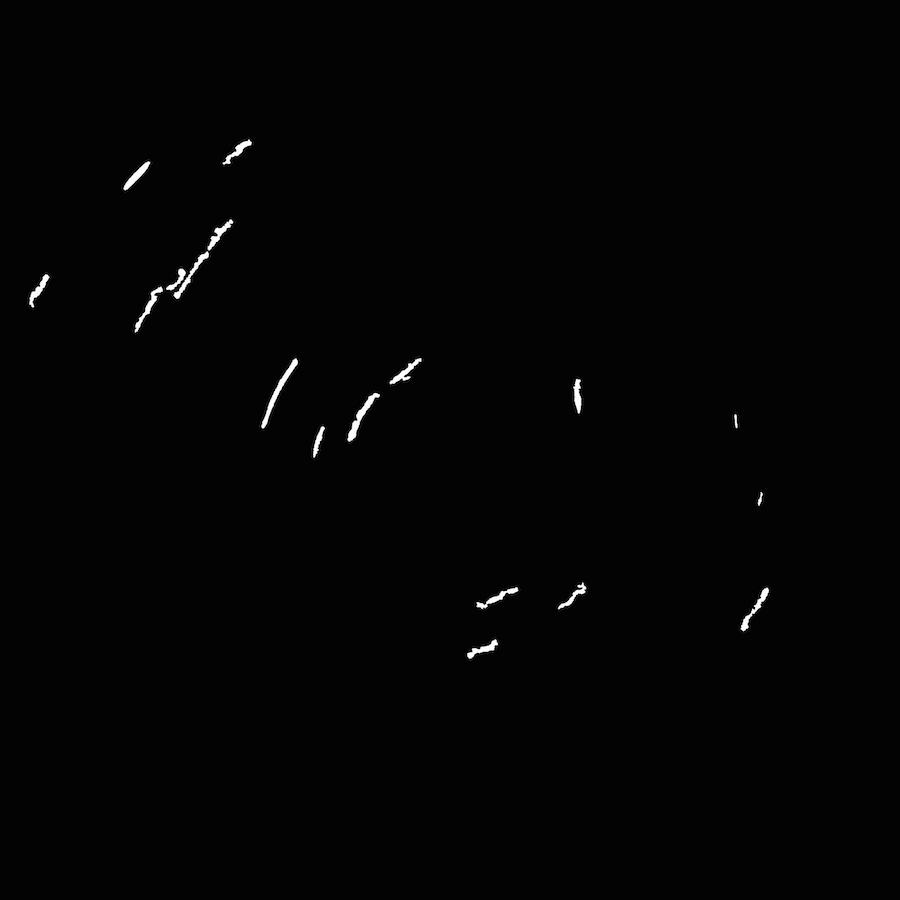}
 \end{minipage}

\caption{Left panel shows the detection images of five CLASH clusters; 
right panel shows the raw output maps produced by the arcfinder with $l/w > 7$. }
\label{fig6_2}
\end{figure*}

\begin{deluxetable}{llllc}
\tabletypesize{\scriptsize}
\tablewidth{0pt}
\tablecaption{The CLASH cluster sample \label{tab1}}
\tablehead{
\colhead{Cluster\tablenotemark{a}} &
\colhead{$\alpha_{\rm J2000}$ } & 
\colhead{$\delta_{\rm J2000}$ } &
\colhead{$z_{\rm Clus}$ }  &
\colhead{$M_{200c}$}\\
\colhead{ }  &   
\colhead{ }  &    
\colhead{ }  &   
\colhead{ }  & 
\colhead{$[10^{15} M_{\odot}/h]$} }
\startdata
X-ray Selected Clusters:  \\
   ~~Abell 209                       & 01:31:52.57 & $-$13:36:38.8               &  ~0.206   &  $0.95 \pm 0.07$   \\
   ~~Abell 383                       & 02:48:03.36 & $-$03:31:44.7               &  ~0.187   &  $0.87 \pm 0.07$    \\
   ~~MACS0329.7-0211                 & 03:29:41.68 & $-$02:11:47.7               &  ~0.450   &  $0.73 \pm 0.10$     \\
   ~~MACS0429.6-0253                 & 04:29:36.10 & $-$02:53:08.0               &  ~0.399   &  $0.80 \pm 0.14$    \\
   ~~MACS0744.9+3927                 & 07:44:52.80 & +39:27:24.4                 &   ~0.686  &  $0.70 \pm 0.04 $     \\
   ~~Abell 611                       & 08:00:56.83 & +36:03:24.1                 &   ~0.288  &  $0.85 \pm 0.05$   \\
   ~~MACS1115.9+0129                 & 11:15:52.05  &+01:29:56.6                 &   ~0.352  &  $0.90 \pm 0.09$  \\
   ~~Abell 1423                      & 11:57:17.26 &+33:36:37.4                  & ~0.213   &   $...$ \\
   ~~MACS1206.2-0847                 & 12:06:12.28 & $-$08:48:02.4               &   ~0.440 &   $0.86 \pm 0.11$    \\
   ~~CLJ1226.9+3332                  & 12:26:58.37 & +33:32:47.4                 &   ~0.890 &   $1.56 \pm 0.10$  \\
   ~~MACS1311.0-0310                 & 13:11:01.67 & $-$03:10:39.5               &   ~0.494 &   $0.46 \pm 0.03$   \\
   ~~RXJ1347.5-1145                  & 13:47:30.59 & $-$11:45:10.1               &   ~0.451 &   $1.16 \pm 0.19$ \\
   ~~MACS1423.8+2404                 & 14:23:47.76 & +24:04:40.5                 &   ~0.545 &   $0.57 \pm 0.10$   \\
   ~~RXJ1532.9+3021                  & 15:32:53.78 & +30:20:58.7                 &  ~0.345  &   $0.53 \pm 0.08$   \\
   ~~MACS1720.3+3536                 & 17:20:16.95 & +35:36:23.6                 &  ~0.391  &   $0.75 \pm 0.08$     \\
   ~~Abell 2261                      & 17:22:27.25 & +32:07:58.6                 &  ~0.224  &   $1.42 \pm 0.17$ \\
   ~~MACS1931.8-2635                 & 19:31:49.66 & $-$26:34:34.0               & ~0.352   &   $0.69 \pm 0.05$   \\
   ~~RXJ2129.7+0005                  & 21:29:39.94 & +00:05:18.8                 &  ~0.234  &   $0.61 \pm 0.06$   \\
   ~~MS2137-2353                     & 21:40:15.18 & $-$23:39:40.7               &  ~0.313  &   $1.04 \pm 0.06$   \\
   ~~RXJ2248.7-4431 (Abell 1063S)    & 22:48:44.29 & $-$44:31:48.4               &  ~0.348  &   $1.16 \pm 0.12$ \\
 ~~\\
   High Magnification Clusters:  \\
   ~~MACS0416.1-2403                 & 04:16:09.39 & $-$24:04:03.9                  & ~0.420  &  $...$    \\
   ~~MACS0647.8+7015                & 06:47:50.03 & +70:14:49.7                     & ~0.584  &  $...$  \\
   ~~MACS0717.5+3745                & 07:17:31.65 & +37:45:18.5                     & ~0.548   & $...$  \\
   ~~MACS1149.6+2223                & 11:49:35.86  & +22:23:55.0                    & ~0.544   & $...$   \\
   ~~MACS2129.4-0741                & 21:29:26.06$^a$ & $-$07:41:28.8$^a$           & ~0.570   & $...$    \\
\enddata
\tablenotetext{a}{Central cluster coordinates derived from optical image.}
\end{deluxetable}
\clearpage

\subsection{The Arc Redshift Distribution}
 
\begin{figure}
\centering
 \includegraphics[width = 0.7\linewidth]{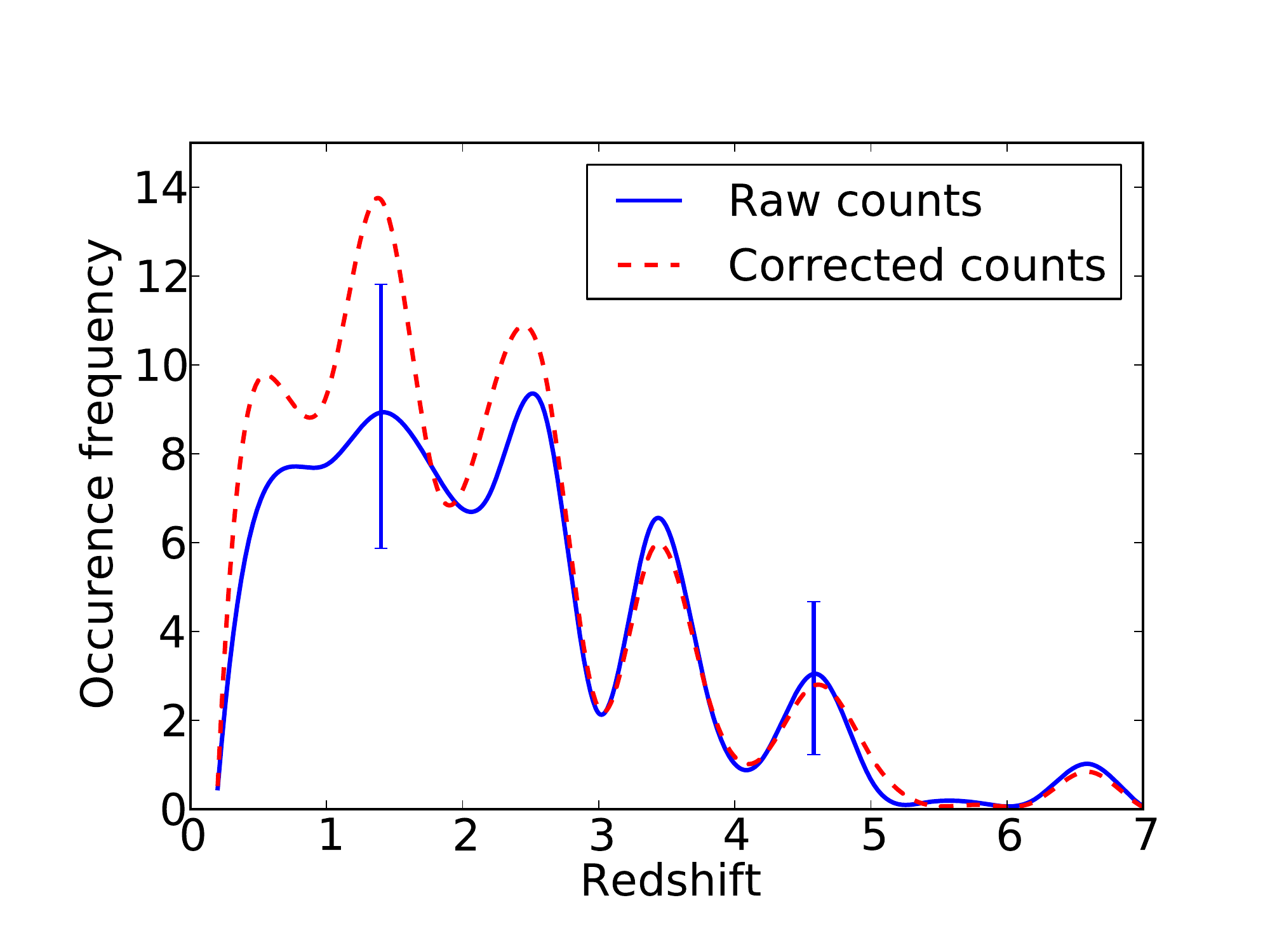}
\caption{The photometric redshift distribution of the detected arcs in the CLASH
X-ray selected sample.
The blue solid line denotes the redshift distribution of the raw data counts, 
which is computed based on the full posterior probability distribution of the 
detected arcs; the red dashed line denotes the redshift distribution after the 
elongation bias, incompleteness and false positive correction, which is computed
based on the corrected full posteriro probability distribution. The errorbar 
represents the $1\sigma$ Poisson error. }
\label{fig6a}
\end{figure}

\begin{figure}
\centering
\subfigure[$S/N$ ratio]{
 \label{fig7:subfig:a}
 \includegraphics[width = 0.45\linewidth]{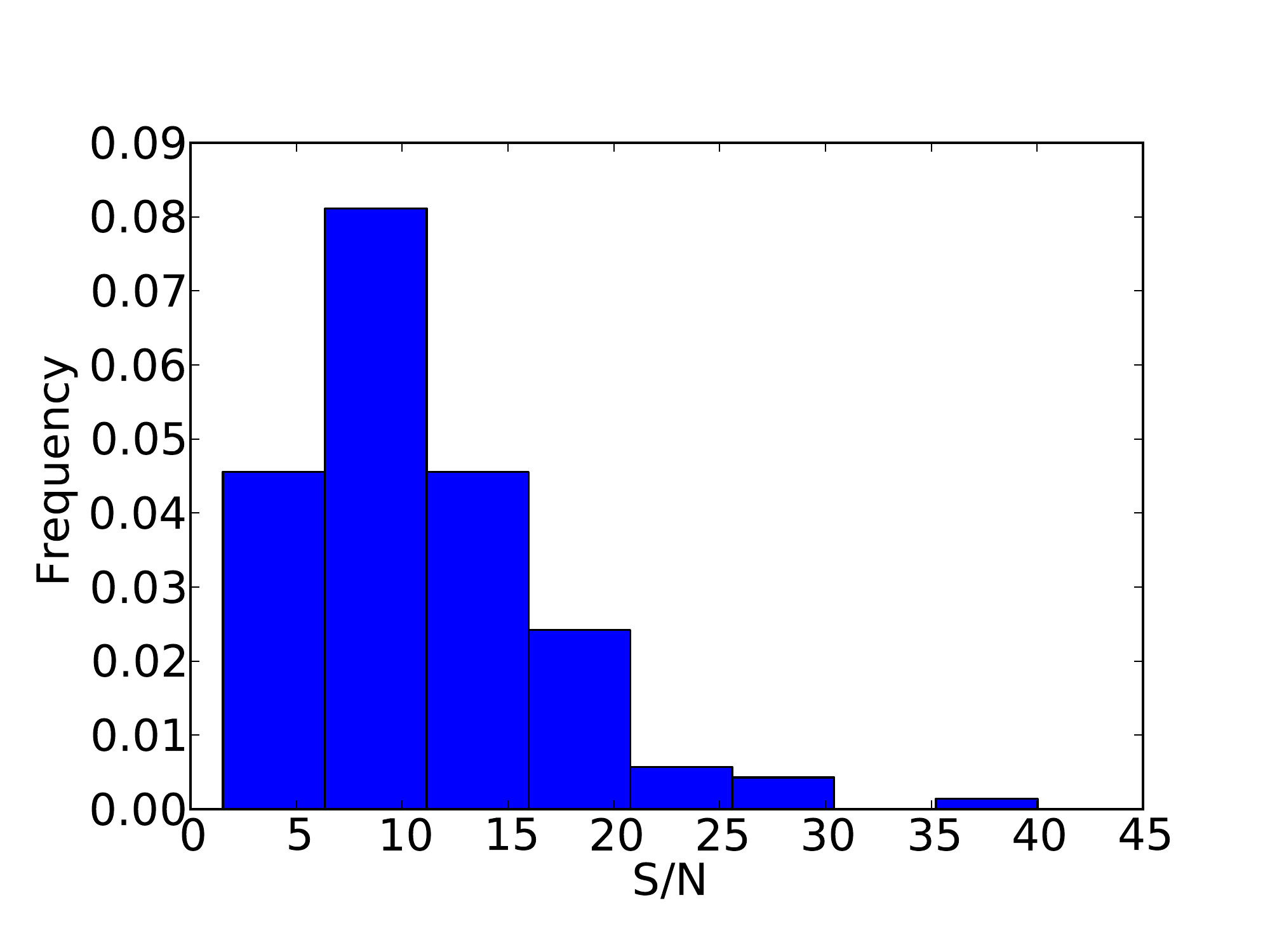}
}
\centering
\subfigure[F814W AB magnitude]{
 \label{fig7:subfig:b}
 \includegraphics[width = 0.45\linewidth]{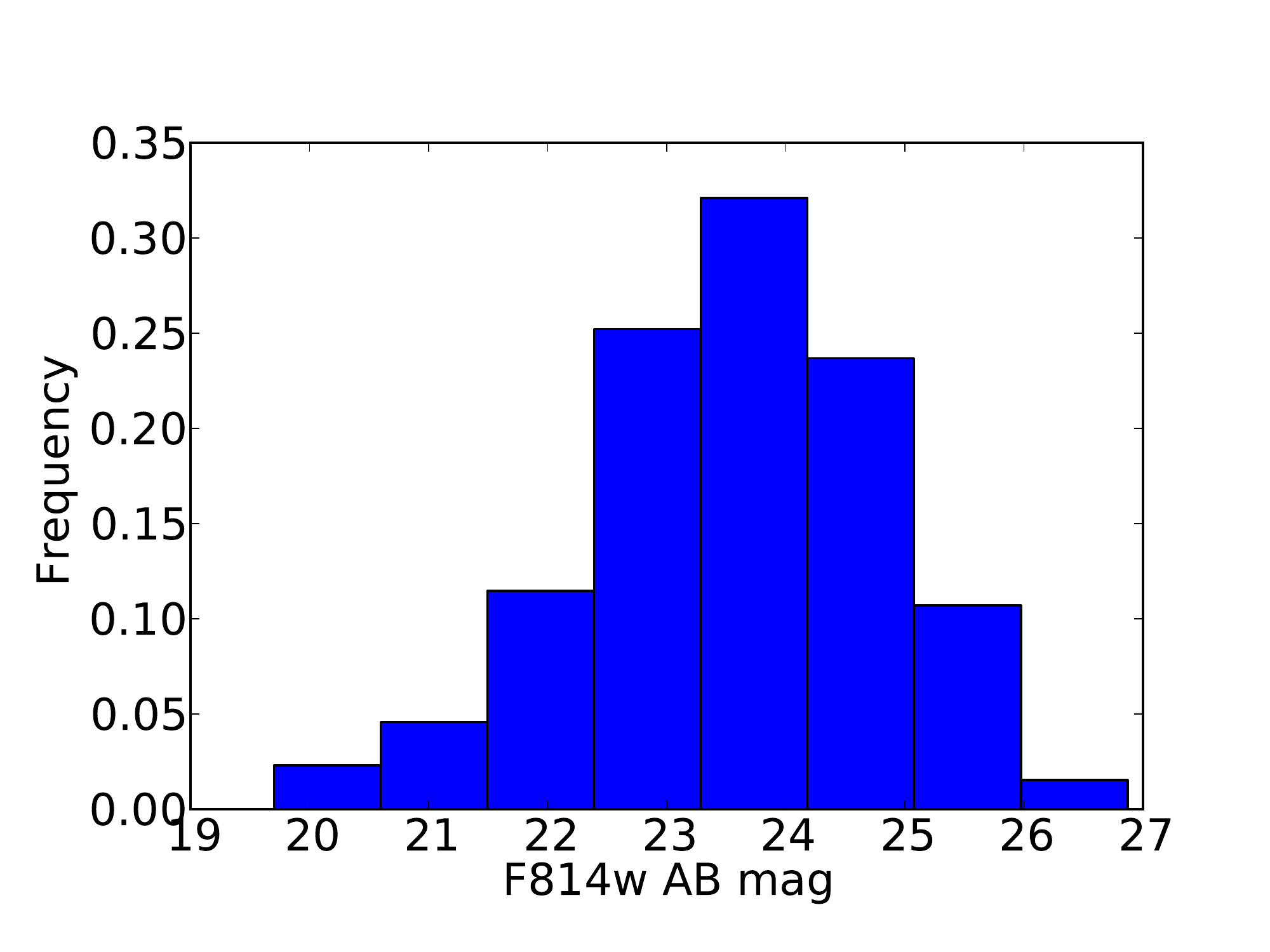}
}
\centering
\subfigure[$l/w$ ratio]{
 \label{fig7:subfig:c}
 \includegraphics[width = 0.45\linewidth]{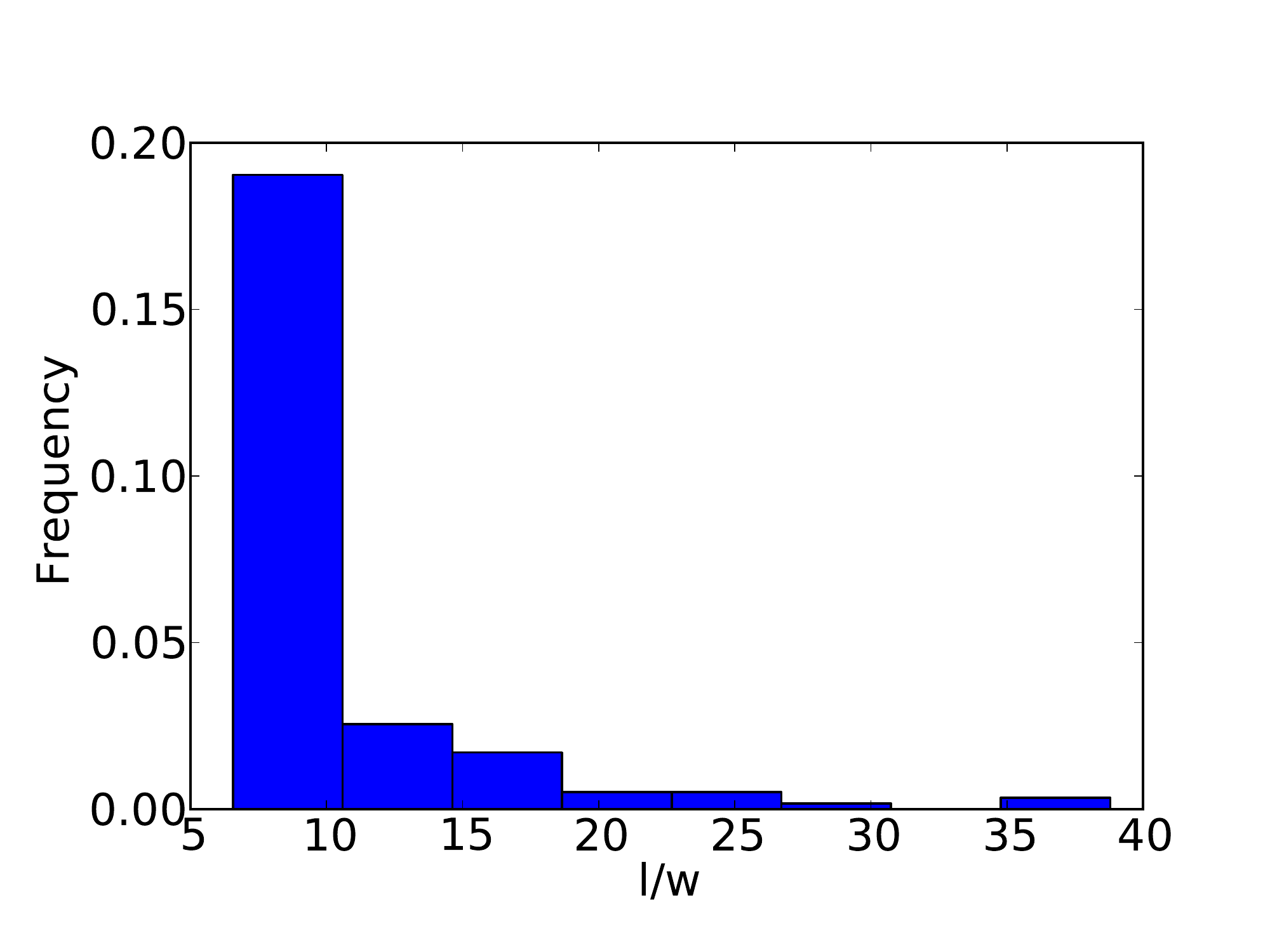}
}
\centering
\subfigure[Angular distance from the cluster center]{
 \label{fig7:subfig:d}
 \includegraphics[width = 0.45\linewidth]{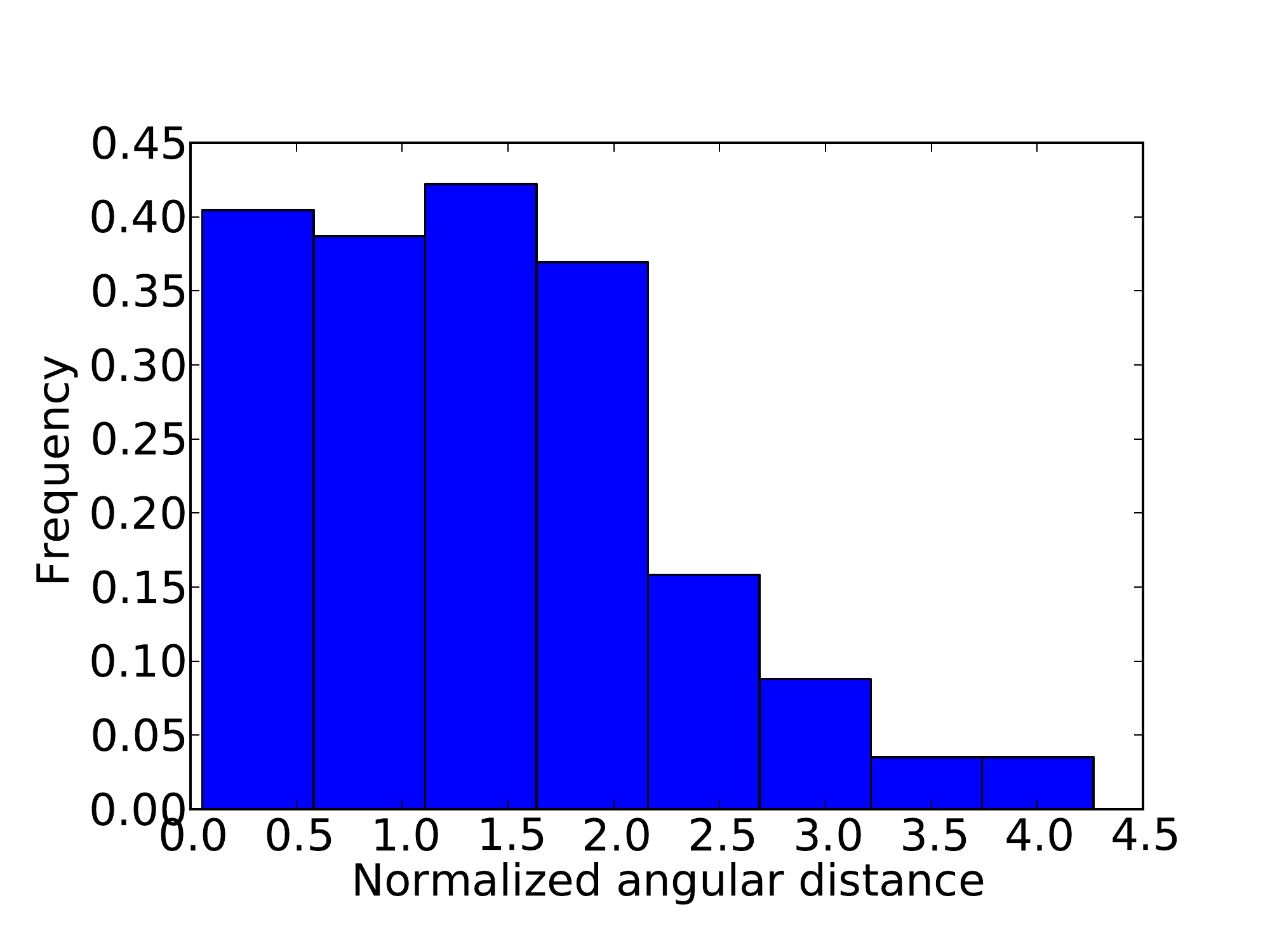}
}
\caption{Panel (a),(b),(c),(d) show the distribution of $S/N$ ratio, AB magnitude, 
$l/w$ ratio and normalized angular distance $RD/r_{200}$ of all the detected arcs in the CLASH sample.} 
\label{fig7:subfig}
\end{figure}

We determine the photometric redshift distribution of the lensed background galaxies
detected by our algorithm using the photometric redshifts derived with the 
Bayesian-based BPZ package (BPZ; \citet{ben00,ben04,coe06}). Spectral energy 
distribution (SED) templates are redshifted and fit to the observed photometry.
The BPZ code adopts a prior that the empirical likelihood of redshift is a 
function of both galaxy magnitude and galaxy morphological type (e.g., bright 
and/or elliptical galaxies are rare at high redshift). We used 11 SED 
templates originally from PEGASE \citep{fio97} that have been recalibrated based
on photometry and spectroscopic redshifts of galaxies in the FIREWORKS catalog 
\citep{wuy08}. We obtain the photometric redshift distribution of all the 
detected arcs and find that they have a median photometric redshift 
$z_{s} = 1.9$. We also find that there is a significant fraction of arcs with 
$z_s \sim 3$ (34\% of the detected arcs have photometric redshift larger than 
3). Figure~\ref{fig6a} shows the  arc number counts as a function of redshift 
before and after correcting for the measurement bias, incompleteness and false 
positive rate. 
To compute the photometric redshift distribution of our arc sample, we sum up the  
individual posterior redshift probability distributions of each detected arc. 
The mean uncertainty of the photometric redshifts in CLASH is $\sigma_z \sim 0.03(1+z)$ and, thus,
we sample the probability distribution using the bin size $\Delta z = 0.4$
which is twice as large as the uncertainty of the arc with highest redshift. The summed distribution prior
to correction for our selection function and elongation bias is shown as the 
blue line in Figure~\ref{fig6a}. To correct for the selection bias, incompleteness and false
positive rate, we re-sum the probability distribution for each arc after first multiplying by the appropriate correction factors. 
The fully corrected redshift distribution, derived in this way, is 
shown by the red line in in Figure~\ref{fig6a}.
Figure~\ref{fig7:subfig} also lists the distribution of arc $S/N$ ratio, arc AB 
magnitude in F814W filter, arc $l/w$ ratio, and the normalized angular distance 
of the arc from the cluster center. Table~\ref{tab2} lists the properties of 
all the detected arcs in 20 X-ray selected sample, including the equatorial and 
pixel coordinates, length, $l/w$ ratio, radial distance from the arc center to 
the cluster center, the normalized radial distance by $r_{200}$, the photometric
redshift and the AB magnitude in the F814W band. In Table~\ref{tab2}, we do not 
exclude the objects with photometric redshifts that are significantly smaller 
than the corresponding cluster redshift. Such probable foreground sources are 
considered to be false positive detections. We eliminate false positive 
detections statistically when we calculate the arc redshift distribution in the
CLASH sample.

\begin{deluxetable}{lccccccccccc}
\tabletypesize{\scriptsize}
\tablewidth{0pt}
\tablecaption{Detected arcs and properties \label{tab2}}
\tablehead{
\colhead{Cluster}    &   \colhead{Arc ID}  &   \colhead{RA} & 
\colhead{DEC}        &   \colhead{$x$}       &   \colhead{$y$}  & 
\colhead{$l$ ($''$)} &   \colhead{$l/w$}   &   \colhead{RD ($''$)\tablenotemark{a}} &
 \colhead{RD/$r_{200}$} & \colhead{$z$}       &   \colhead{AB mag (F814W)}}
\startdata
Abell1423	&	1	&	179.33	&	33.60	&	2204.00	&	1905.00	&	 6.82	&	 9.67	&	                43.09	&	 1.09	&	 0.62	&	22.61	\\
\nodata         &  \dag	2 	&	179.30	&	33.62	&	3553.00	&	3227.00	&	 6.14	&	 7.61	&	                82.69	&	 2.10	&	 0.00	&	24.29	\\
Abell209	&	1	&	22.96	&	-13.61	&	2760.00	&	2738.00	&	11.90	&	10.00	&	                23.16	&	 0.62	&	 3.50	&	21.73	\\
Abell2261 	&  \dag	1 	&	260.59	&	32.12	&	3542.00	&	1698.00	&	 7.08	&	 8.89	&	                84.98	&	 1.84	&	 0.00	&	24.39	\\
\nodata	        &  \dag	2 	&	260.61	&	32.12	&	2505.00	&	1994.00	&	 8.46	&	10.55	&	                32.18	&	 0.70	&	 0.33	&	23.01	\\
\nodata	        &	3	&	260.62	&	32.13	&	2165.00	&	2317.00	&	 6.72	&	 7.92	&	                24.54	&	 0.53	&	 3.54	&	23.75	\\
\nodata	        &	4	&	260.60	&	32.13	&	2902.00	&	2395.00	&	10.55	&	14.29	&	                26.77	&	 0.58	&	 1.80	&	22.07	\\
\nodata	        &\dag	5	&	260.64	&	32.13	&	1324.00	&	2620.00	&	13.06	&	23.34	&	                76.98	&	 1.67	&	 0.27	&	23.79	\\
\nodata	        &	6	&	260.61	&	32.15	&	2577.00	&	3516.00	&	 6.64	&	10.20	&	                66.94	&	 1.45	&	 1.35	&	23.22	\\
Abell383	&	1	&	42.03	&	-3.54	&	1620.00	&	2246.00	&	 6.80	&	 7.03	&	                54.43	&	 1.30	&	 0.73	&	25.74	\\
\nodata	        &	2	&	42.02	&	-3.53	&	2181.00	&	2462.00	&	12.36	&	14.63	&	                17.70	&	 0.42	&	 4.22	&	22.94	\\
\nodata	        &	3	&	42.01	&	-3.53	&	2553.00	&	2616.00	&	19.23	&	22.41	&	                15.27	&	 0.36	&	 0.89	&	20.15	\\
\nodata	        &	4	&	42.02	&	-3.53	&	2114.00	&	2579.00	&	 7.32	&	 7.91	&	                15.79	&	 0.38	&	 3.12	&	23.20	\\
\nodata	        &	5	&	42.03	&	-3.53	&	1566.00	&	2735.00	&	 6.33	&	 8.56	&	                49.75	&	 1.18	&	 2.46	&	25.06	\\
\nodata	        &\dag	6	&	42.01	&	-3.53	&	2314.00	&	2691.00	&	 8.43	&	 9.01	&	                 1.04	&	 0.02	&	 0.30	&	21.93	\\
\nodata	        &	7	&	42.00	&	-3.53	&	2828.00	&	2893.00	&	 7.48	&	17.11	&	                34.96	&	 0.83	&	 6.31	&	23.92	\\
\nodata	        &	8	&	42.01	&	-3.52	&	2281.00	&	3035.00	&	20.42	&	31.37	&	                22.65	&	 0.54	&	 5.00	&	23.33	\\
\nodata	        &	9	&	42.01	&	-3.52	&	2281.00	&	3024.00	&	 7.87	&	16.59	&	                21.94	&	 0.52	&	 3.24	&	24.81	\\
Abell611	&	1	&	120.24	&	36.06	&	2335.00	&	2665.00	&	25.66	&	30.64	&	                15.35	&	 0.40	&	 1.12	&	20.44	\\
\nodata	        &\dag	2	&	120.26	&	36.06	&	1358.00	&	2727.00	&	 7.02	&	 7.37	&	                75.89	&	 2.00	&	 0.27	&	19.75	\\
CLJ1226	        &	1	&	186.74	&	33.54	&	2838.00	&	2253.00	&	 6.30	&	 9.52	&	                26.72	&	 1.52	&	 2.79	&	25.90	\\
\nodata	        &	2	&	186.75	&	33.54	&	2164.00	&	2394.00	&	14.22	&	12.37	&	                23.10	&	 1.31	&	 2.30	&	24.96	\\
\nodata	        &	3	&	186.75	&	33.55	&	2144.00	&	2742.00	&	 8.47	&	19.19	&	                28.47	&	 1.62	&	 3.45	&	23.79	\\
MACS0329 	&	1	&	52.41	&	-2.19	&	3049.00	&	2869.00	&	 7.96	&	 9.55	&	                43.27	&	 1.56	&	 1.03	&	23.28	\\
\nodata	        &	2	&	52.42	&	-2.18	&	2685.00	&	3153.00	&	 6.20	&	 9.45	&	                44.66	&	 1.61	&	 3.35	&	24.25	\\
MACS0429	&	1	&	67.40	&	-2.90	&	2752.00	&	1709.00	&	 6.48	&	 8.01	&	                52.93	&	 2.08	&	 1.67	&	24.22	\\
\nodata	        &	2	&	67.40	&	-2.89	&	2499.00	&	2200.00	&	10.69	&	12.24	&	                18.83	&	 0.74	&	 1.35	&	21.79	\\
MACS0744  	&\dag	1	&	116.22	&	39.44	&	2682.00	&	1488.00	&	 7.00	&	 7.56	&	                66.12	&	 3.43	&	 0.41	&	20.22	\\
\nodata	        &	2	&	116.23	&	39.45	&	2128.00	&	2078.00	&	 6.41	&	 7.74	&	                36.39	&	 1.89	&	 4.79	&	23.85	\\
\nodata	        &\dag	3	&	116.20	&	39.45	&	3147.00	&	2191.00	&	 7.32	&	 7.32	&	                45.92	&	 2.38	&	 0.14	&	18.48	\\
\nodata	        &	4	&	116.23	&	39.46	&	1969.00	&	2581.00	&	 6.75	&	 9.40	&	                35.47	&	 1.84	&	 4.73	&	23.99	\\
\nodata	        &	5	&	116.23	&	39.46	&	2033.00	&	2625.00	&	 6.10	&	 8.37	&	                32.04	&	 1.66	&	 4.41	&	24.43	\\
\nodata	        &	6	&	116.20	&	39.46	&	3477.00	&	2652.00	&	 7.56	&	14.86	&	                63.92	&	 3.32	&	 4.11	&	23.76	\\
\nodata	        &	7	&	116.21	&	39.46	&	2839.00	&	2632.00	&	 6.77	&	 7.71	&	                23.47	&	 1.22	&	 1.17	&	20.34	\\
MACS1115	&	1	&	168.96	&	 1.49	&	2602.00	&	2228.00	&	14.44	&	18.05	&	                18.34	&	 0.58	&	 2.46	&	23.08	\\
\nodata	        &	2	&	168.98	&	 1.50	&	1586.00	&	2402.00	&	 9.53	&	15.25	&	                60.03	&	 1.90	&	 1.76	&	24.83	\\
\nodata	        &\dag	3	&	168.97	&	 1.50	&	2500.00	&	2374.00	&	 7.76	&	 7.60	&	                 7.74	&	 0.25	&	 0.42	&	20.91	\\
\nodata	        &	4	&	168.97	&	 1.50	&	2355.00	&	2371.00	&	10.65	&	12.58	&	                12.57	&	 0.40	&	 4.21	&	21.72	\\
\nodata	        &	5	&	168.96	&	 1.51	&	2896.00	&	2911.00	&	14.52	&	16.16	&	                37.20	&	 1.18	&	 4.12	&	22.91	\\
\nodata	        &	6	&	168.96	&	 1.51	&	2578.00	&	3022.00	&	 7.14	&	 7.11	&	                34.71	&	 1.10	&	 3.25	&	24.41	\\
MACS1206  	&\dag	1	&	181.55	&	-8.81	&	2247.00	&	2078.00	&	 6.69	&	 7.04	&	                28.16	&	 0.99	&	 0.55	&	20.26	\\
\nodata	        &	2	&	181.54	&	-8.80	&	2790.00	&	2454.00	&	14.27	&	15.73	&	                19.47	&	 0.68	&	 1.05	&	19.76	\\
\nodata	        &	3	&	181.54	&	-8.80	&	3292.00	&	2420.00	&	 6.54	&	 7.51	&	                52.08	&	 1.83	&	 2.41	&	24.64	\\
\nodata	        &\dag	4	&	181.55	&	-8.80	&	2477.00	&	2438.00	&	11.97	&	10.88	&	                 0.92	&	 0.03	&	 0.49	&	23.77	\\
\nodata	        &	5	&	181.57	&	-8.80	&	1618.00	&	2471.00	&	 8.39	&	12.32	&	                56.79	&	 1.99	&	 1.56	&	23.19	\\
\nodata	        &	6	&	181.53	&	-8.79	&	3484.00	&	3134.00	&	 8.36	&	 8.57	&	                78.90	&	 2.77	&	 0.72	&	19.30	\\
MACS1311	&	1	&	197.75	&	-3.17	&	2881.00	&	2742.00	&	 6.25	&	 7.66	&	                29.33	&	 1.30	&	 2.83	&	24.77	\\
MACS1423  	&\dag	1	&	215.93	&	24.07	&	3433.00	&	2050.00	&	 9.84	&	17.73	&	                66.78	&	 3.12	&	 0.00	&	23.74	\\
\nodata	        &	2	&	215.95	&	24.07	&	2606.00	&	2128.00	&	 6.13	&	 7.80	&	                24.49	&	 1.14	&	 1.47	&	23.92	\\
\nodata	        &	3	&	215.94	&	24.07	&	2804.00	&	2120.00	&	 8.80	&	15.30	&	                30.97	&	 1.45	&	 2.57	&	23.94	\\
\nodata	        &	4	&	215.95	&	24.08	&	2378.00	&	2773.00	&	 8.18	&	11.48	&	                20.10	&	 0.94	&	 1.79	&	22.72	\\
\nodata	        &	5	&	215.95	&	24.09	&	2576.00	&	3211.00	&	 7.70	&	11.46	&	                47.03	&	 2.20	&	 3.16	&	24.47	\\
MACS1720	&	1	&	260.06	&	35.60	&	2757.00	&	2042.00	&	 8.33	&	11.23	&	                33.32	&	 1.20	&	 4.38	&	24.29	\\
\nodata	        &	2	&	260.07	&	35.60	&	2621.00	&	2346.00	&	 8.60	&	10.80	&	                11.90	&	 0.43	&	 0.82	&	23.23	\\
MACS1931	&	1	&	292.96	&	-26.59	&	2559.00	&	1917.00	&	 9.71	&	11.49	&	                37.90	&	 1.38	&	 3.55	&	24.23	\\
MS2137	        &	1	&	325.06	&	-23.66	&	2422.00	&	2713.00	&	14.81	&	12.99	&	                14.59	&	 0.46	&	 1.77	&	21.78	\\
\nodata	        &	2	&	325.07	&	-23.65	&	2181.00	&	2903.00	&	10.75	&	14.08	&	                33.37	&	 1.05	&	 1.71	&	23.95	\\
\nodata	        &	3	&	325.07	&	-23.65	&	2215.00	&	3046.00	&	10.93	&	12.95	&	                39.93	&	 1.26	&	 1.97	&	23.86	\\
RXJ1347	        &	1	&	206.87	&	-11.77	&	3032.00	&	1776.00	&	12.76	&	21.51	&	                55.55	&	 1.82	&	 1.64	&	23.58	\\
\nodata	        &	2	&	206.87	&	-11.76	&	3141.00	&	2064.00	&	 9.07	&	11.79	&	                46.58	&	 1.52	&	 2.43	&	21.57	\\
\nodata	        &	3	&	206.87	&	-11.75	&	3117.00	&	2872.00	&	 7.02	&	10.12	&	                43.42	&	 1.42	&	 4.28	&	24.63	\\
\nodata	        &	4	&	206.88	&	-11.75	&	2521.00	&	2951.00	&	 6.82	&	 7.51	&	                29.92	&	 0.98	&	 0.78	&	21.46	\\
\nodata	        &	5	&	206.88	&	-11.74	&	2549.00	&	3162.00	&	 7.83	&	13.50	&	                43.50	&	 1.42	&	 3.78	&	24.22	\\
RXJ1532	        &\dag	1	&	233.22	&	30.34	&	2841.00	&	2088.00	&	 7.75	&	 9.30	&	                34.16	&	 1.25	&	 0.27	&	22.25	\\
RXJ2129	        &	1	&	322.41	&	 0.08	&	2890.00	&	2044.00	&	 7.42	&	 7.79	&	                38.34	&	 1.36	&	 3.17	&	23.61	\\
\nodata	        &\dag	2	&	322.44	&	 0.09	&	1394.00	&	2310.00	&	 6.26	&	 9.57	&	                73.09	&	 2.60	&	 0.00	&	24.96	\\
\nodata	        &	3	&	322.42	&	 0.09	&	2295.00	&	2528.00	&	 7.20	&	 7.83	&	                13.81	&	 0.49	&	 1.55	&	22.65	\\
RXJ2248	        &	1	&	342.18	&	-44.54	&	2613.00	&	1950.00	&	 6.15	&	 6.89	&	                36.42	&	 1.31	&	 3.08	&	24.18	\\
\nodata	        &	2	&	342.16	&	-44.54	&	3371.00	&	2095.00	&	 8.91	&	11.34	&	                62.35	&	 2.25	&	 3.09	&	24.02	\\
\nodata	        &	3	&	342.18	&	-44.54	&	2622.00	&	2084.00	&	 8.06	&	10.83	&	                28.10	&	 1.01	&	 1.64	&	23.35	\\
\nodata	        &	4	&	342.17	&	-44.54	&	2830.00	&	2074.00	&	 6.61	&	 9.53	&	                34.94	&	 1.26	&	 2.75	&	24.42	\\
\nodata	        &	5	&	342.19	&	-44.53	&	2227.00	&	2545.00	&	10.12	&	10.26	&	                18.06	&	 0.65	&	 1.41	&	22.36	\\
\nodata	        &	6	&	342.19	&	-44.53	&	2039.00	&	2697.00	&	13.95	&	17.47	&	                32.67	&	 1.18	&	 1.38	&	21.32	\\
\nodata	        & 	7	&	342.17	&	-44.53	&	2853.00	&	2642.00	&	 6.19	&	 7.28	&	                24.70	&	 0.89	&	 1.41	&	22.35	\\
\nodata	        &	8	&	342.19	&	-44.53	&	2343.00	&	2684.00	&	 7.32	&	 8.80	&	                15.81	&	 0.57	&	 3.76	&	24.48	\\
\nodata	        &	9	&	342.20	&	-44.52	&	1664.00	&	2902.00	&	 9.44	&	13.06	&	                60.38	&	 2.18	&	 1.36	&	25.23	\\
\nodata	        &\dag	10	&	342.21	&	-44.52	&	1358.00	&	2943.00	&	 6.42	&	 7.33	&	                79.70	&	 2.87	&	 0.00	&	21.79	\\
\nodata	        &	11	&	342.20	&	-44.52	&	1743.00	&	2930.00	&	10.16	&	14.05	&	                56.68	&	 2.04	&	 2.79	&	24.33	\\
\nodata	        &	12	&	342.20	&	-44.52	&	1856.00	&	3108.00	&	 6.95	&	 7.97	&	                57.66	&	 2.08	&	 1.96	&	24.89	\\
\nodata	        &\dag	13	&	342.21	&	-44.52	&	1629.00	&	3261.00	&	 6.73	&	 6.85	&	                75.27	&	 2.71	&	 0.40	&	19.07	\\
\nodata	        &	14	&	342.20	&	-44.52	&	1909.00	&	3340.00	&	 6.49	&	 7.68	&	                66.85	&	 2.41	&	 0.85	&	24.11	\\

\enddata
\tablenotetext{a}{RD = radial distance from the arc center to the cluster 
center in the unit of arcsecond; IDs with $\dag$ denote the false positive detection.}
\end{deluxetable}
\clearpage

\section{MOKA Lensing Simulations}
\label{s5}

\subsection{The MOKA Simulated Cluster Sample}

In order to confirm or resolve the arc statistics problem, we require realistic 
model predictions to compare with the observed CLASH arc counts. In 
previous studies,  mock clusters were selected from N-body simulations 
using either dark matter only \citep{wam04,hil07,hor10} or dark matter with 
other ingredients \citep{puc05,got07,roz08,hil08,men10}. The simulated clusters
were then projected onto the plane of the sky as viewed from various directions 
to create the 2-D mass models. However, the total number of clusters and/or 
their mass and redshift ranges used in these prior studies are not optimally matched to the CLASH dataset.
Therefore, we generate a simulated cluster sample by running the publicly 
available MOKA package \citep{gio12}. MOKA uses simulation-calibrated analytical
relations to describe the dark matter and baryonic content of clusters, which 
allows one to incorporate all the cluster properties that are relevant for 
strong cluster lensing. For example, for each halo, a triaxial NFW profile and 
a random orientation are assigned. The axial ratios are generated from the 
prescriptions of \citet{jin02}. The halo concentration, $c$, and its dependence
on cluster mass, $M$, and redshift are modeled based on the $c-M$ relation of 
\citet{bha13} The joint weak lensing + strong lensing analysis by 
\citet{mer15,ume14} indicates that the observed $c-M$ relation derived from the 20 
X-ray selected CLASH clusters agrees with the relation presented in \citet{bha13}. 
The scatter in the 
concentration value at a fixed mass is well-described by a Gaussian distribution instead of 
a log-normal distribution, with rms $\sim$ 0.33. We adopt this scatter in our MOKA simulations. 
The dark matter substructures, 
the central brightest cluster galaxy (BCG) and adiabatic contraction are also 
incorporated into the MOKA generated models. MOKA is  computationally efficient 
and is able to create a single simulated cluster lens model in a few CPU seconds on a 
personal computer by using a fast semi-analytic approach. The details of the 
code and its implementation can be found in \citet{gio12}. 

For our study, we create 640 mock clusters with the same mass and redshift 
range as the 20 X-ray selected CLASH clusters (32 different realizations for 
each corresponding mass and redshift). In particular, the density profile of 
the main halo follows a NFW profile while the density profile of the subhalos is chosen 
to be truncated Singular Isothermal Sphere (SIS) profile; the spatial density 
distribution of the subhalos follows the measurement from numerical simulations by \citet{gao04}; 
the mass resolution of the subhalos is $10^{10} M_{\odot}$. We
calculate the deflection angle, convergence and shear fields for each projected 
mass distribution. The angular resolution of the simulated cluster images is 
$0.065''$, which matches the pixel scale in the CLASH images.

\subsection{Background Source Images and Ray-Tracing Method}

To create the sky scene from the MOKA mass models, we follow a methodology similar to that in \citet{hor11}: 
we choose galaxies from the F775W UDF image as the sample of sources to be lensed by our
simulated cluster mass models.
This ensures we have a realistic background field that incorporates the observed distributions of galaxy 
morphologies, redshifts, luminosities, angular sizes, and ellipticities directly into our 
simulation. We then simulate the lensed UDF images via ray-tracing, as 
briefly described in Section~\ref{s3.1}. Adopting the thin lens approximation, 
the lensing can be described by the lens equation,
\begin{equation}
\beta = \theta - \alpha(\theta,z_s),
\end{equation} 
where $\theta$ is the image position, $\beta$ is the source position in the 
source plane, and $\alpha$ is the deflection angle which has a weak dependence 
of source redshift.  \citet{coe06} have produced a UDF photometric redshift 
catalogue and a corresponding segmentation map containing 9821 objects detected 
above a $8\sigma$ level. Based on the redshift catalog, we assign all the UDF 
sources among 20 redshift bins with bin widths of $\Delta z = 0.3$. In each redshift bin ($\alpha$ is 
then fixed) we perform the ray-tracing to generate the simulated lensed image 
and combine each of the simulated lensed objects from all bins into a final 
image. Finally, we match the noise levels in the simulations to that in the CLASH images.

\section{Comparison Between Simulated Images and Real Observations}\label{s6}

\begin{figure}
\centering
 \includegraphics[width = 0.7\linewidth]{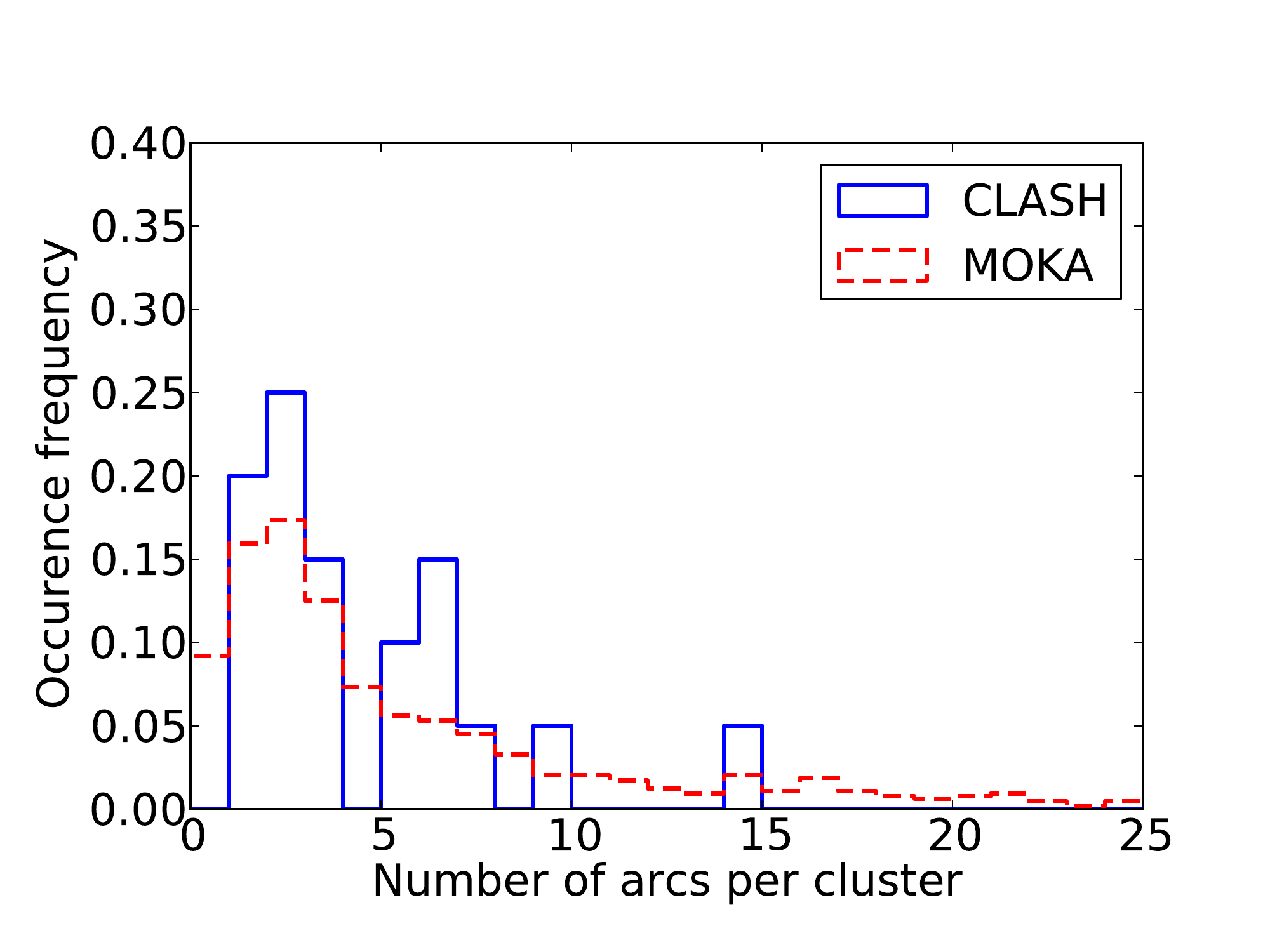}
\caption{The comparison of distribution of arc number per cluster 
between the X-ray selected CLASH sample and the MOKA simulated sample with same 
mass and redshift range.
}
\label{fig9}
\end{figure}

\begin{figure}
\centering
\subfigure[$z_{l} \le z_{median}$ subsample]{
 \label{fig10b:subfig:a}
 \includegraphics[width = 0.45\linewidth]{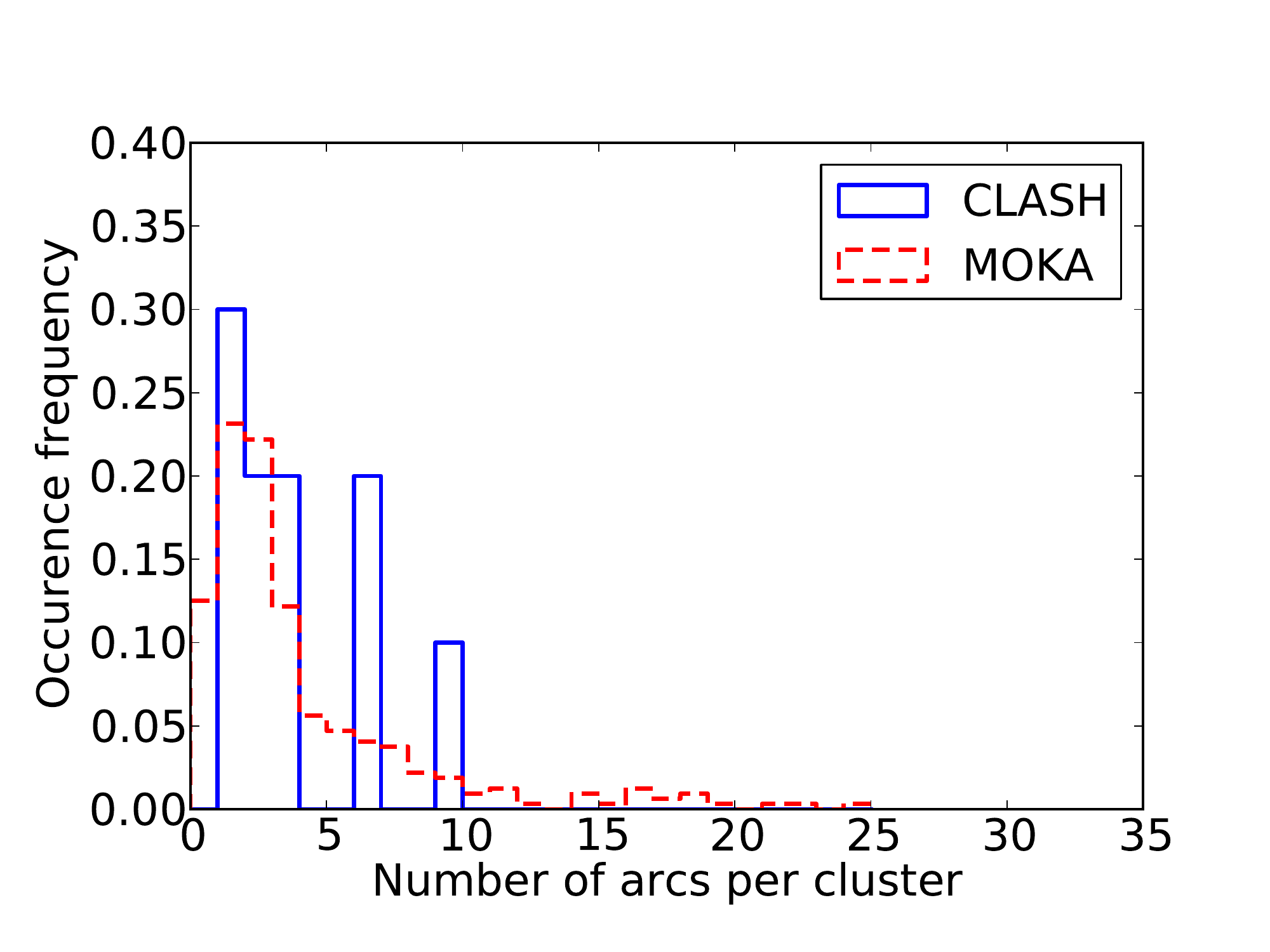}
}
\centering
\subfigure[$z_{l} > z_{median}$ subsample]{
 \label{fig10b:subfig:b}
 \includegraphics[width = 0.45\linewidth]{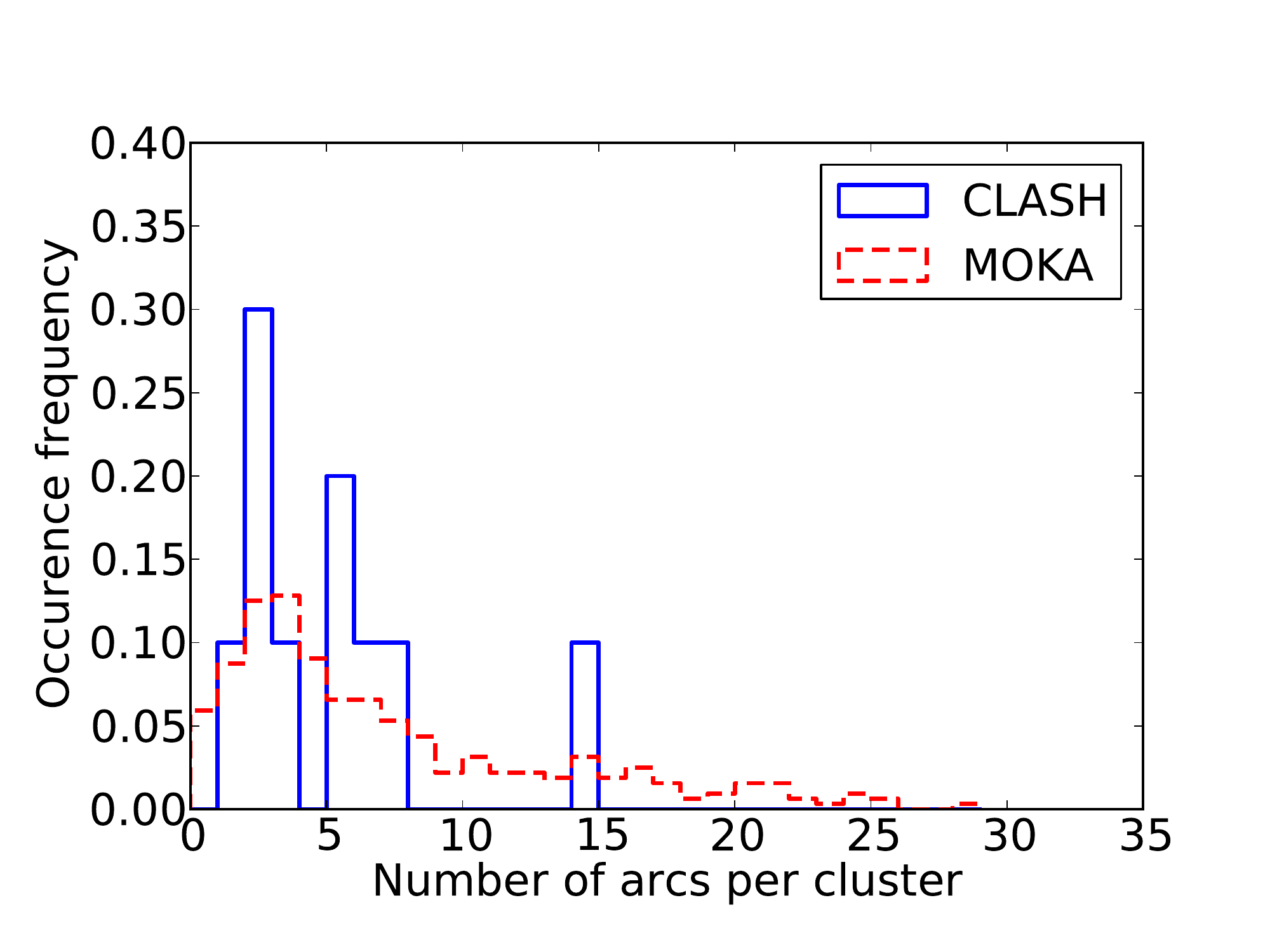}
}

\caption{The lensing efficiency as a function of cluster redshift. The 20 X-ray
selected CLASH clusters are divided into two sub-samples by their cluster 
redshift: $z_{CL} \le z_{median}$ and $z_{CL} > z_{median}$, where 
$z_{median} = 0.352$. (a) and (b) list the comparison of the number distribution
of the sub-samples between the observation and simulation. 
}
\label{fig10b:subfig}
\end{figure}

We run the arcfinder on all 640 simulated images. A raw total of 3304 arcs with 
$l/w \ge 6.5$ and $l \ge 6''$ are detected in 640 simulated 
realizations. We correct this total number of arcs for elongation bias and 
incompleteness and obtain $3585 \pm 165$ arcs, giving a mean of 
$4 \pm 1$ arcs per cluster after applying the false positive correction. This value matches the 
observed lensing efficiency of $4 \pm 1$ precisely. There is no significant difference  
between the arc abundance detected in the observations with that detected in the MOKA simulations. 
Examining the observed and simulated 
distribution of number of arcs per cluster (Figure~\ref{fig9}), a 
Kolmogorov-Smirnov test\footnote{The K-S test performed here use the 
{\tt ks$\_$2samp} routine from the SciPy package.} yields a p-value = 0.92, 
indicating that the null hypothesis that both distributions are drawn from the 
same parent distribution cannot be strongly rejected. We further test the 
lensing efficiency as a function of cluster redshift by assigning the observed 
and simulated samples into two sub-samples by their redshift: 
$z_{CL} \le z_{median}$ and $z_{CL} > z_{median}$, where $z_{median}=0.352$. 
For each sub-sample, we compare the observed and simulated number distribution 
(see Figure~\ref{fig10b:subfig}) of the lensing efficiency. On average, the 
higher redshift clusters are slightly more efficient lenses than the lower 
redshift clusters but the differences are all at marginal statistical 
significance. The K-S tests indicates that, in both redshift bins, the observed 
and simulated distributions of the lensing efficiency are consistent with being drawn from
a common population (p-values are 0.99 and 0.65 for the lower and higher redshift bins, respectively). 
We summarize our arc statistics results
for the observations and simulations in Table~\ref{tab3}. The second and third 
columns in Table~\ref{tab3} denote the lensing efficiency of the observed and 
simulated samples, respectively; the fourth column is the p-value of the K-S test on the 
observed and simulated distributions.

\begin{figure}
\centering
\subfigure[]{
 \includegraphics[width = 0.45\linewidth]{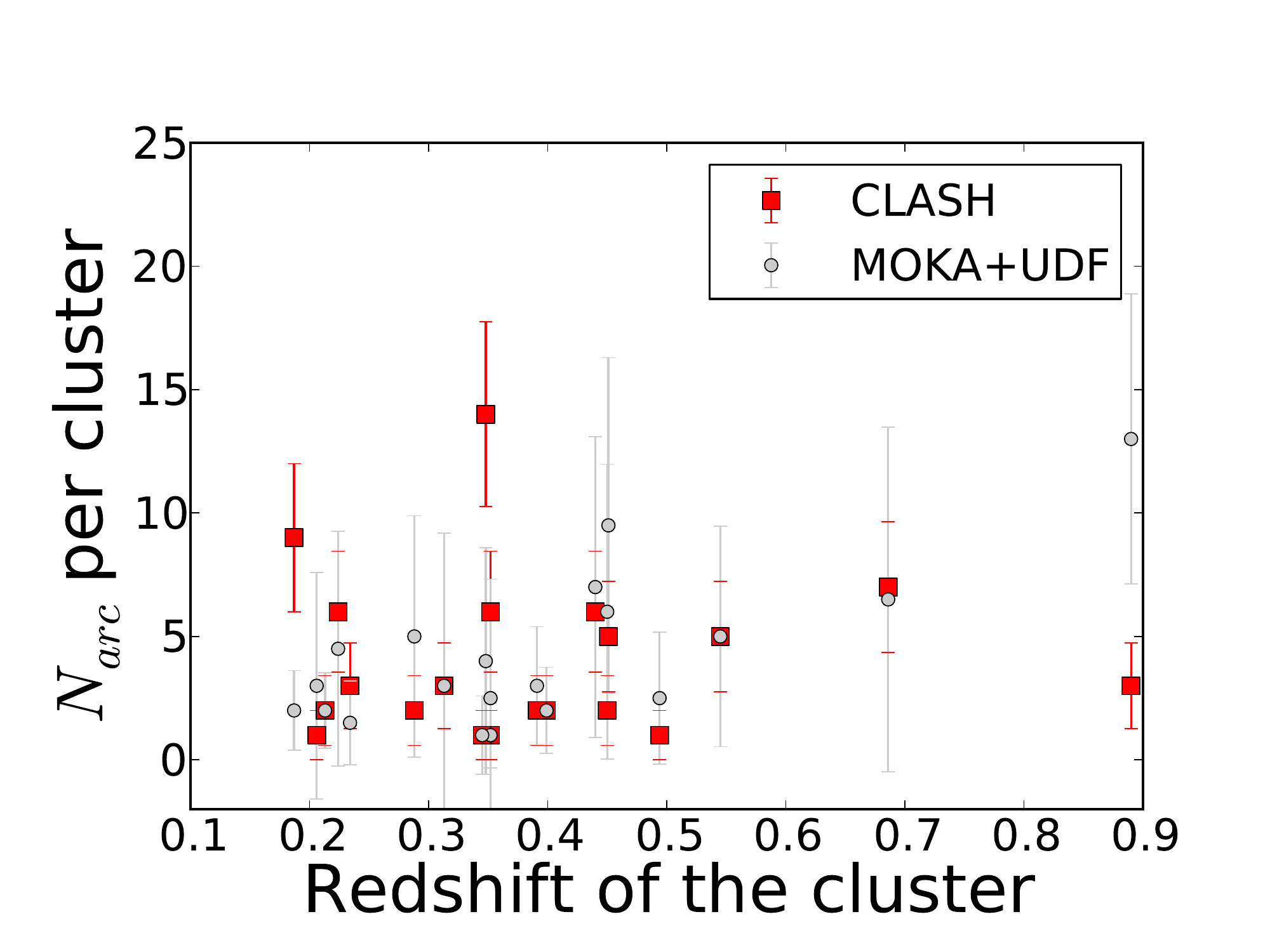}
\label{fig11:subfig:a}
}
\centering
\subfigure[]{
 \includegraphics[width = 0.45\linewidth]{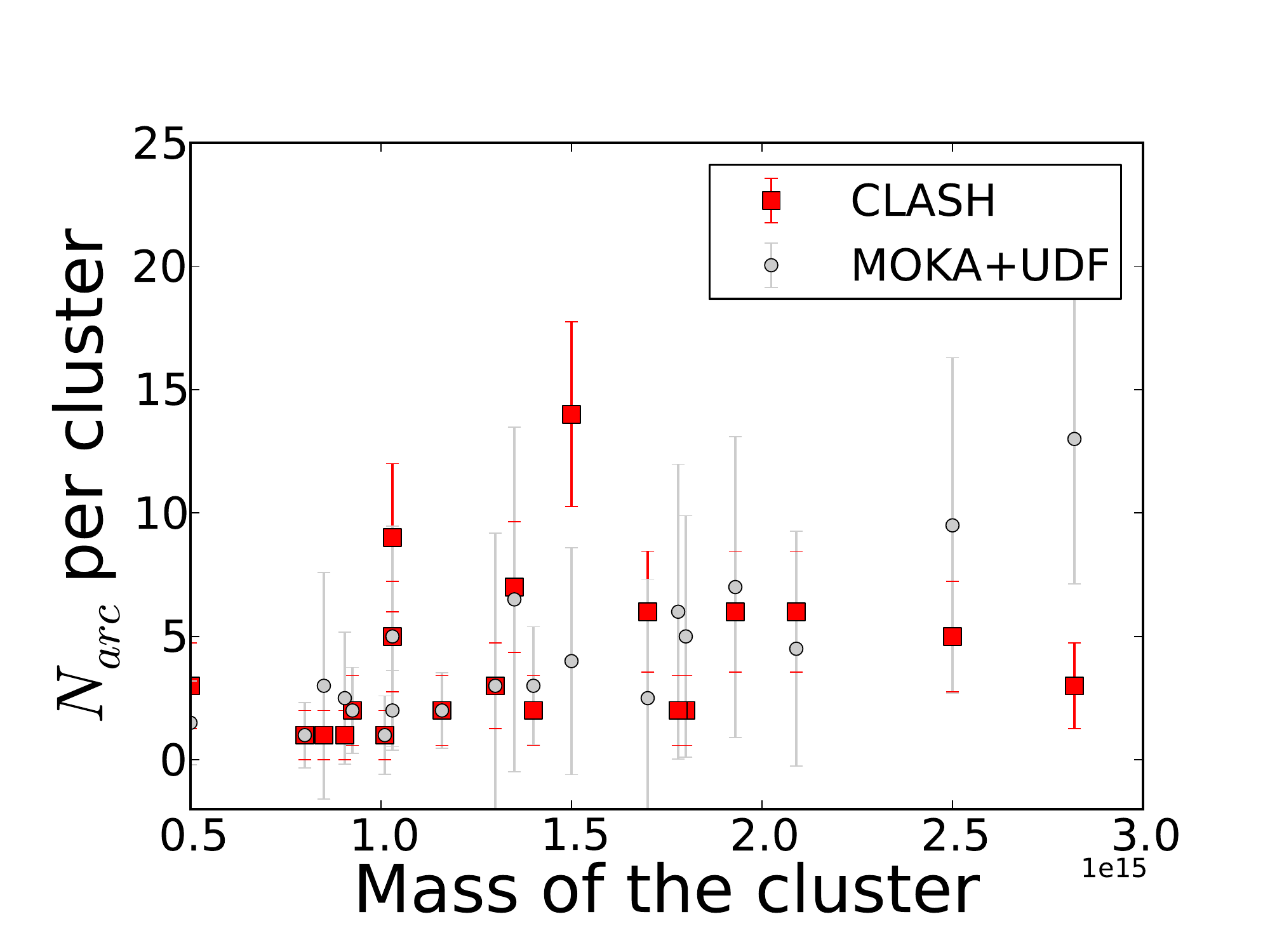}
\label{fig11:subfig:b}
}
\centering
\subfigure[]{
 \includegraphics[width = 0.45\linewidth]{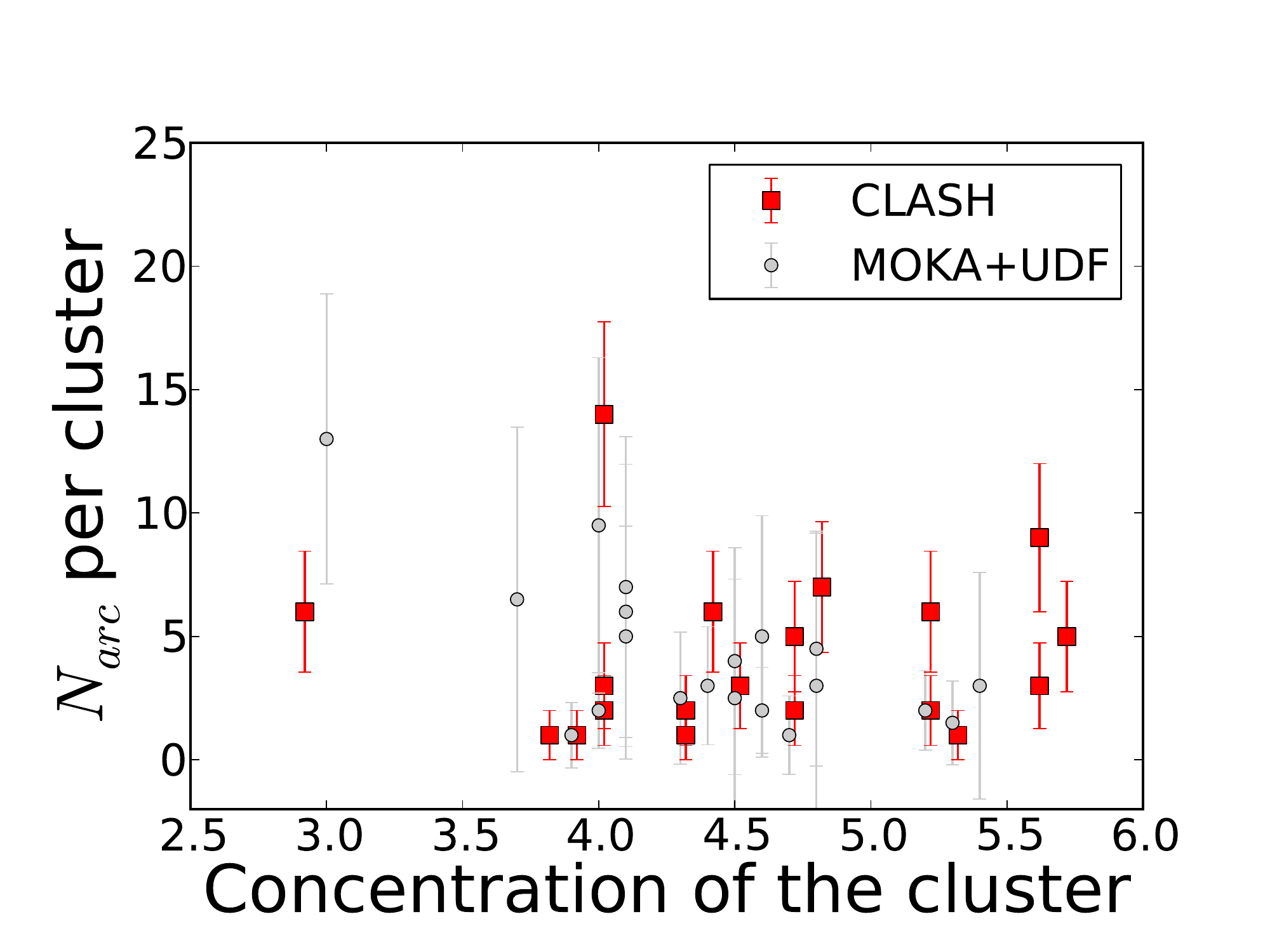}
\label{fig11:subfig:c}
}
\centering
\subfigure[]{
 \includegraphics[width = 0.45\linewidth]{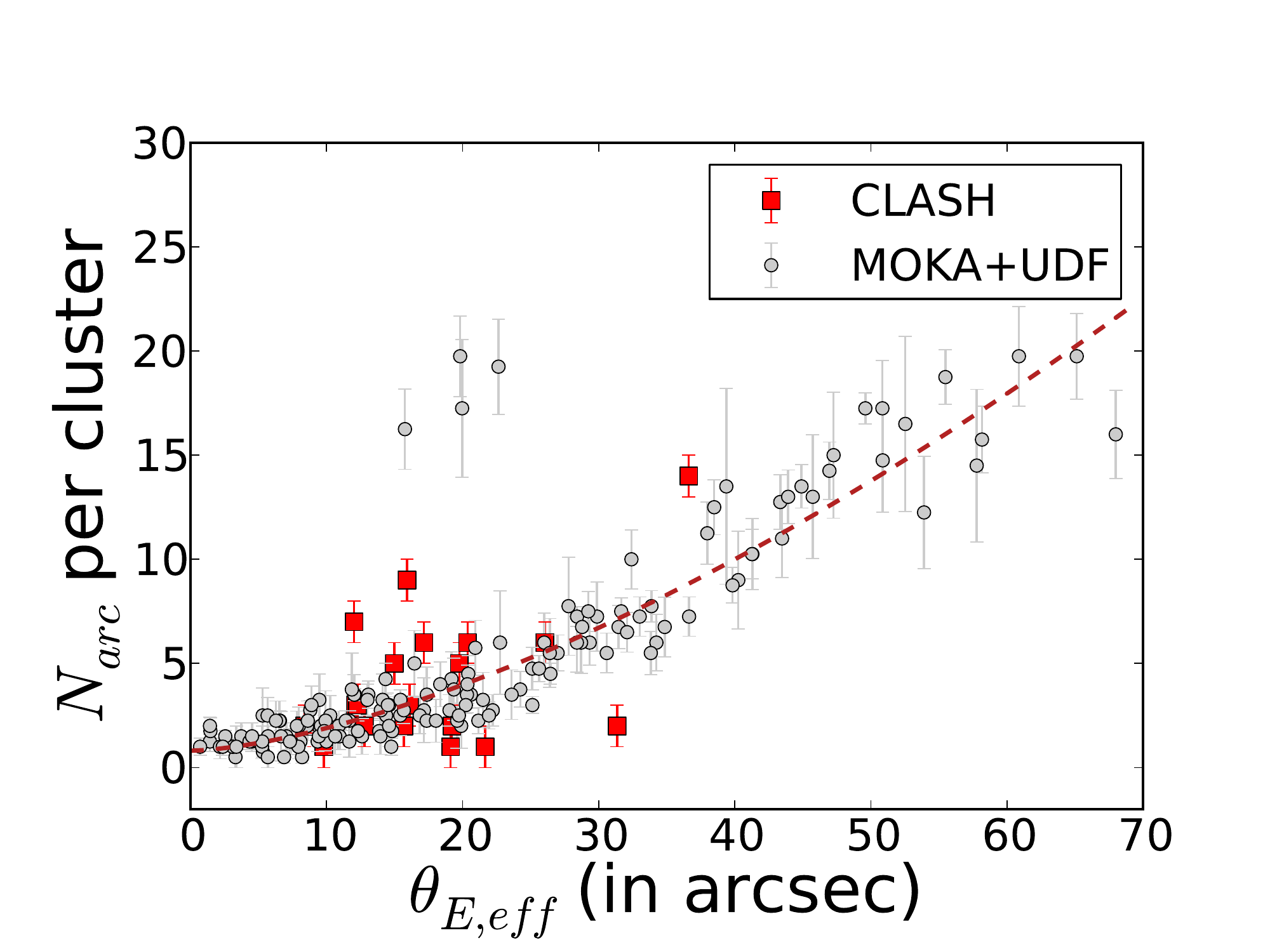}
\label{fig11:subfig:d}
}

\caption{Figure (a), (b), (c) show the comparisons of the lensing efficiency
between the CLASH and MOKA samples for the corresponding cluster redshift, mass 
and concentration, respectively. Figure (d) shows the relation between the 
lensing efficiency and the effective Einstein radius $\theta_{E,eff}$ for all
the CLASH and MOKA data points. With the upper left outlier excluded, the dashed
line gives the best fitting curve for all the MOKA data points.}
\label{fig11:subfig}
\end{figure}

We now explore the relationships between the lensing 
efficiency and the cluster's redshift, mass, concentration and effective Einstein radius 
$\theta_{E,eff} = \sqrt{A / \pi}$ for CLASH and MOKA samples, where $A$ is the 
area enclosed by the tangential critical curve. Figure~\ref{fig11:subfig} 
shows the lensing efficiencies as functions of cluster redshift, mass, central concentration 
and $\theta_{E,eff}$. Since the CLASH sample does not span very wide range in the cluster 
redshift, mass and concentration, it is perhaps not surprising that there are no clear 
correlations between the lensing efficiency and the redshift, the mass or the 
concentration for both the CLASH and MOKA samples. However, there is a very 
significant correlation between the MOKA lensing efficiency and 
$\theta_{E,eff}$, and the correlation can be described by the following formula:
\begin{equation}
N_{arc} = (0.03 \pm 0.01)\theta_{E,eff}^{1.54 \pm 0.08}[\rm arcsec] + (0.81 
\pm 0.22 ),
\end{equation}
as the dashed line in Figure~\ref{fig11:subfig:d} shows. The non-zero value of the y-intercept  
reflects a contribution from false positive detections (consistent 
with our estimation from simulations) and intrinsic scatter. 

\begin{deluxetable}{lccccc}
\tabletypesize{\scriptsize}
\tablewidth{0pt}
\tablecaption{Comparison of Observed and Simulated Cluster Lensing Efficiencies\label{tab3}}
\tablehead{
\colhead{ Redshift }  &   
\colhead{CLASH }  &    
\colhead{MOKA}  &   
\colhead{p-value of } & \\
\colhead{Range}  &   
\colhead{Observations}  &   
\colhead{Simulations}  &    
\colhead{ K-S test} &}
\startdata
All Clusters             & $4 \pm 1$ &  $4 \pm 1$  &   0.92 \\   
$z_{CL} \le z_{median}$  &  $3 \pm 1$ & $3 \pm 1$  &   0.99 \\
$z_{CL} > z_{median}$    &  $5 \pm 1$ & $6 \pm 1$  &   0.65\\
\enddata
\end{deluxetable}

\section{What is the Dominant Determinant of Cluster Lensing Efficiency?}
\label{s7}

We now assess the relative importance of the redshift distribution
of the lensed sources and the $c-M$ relation of the clusters on the resulting 
giant arc abundance. We accomplish this by conducting a series of simulations 
where we alter either the redshift distribution of the background galaxies or 
the assumed $c-M$ relation. While other effects such as DM substructure, halo 
triaxiality, and the mass profile of the BCG, may also play a role in determining the distribution of arc number counts, we focus 
here on studying impact of the redshift distribution and $c-M$ relation as 
these are potentially the most important effects. As shown below, however, we 
find that the lensing efficiency of CLASH-like clusters is not very sensitive 
to the redshift distribution of the background galaxy population so long as there
is a significant fraction of the source galaxy population that lies at $z > 1$. 
We also find that the lensing efficiency is quite sensitive to the DM concentration 
distribution.

\subsection{Simulated Lenses, Background Sources all at $z_s = 1$ or $z_s = 2$}

We start by testing how the source redshift distribution affects the arc 
abundance. We use the same 160 simulated MOKA clusters but first set all 
the UDF source redshifts to $z_s = 1$ and, in a separate realization, then set all
source redshifts to $z_s = 2$ to see the impact of a delta function redshift distribution (which is
obviously an extreme assumption). We then perform the ray-tracing
to create 640 new simulated images for each case. We run arcfinder on these 
images and detect 1748 and 3764 arcs in total, respectively, when $z_s = 1$ and $z_s = 2$. After applying 
statistical corrections, we find  lensing efficiencies of $2 \pm 1$ ($z_s = 1$) and 
$5 \pm 1$ ($z_s = 2$). The lensing efficiency decreases by a factor about 2 when the 
background redshift distribution is a delta function with all sources at $z_s = 1$. 
However, when putting all sources at $z_s = 2$ one obtains a similar lensing efficiency 
as that obtained when using realistic UDF 
redshift distribution. The distributions of arc number per 
cluster for these 3 cases are shown in Figure~\ref{fig12:subfig:a}.  K-S tests 
indicate that the arc number distributions when using the UDF 
redshift distribution and using a delta function at $z_s = 2$ are consistent (p-value = 0.45). 
The arc number distribution when assuming a  
$z_s = 1$ delta function differs significantly from that with UDF redshift distribution or $z_s = 2$ 
delta function redshift distribution (K-S test p-value = $3.5\times 10^{-6}$).

\subsection{CLASH Mass Models, UDF Redshift Distribution for the Background Galaxies}

Given the CLASH mass models (CLMM), we would like to check if the UDF field is 
representative as a background source for the simulations. We use the publicly 
available mass models of 19 CLASH X-ray selected clusters \citep{zit14} to lens
the UDF source galaxies, and to create 152 simulated images. We detect 656 arcs 
from these images, corresponding to a lensing efficiency of $ 3 \pm 1$. This efficiency differs from
that found for the actual CLASH images ($4 \pm 1$) by $0.7\sigma$. 
The distributions of arc number per cluster are consistent with one another (see Figure~
\ref{fig12:subfig:b}). A K-S test gives the p-value = 0.42.

\subsection{CLASH Mass Models, Background Sources all at $z_s = 1$ or 
$z_s = 2$}

We now assess whether the lensing efficiency is altered significantly when using
the CLASH mass models along with delta function redshift distributions. 
Again, we arbitrarily place all the UDF sources redshift to $z_{s} = 1$ and 
$z_s = 2$, and perform ray tracing through 19 CLASH mass model to create 152 new
simulated images for each case. We detect a total of 414 and 670 
arcs for the $z_s = 1$ and $z_s = 2$ source distributions, respectively. These correspond 
to lensing efficiencies of $2 \pm 1$ and $3 \pm 1$.
Similar to that in the MOKA simulations, the lensing efficiency and distribution of arc 
numbers are similar for simulations with UDF redshift 
distribution and $z_s = 2$ (p-value = 0.5). Whereas the lensing efficiency for 
$z_s = 1$ is again about 2 times lower than that with UDF redshift distribution and 
$z_s = 2$, and the arc number distribution for $z_s = 1$ is also significantly different 
(K-S p-value = $1.8\times 10^{-4}$). Figure~\ref{fig12:subfig:c} shows the 
distributions of arc number per cluster of the three samples.

\subsection{Different $c-M$ Relations, UDF Redshift Distribution for the 
Background Galaxies}

\begin{figure}
\centering
\subfigure[]{
 \label{fig12:subfig:a}
 \includegraphics[width = 0.45\linewidth]{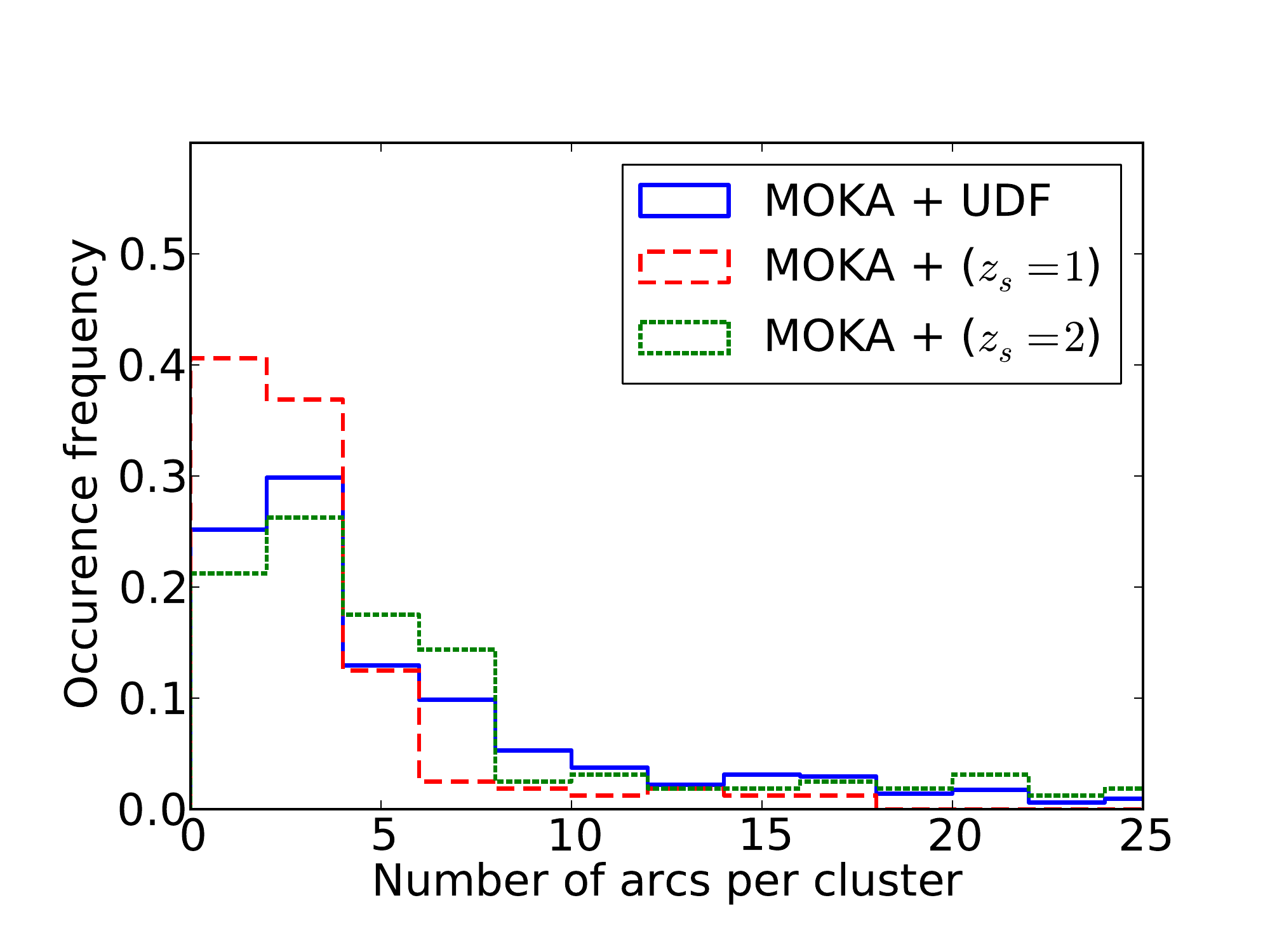}
}
\centering
\subfigure[]{
 \label{fig12:subfig:b}
 \includegraphics[width = 0.45\linewidth]{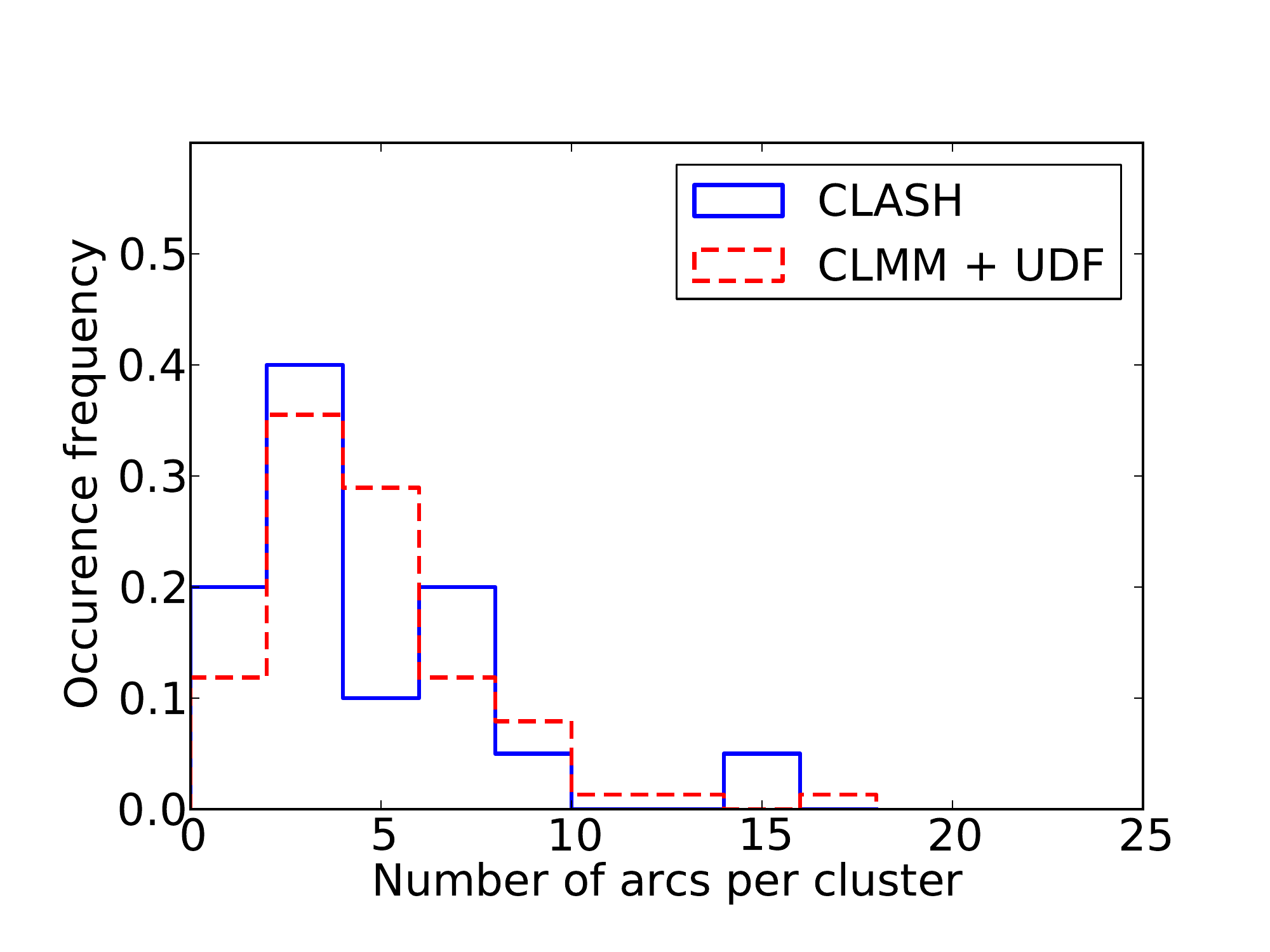}
}
\centering
\subfigure[]{
  \label{fig12:subfig:c}
 \includegraphics[width = 0.45\linewidth]{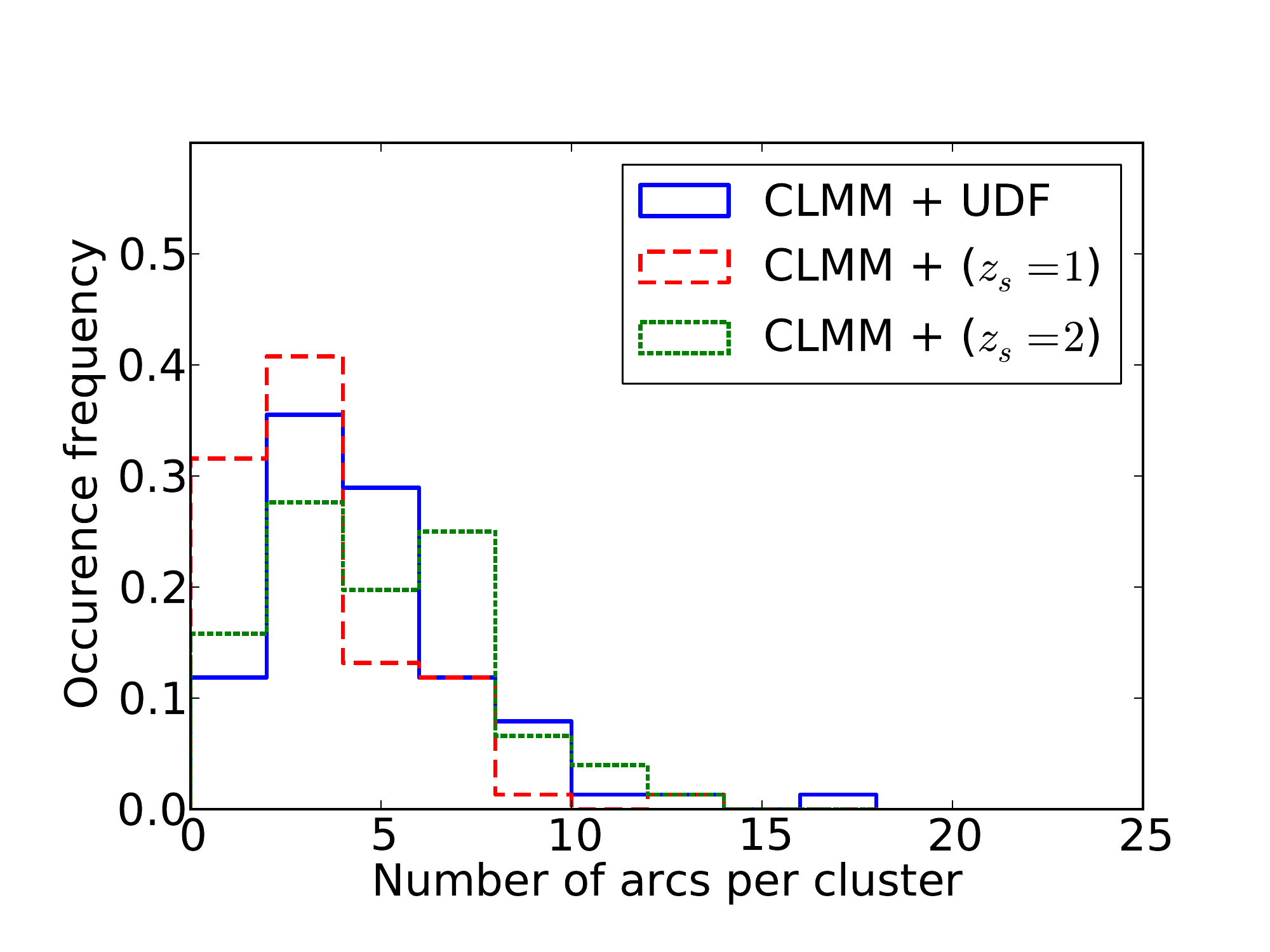}
}
\centering
\subfigure[]{
 \label{fig12:subfig:d}
 \includegraphics[width = 0.45\linewidth]{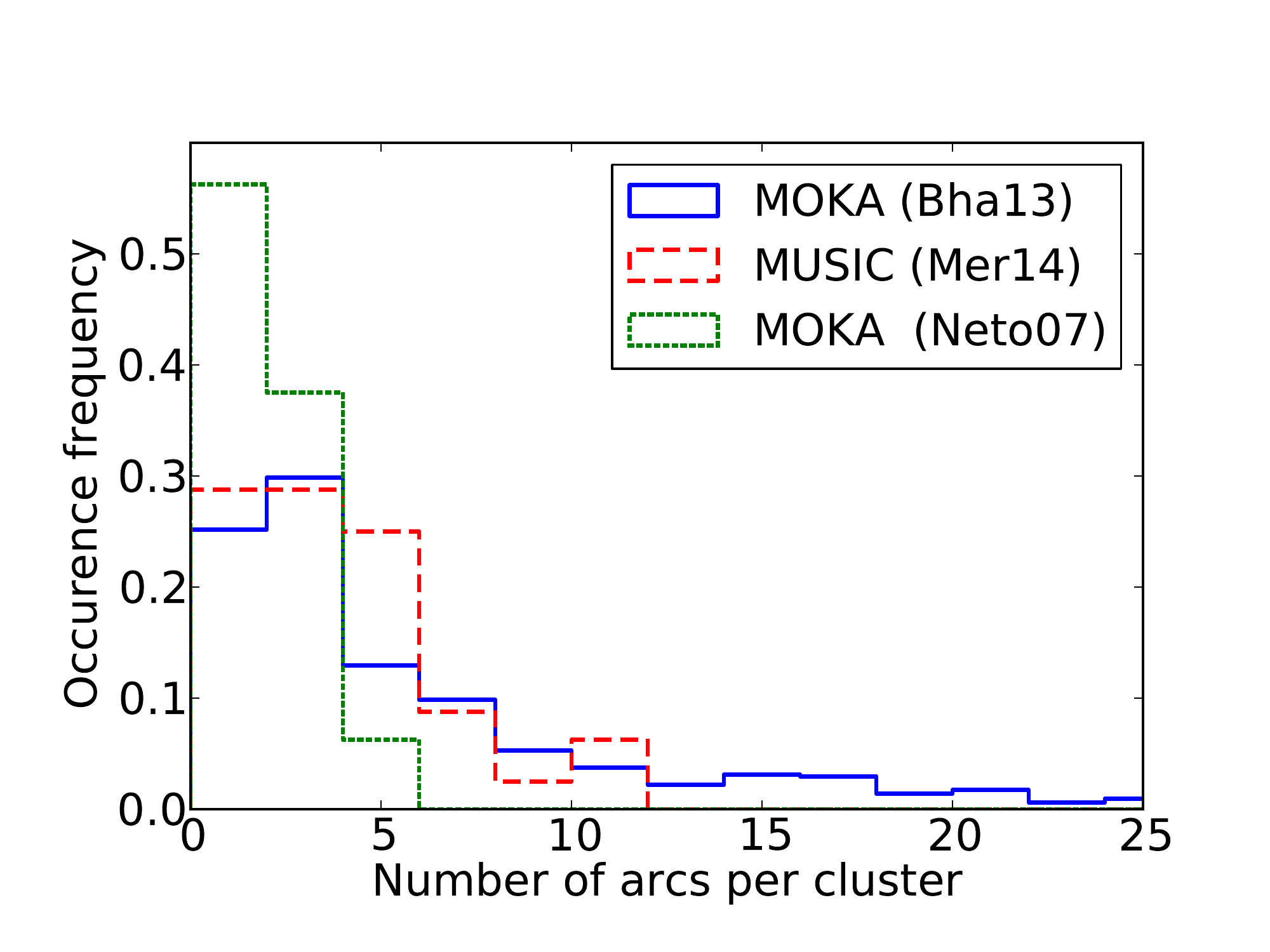}
}

\caption{Comparison of the distribution of arc number per cluster between
diffrent samples. (a) shows the comparison of the arc number distribution between 
samples with MOKA mass models and different source redshift distributions; (b) 
shows the comparison of the arc number distribution between the CLASH sample and 
CLMMs (CLASH mass models) using the UDF redshift distribution; (c) shows the comparison of the 
arc number distribution between samples with CLMMs and different source redshift 
distributions; (d) shows the comparison of the arc number distribution between 
samples with the same source redshift distributions and mass models, but implemented 
with different $c-M$ relations.}
\label{fig12:subfig}
\end{figure}

Here we show how the arc abundance depends on the cluster $c-M$ relation. 
Using the UDF redshift distribution, we re-simulate 160 new clusters and 
simulated images with MOKA by adopting the $c-M$ relation in \citet{net07}, instead of 
\citet{bha13}. We detect 230 arcs from 160 realizations using the \citet{net07} $c-M$ relation,
which, after corrections, yields a lensing efficiency of $1 \pm 1$. The lensing 
efficiency is a factor of 4 lower using the \citet{net07} $c-M$ relation than when we adopt
the \cite{bha13} $c-M$ relation. This arc abundance is seen to be quite
sensitive to the parameters of the $c-M$ 
relation. As above, Figure~\ref{fig12:subfig:d} shows the arc number distributions of three 
samples.

\section{MUSIC Lensing Simulations}
\label{s8}

Although the lensing efficiency in semi-analytic MOKA simulations is in 
excellent agreement with that found in the CLASH observations, it is important 
to make sure this is a robust result. Thus, we study a different suite of 
simulations to determine the arc abundance using simulated clusters drawn 
directly from high resolution, hydrodynamical simulations. For this, we use 
a set of mock clusters taken from the MUSIC-2 N-body/hydrodynamical 
simulations \citep{men14}. The MUSIC-2 sample \citep{sem13,bif14} consists of a 
mass limited sample of re-simulated halos selected from the MultiDark 
cosmological simulation. This simulation is dark-matter only and contains 
$2048^3$ particles in a $(1h^{-1} \rm Gpc)^3$ cube, which was performed in 2010 
using ART \citep{kra97} at the NASA Ames Research center. All these 
simulations are accessible from the online MultiDark Database2 . The run was
using the best-fitting cosmological parameters to WMPA7+BAO+SNI ($\Omega_M$ = 
0.27, $\Omega_b$ = 0.0469, $\Omega_{\Lambda}$ = 0.73, $\sigma_8$ = 0.82, $n$ = 
0.95, $h$ = 0.7). There were 282 cluster-scale halos in the simulation box which are more massive 
than $10^{15} h^{-1} M_{\odot}$ at redshift $z = 0$ and are selected to construct our sample. 
All these massive clusters were re-simulated both with and without 
radiative physics. The initial conditions for the re-simulations were 
generated in a finer mesh of size $4096^3$, by following the zooming technique 
described in \citet{kly01}. By doing so, the mass resolution of the 
re-simulated objects corresponds to $m_{DM} =9.01\times10^8 h^{-1} M_{\odot}$ 
and to $m_{SPH} =1.9\times10^8 h^{-1} M_{\odot}$, which was improved by a 
factor of 8 with respect to the original simulations. The parallel TREEPM+SPH 
GADGET code \citep{spr05} was used to run all the re-simulations. Snapshots for
15 different redshifts in the range $0 \le z \le 9$ are stored for each 
re-simulated object. The snapshots which overlap with the redshifts of the 
CLASH clusters are at $z = 0.250, 0.333, 0.429$ and $0.667$.

These re-simulated cluster halos were originally used to estimate the expected
concentration-mass ($c-M$) relation for the CLASH cluster sample \citep{mer15, men14}. 
As in these works, we use the X-ray image simulator X-MAS \citep{gar04}
to produce simulated Chandra observations of the halos, and use them to further
identify objects that match the X-ray morphologies and masses of the X-ray 
selected CLASH clusters. The $c-M$ relation from our X-ray selected 
set of simulated clusters agrees with that directly derived from the CLASH 
data at the 90\% confidence level \citep{mer15} and is fully consistent with 
the stacked weak-lensing signal derived from the ground-based wide-field 
observations \citep{ume14}. We perform ray-tracing through these X-ray selected
simulated clusters (BCG and radiative physics are not included) 
to lens the UDF sources and create 100 simulated CLASH images. 

\subsection{Lensing Statistics of MUSIC Simulated Samples and Comparison with 
Real Observations }

\begin{figure}
\centering
\subfigure[]{
\label{fig14:subfig:a}
 \includegraphics[width = 0.45\linewidth]{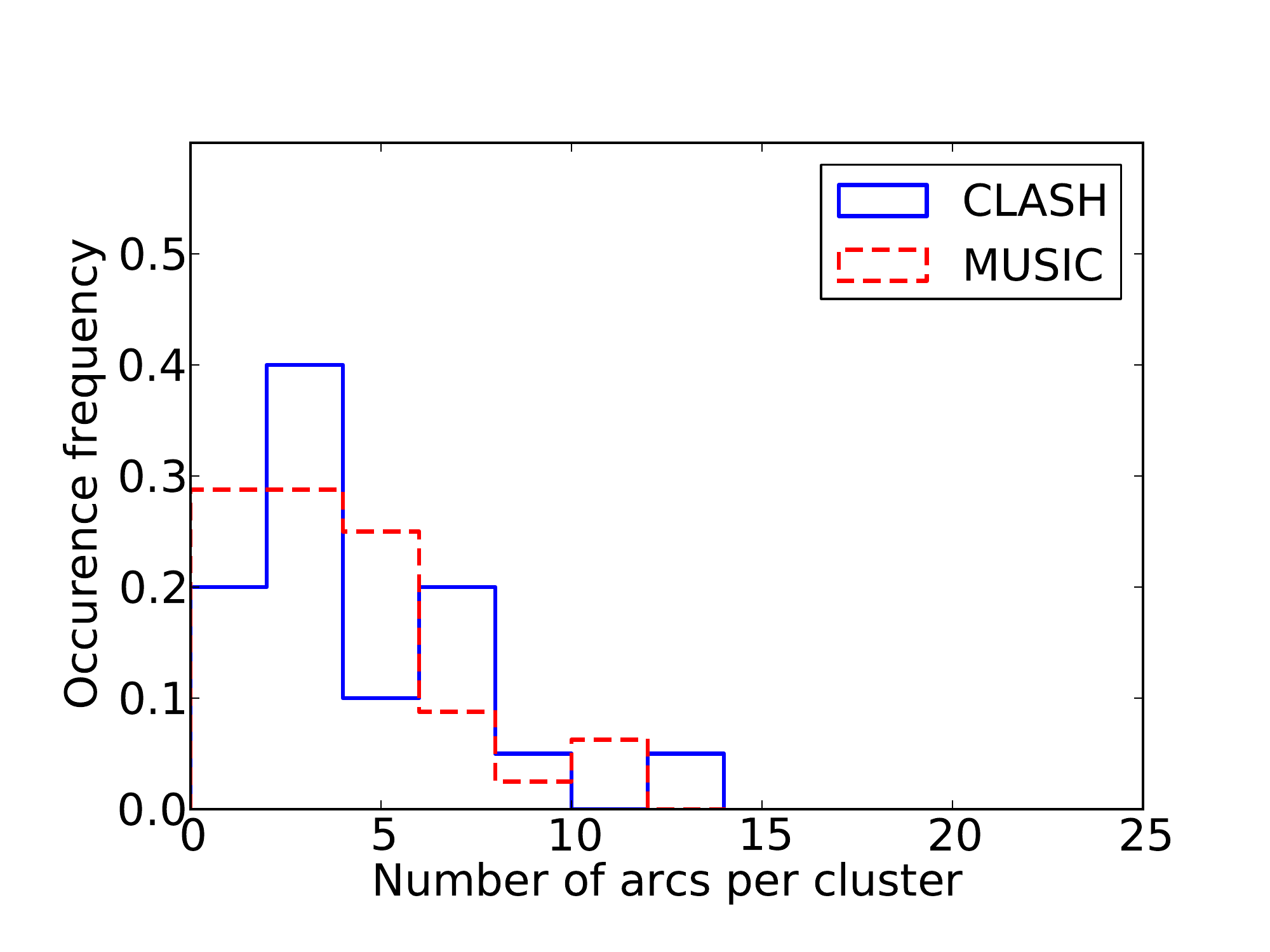}
}
\centering
\subfigure[]{
\label{fig14:subfig:b}
 \includegraphics[width = 0.45\linewidth]{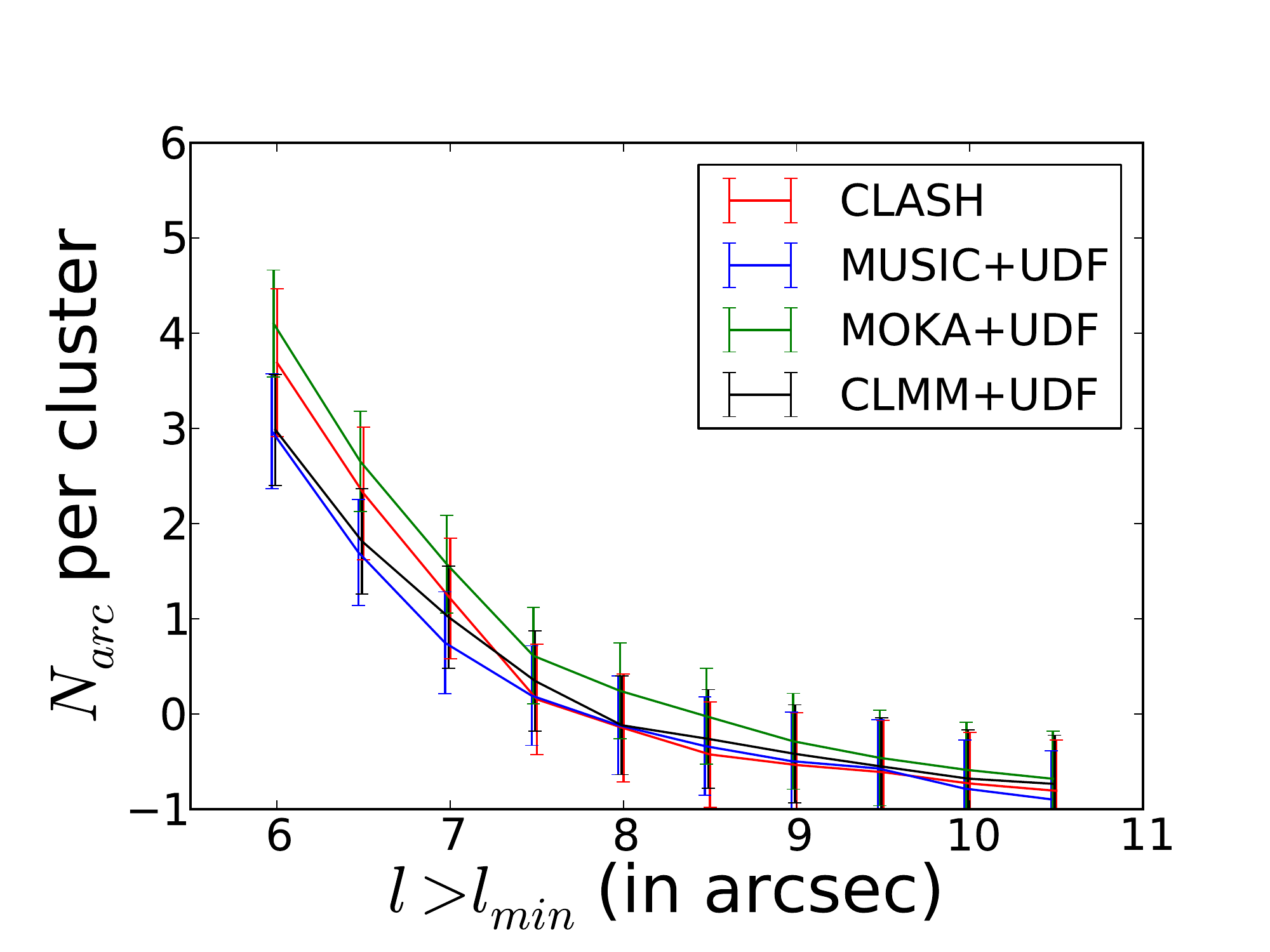}
}
\centering
\subfigure[]{
\label{fig14:subfig:c}
 \includegraphics[width = 0.45\linewidth]{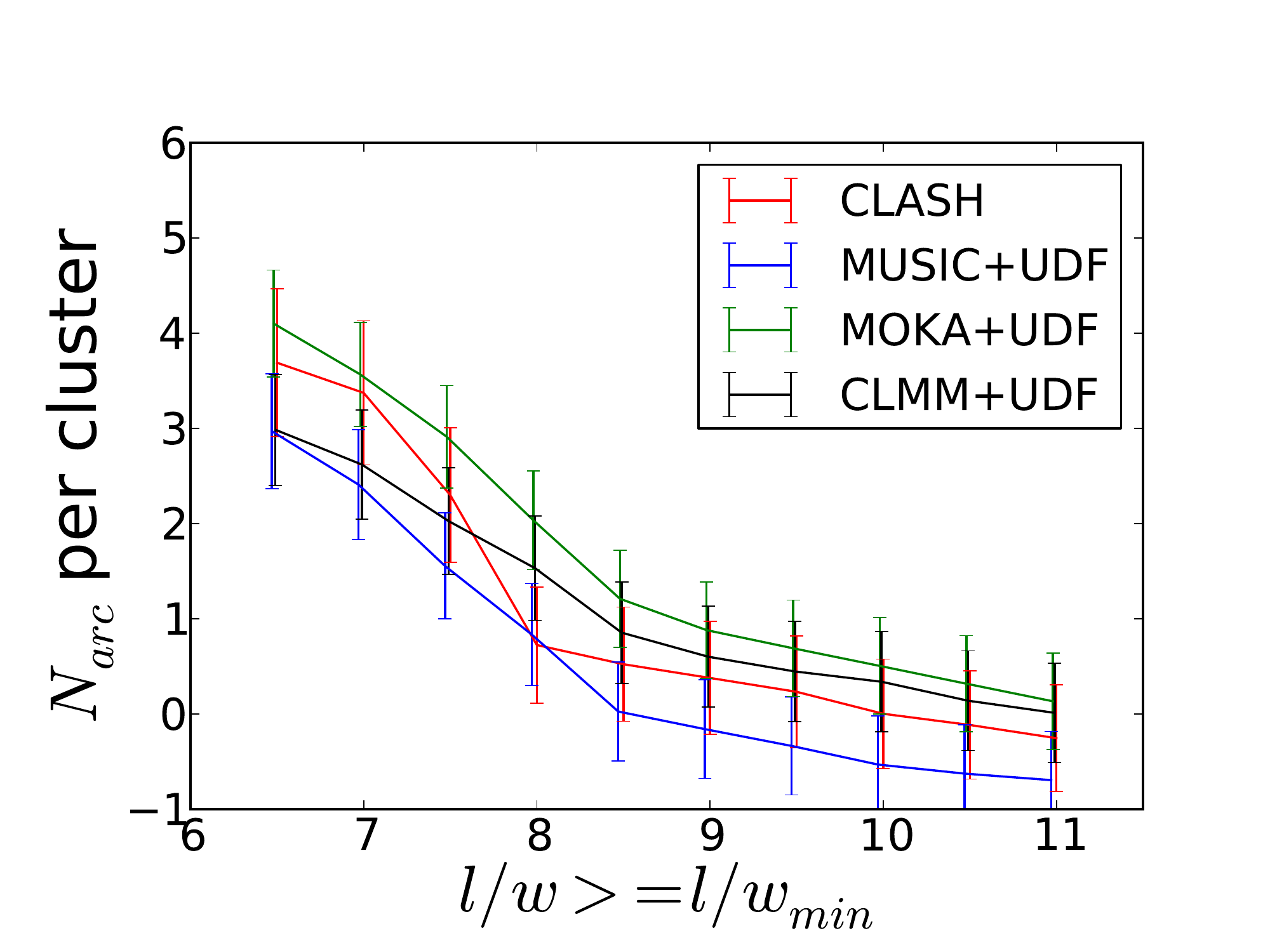}
}
\caption{(a) shows the comparison of the number distribution between 
CLASH sample and MUSIC sample; (b) shows the lensing efficiency as a function of 
$l_{min}$ for arcs with $l/w \ge 7$ for different samples; (c) shows the 
lensing efficiency as a function of $l/w_{min}$ for arcs with $l \ge 6''$ for 
different samples;  }
\label{fig14:subfig}
\end{figure}

We run the arcfinder on the 100 MUSIC simulated images and detect a total of 343 arcs with 
$l/w \ge 7$ and $l \ge 6''$. We correct the total number of
arcs for the elongation bias and incompleteness, yielding a final number of $447 \pm 24$ 
arcs, which corresponds to a mean value of $3 \pm 1$ arcs per cluster 
after application of the false positive correction. The MUSIC lensing efficiency is fully consistent
with the lensing efficiency of the observed CLASH X-ray selected sample ($4 \pm 1$). 
Figure~\ref{fig14:subfig:a} shows 
the observed and simulated distributions of arc number per cluster. A K-S test
between these two distributions has a p-value = 0.95.
We also explored the dependence of the lensing efficiency on the $l/w_{min}$
and $l_{min}$  in the MUSIC simulations (Figure~\ref{fig14:subfig:b},~\ref{fig14:subfig:c}). 
The lensing efficiency decreases with increasing $l/w_{min}$ and $l_{min}$ values, which is consistent
with the behavior seen in the CLASH observations.  
We summarize the main arc statistics results of this paper in Table~\ref{tab4}:
the second column in Table~\ref{tab4} is the rounded-off value of the mean lensing
efficiency (number of arcs per cluster); the third column is the
significance of difference in lensing efficiency between the specific simulation sample and that derived for 
the observed the CLASH X-ray selected sample. 
As with the MOKA simulations, the MUSIC simulated clusters yield
cluster lensing efficiencies that match that seen in the observations when the simulations adopt a $c-M$ 
relationship and a source redshift distribution that matches the observations.

\begin{table}
\centering
\caption{Comparison of lensing efficiency between observation and simulation}
\begin{tabular}[]{lccc}
\hline \hline
                        & Lensing efficiency    &    Difference relative to \\
                        &                       &    CLASH X-ray selected sample  \\
\hline
Observation (X-ray selected sample)                   &  $4 \pm 1$    &      \\ 
\hline
Observation (high-magnification sample)           &  $5 \pm 1$    &   $0.7\sigma$   \\
\hline
CLMM + UDF z-distn     &  $3 \pm 1$    &  $0.7\sigma$     \\
\hline
CLMM + ($z_s = 1$)           &  $2 \pm 1$    &  $1.4\sigma$    \\
\hline
CLMM + ($z_s = 2$)           &  $3 \pm 1$    &  $0.7\sigma$    \\
\hline
MOKA + UDF z-distn           &  $4 \pm 1$    &     \\
\hline
MOKA + ($z_s = 1$) + (B13) $c-M$           &  $2 \pm 1$    &   $2.2\sigma$   \\
\hline
MOKA + ($z_s = 2$) + (B13) $c-M$           &  $5 \pm 1$    &   $0.7\sigma$  \\
\hline
MOKA + UDF z-distn + (N07) $c-M$           &  $1 \pm 1$    &   $2.2\sigma$   \\
\hline
MUSIC + UDF z-distn      & $3 \pm 1$ &   $0.7\sigma$    \\
\hline

\end{tabular}
\label{tab4}
\end{table}

\section{Discussion}
\label{s9}

Since the arc statistics was originally proposed as a cosmological probe, many
previous studies have investigated the sensitivity of the arc abundance on
various cosmological effects. Cosmology enters the strong lensing properties of
the galaxy clusters in two ways: first, the arc abundance depends on the
angular-diameter distance and volumn which are determined by the cosmological
expansion; second, the arc abundance depends on the cluster abundance and
internal structure which are cosmological sensitive. N-body simulations and 
semi-analytic approaches have been utilized in earlier studies to explore the 
sensitivity of arc abundance on $\sigma_8$ \citep{wam04,li06,fed08} and an 
increaseing function of arc abundance with $\sigma_8$ has been observed, though 
whether the large increments in arc abundance when increasing the $\sigma_8$ are
quantitatively reliable is not clear; \citet{bol15} has studied the arc 
abundance dependence on $\sigma_8$ and $\Omega_m$ for a given survey area. They 
use MOKA to generate mock clusters with different mass and redshift and populate
them into the light cones spanned by the survey region. They identify the 
increasing functions of arc abundance with both parameters. The arc abundance 
seems more sensitive to $\sigma_8$ than $\Omega_m$, because $\sigma_8$ has an 
effects on the cluster formation time, which in turn affects the cluster 
internal lensing properties such as triaxiality and concentration. However, the 
degeneracy between two parameters for the arc abundance limit its ability to 
distinguish different cosmologies. The arc abundance sensitivity on various dark
energy models has also been studied, which includes a constant equation-of-state
parameter $w \neq 1$ \citep{bar03} and time varying $w$ parameter \citep{men05},
the arc abundance could change by at most a factor of a even with substantial 
change in $w$. \citet{jul10} and \citet{dal11b} studied how cosmology affects 
the arc statistics through geometry effects. They found the expansion function 
thus the cosmological models can be constrained from the ratio of the lensing 
efficiencies at different redshift. To achieve competitive results, however, the
mass distribution of the clusters must be determined with very high precision, 
and a sample of about ten clusters contaning about 20 arc families each are 
needed. Moreover, the arc statistics could even change by $30\%$ with different 
non-Gaussianity parameters based on theoretical framework \citet{dal11a}. 
Therefore, these studies may indicate that, amongest all the cosmological 
parameters, the arc abundance seems to be most sensitive to $\sigma_8$. 
Interestingly, most of the simulations in early arc statistics works have adoped
a typically higher $\sigma_8$ value $\sim 0.9-0.95$ ($\sigma_8 = 1.12$ was 
adopted in B98) than the current concensus from WMAP7 and PLANCK, which could 
have made the discrepancy between the simulations and the observations even 
larger. It implies that at least the deficit of cluster abundance under 
different cosmologies might not be the main solution to ``arc statistics 
problem'' in the first place. Since the dependence of the cluster internal 
lensing properties on $\sigma_8$ is still not well known, we simply adopt the  
value of $\sigma_8 = 0.83$ along with other cosmological parameters from the 
Planck results \citep{pla14}. We believe that our conclusion would not change 
significantly unless there is large revision in the the Planck value for 
$\sigma_8$.

As one of the promising candidate solutions to the arc statistics problem,
the impact of source redshift distribution on the arc abundance has been 
emphasized by many previous studies. \citet{wam04} studied the magnification 
probability for light rays propagating cross a cosmological scale and found that
the probability of high magnification events highly depends on the source 
redshift. They concluded that the arc abundance should have a steep increase 
with source redshift because the number of halos suitable for strong lensing 
increases exponentially with redshift, and they suggest this as the solution for
the arc statistics problem. \citet{bay11,bay12} has established a large sample 
of arcs (105) from the Sloan Giant Arcs Survey (SGAS) and from the Second Red 
Sequence Cluster Survey (RCS2) and study the redshift distribution of the arcs. 
They find that arcs with $g\le24$ have a median redshift of $z_{s} \sim 2$. 
\citet{bay12} claim the arc statistics problem can be solved by adopting their 
measured redshift distribution and using the scaling of the optical depth given
in \citet{wam04}. However, \citet{li05} and \citet{fed06} show that the scaling 
of the optical depth is very different from  what assumed by \citet{wam04} and 
that the \citet{wam04} assumption that the magnification is a good measure for 
the $l/w$ ratio of an arc is not justified in detail. Furthermore, it is unclear
if the arc identification used by \citet{bay12} (e.g. by curvature radius of 
arcs and by visual inspection) might bias the selection in favor of luminous 
and highly curved arcs. If so, the corresponding arc abundance and redshift 
distribution could also possibly be biased.

Our results show that the simulations performed either with a UDF redshift distribution or a delta function
redshift distribution at $z_s = 2$ give very similar arc abundances. When we change the redshift 
distribution of the background sources to a delta function at $z_s = 1$ the arc abundance
drops by a factor of 2 rather than the 
order of magnitude change in the arc abundance noted in some previous 
studies (e.g., \citet{wam04}). The factor of 2 change is consistent with \citet{hor05,hor11}, 
who also used UDF images
as background sources to perform the ray-tracing. \citet{hor05} used the same 
simulated clusters at $z_{c} = 0.2$ as used in B98 to lens the UDF sources and 
found an arc abundance that was 3 times higher than that in B98. They 
attributed this over-abundance to the use of a source number density that was 
3.2 times higher than that in B98. They found that changing the source 
redshift distribution from a delta function at $z_{s} = 1$ to a realistic UDF distribution results in 
only a small change in the final arc abundance. These results suggest that 
the redshift distribution does not have a major impact on the final arc abundance unless one
selects a distribution that significantly underpopulates galaxies in the $z > 1$ range. 

The MUSIC simulated halos do not have BCGs at the center and do not implement
complex gaseous physics. However, \citet{kil12} has compared the arc production 
efficiency of the adiabatic simulations with some more sophisticated simulations
which include the effects such as gas cooling, star formation, feedback from 
AGNs and SN+galactic winds, etc. The comparable results indicates that the 
implementation of baryonic physics will probably not lead to a significant 
change in the arc abundance derived from simulations without such processes.

Previous studies have already revealed the correlation between the lensing cross section 
and the Einstein radius, $\theta_{E,eff}$, from N-body simulation \citep{men11} and the semi-analytic
calculations \citep{red12}. Our study confirms this correlation as reflected by the 
dependence of  the number of arcs per cluster on $\theta_{E,eff}$, as shown in
Figure~\ref{fig11:subfig:d} shows. The relation between the MOKA cluster lensing 
efficiency and  $\theta_{E,eff}$ in our study is well fit by
a linear relation in log-log plane with a slope of $1.54 \pm 0.08$, which is 
flatter than the slope $1.79 \pm 0.04$ in \citet{men11} and $2.4 \pm 0.04$ in 
\citet{red12}. The detection of this correlation in our MOKA simulations is due to 
the relatively large size of the MOKA cluster sample (640 simulated 
clusters), whereas the CLASH sample is too small to robustly unveil this correlation. For the arc abundance 
$\sim 5$ per cluster, the fractional error for an ensemble of 32 realizations is
 $1 / \sqrt{5\times 32} \sim 8\% $. Therefore, to measure the correlation 
observationally to 10\% and to detect a $\sim 15 \%$ deviation from such 
correlation, we need $\frac{\times (1/0.1)^2}{5} = 20$ clusters in each mass 
bin, and we probably need a cluster sample with size $\sim$200 if 10 different 
mass bins are expected.  

We are able to identify the relative significance of several key
physical effects which contribute to the arc abundance enhancement. 
As Table~\ref{tab4} shows, varying the source redshift distribution leads to, 
at most, a factor of 2 variation in the arc abundance. Variation of the $c-M$ 
relations will affect the matter distribution of the inner cluster core and, hence,
lead to variations in the arc abundance. Using several recent estimates of the $c-M$ relation
\citep{net07,bha13} results in variations of the arc abundance by up to a factor of $\sim 4 - 5$.
Using the most recent estimates of the $c-M$ relation in simulations appears to produce
excellent agreement with the observed arc abundance.
However, quantities such as mass and concentration alone 
are not sufficient to reflect the likely complex 
dependencies of the arc abundance on various effects. As shown in 
Figure~\ref{fig11:subfig:b} and \ref{fig11:subfig:c}, the arc abundance fails to
exhibit a strong dependence on either the concentration or the cluster mass alone, 
for both the CLASH and MOKA samples. By contrast, the effective Einstein radius, 
$\theta_{E,eff}$, is a good indicator of the lensing efficiency. 

Given our results, even without fully understanding the cosmological dependence
of the arc abundance, we could still conclude that the initial ``arc 
statistics problem'' appears to have been largely due to 
inadequate modeling of the mass distributions of the clusters and, secondarily, 
due to inadequate modeling of the background source number density and redshift 
distribution. In addition, the previous use of mostly visual identification of 
arcs may have resulted in an inadequate modeling of the false positive 
contamination rate and completeness corrections. We can divide the 
contributions from different physical effects on cluster lensing efficiency into 
three general categories: the cluster abundance, the background source redshift
distribution, and the individual cluster lensing cross section. Our study would suggest 
that the lensing efficiency is more strongly dependent on the individual cluster lensing cross sections 
than on the source redshift distribution. However, different cosmology
could alter both the cluster abundance and the individual cluster lensing cross
sections and the relative significance of such factors has not been explored 
in this study given the small cluster sample size. Future large cluster 
surveys (e.g., DES, LSST, Euclid, WFIRST) will definitely help to answer this question. We suspect that 
two other related problems in lensing,  the over-concentration problem and 
Einstein radii problem,  where it has been found that some real clusters at 
intermediate redshift have denser cores than clusters of similar mass produced 
in simulations \citep{bro08,ogu09,ric10,ser10,mer15} and where some real clusters have 
larger Einstein radii than expected in standard $\Lambda CDM$ cosmology, may  
well be due to a combination of insufficiently accurate cluster simulations and 
observational sample selection effects. 

\section{Summary}
\label{s10}

We have carried out an observational and theoretical study of the arc statistics
problem in clusters of galaxies. 
We have devised an automated arcfinder to efficiently and objectively detect arcs. 
We test our arcfinder 
using a large number of simulated cluster images and have quantified the incompleteness and 
false positive rate in arc detection. We also investigate how image noise affects 
the shape determination of the arcs and statistically correct for the observed elongation bias. 
We run our arcfinding algorithm on 20 X-ray selected CLASH clusters and 5 
high-magnification CLASH clusters. After correcting for arc shape elongation 
bias, incompleteness and false positive rate we find a large arc 
($l/w > 6.5$ and $l \ge 6''$ ) lensing efficiency of $4 \pm 1$ arcs per cluster and $5 \pm 1$
arcs per cluster, 
respectively, for the X-ray selected and high-magnification selected CLASH 
samples.

We simulate mock clusters using both the MOKA semi-analytic cluster generator and
the MUSIC-2 N-body results. In both cases, we focus on simulated clusters that have the
same mass and redshift range as the CLASH clusters. For the MOKA simulations, we use ray-tracing 
to create 640 simulated cluster realizations with the F775W UDF 
image as the background source. 
For the simulations extracted from the
high resolution, hydrodynamical simulations (MUSIC), we identify halos that, in addition to having
similar redshifts and Virial masses as the CLASH clusters, are also selected to have similar 
X-ray morphologies as the CLASH clusters. 
We find a lensing efficiency of $4 \pm 1$ arcs
per cluster in the MOKA sims and $3 \pm 1$ arcs per cluster in the MUSIC sims. These lensing
efficiencies both match the observed lensing efficiency of $4 \pm 1$ arcs per 
cluster. We also study the arc abundance dependence on the cluster 
redshift by splitting the sample into two bins divided at the median sample redshift of $z_{median} = 0.352$
and find no significant differences in either the overall lensing efficiency and arc redshift distributions.
The dependence of the MOKA and MUSIC lensing efficiencies on $l_{min}$ and $l/w_{min}$ 
also match that seen in the
observed CLASH ones. 

For the future, the relative short running time (less than 5 minutes for images 
with $3000\times 3000$ pixels) of our arcfinder allows us to 
perform large-scale ``blind'' searches for giant arcs in various other surveys, especially those with moderately
high-angular resolution such as WFIRST and Euclid.  Moreover, continued study of the correlation
between the arc abundance and the $\theta_{E,eff}$ should be conducted to
 assess just the reliability of using arc abundance (which is an observable) 
as a predictor of $\theta_{E,eff}$.

\acknowledgments 
We thank the referee for providing helpful comments and suggestions that significantly improved the paper.
We thank Carlo Giocoli for making the MOKA code accessible and for his generous help in 
simulating the clusters. We thank the MUSIC group for also providing us with simulated cluster 
data sets. We acknowledge Matthias Bartelmann, Dan Coe, Colin Norman and Brice 
Menard for many useful discussions. 
Bingxiao Xu is supported by NASA funding received for the CLASH Multi-Cycle Treasury Program (HST-GO-12065).
Adi Zitrin is supported by NASA through Hubble Fellowship
grant \#HST-HF2-51334.001-A awarded by STScI. Julian Merten is supported by the People 
Programme (Marie Curie Actions) of the European Union Seventh Framework 
Programme (FP7/2007-2013) under REA grant agreement number 627288. 
The results in this paper are based on observations made with the NASA/ESA Hubble Space Telescope.
The Space Telescope Science Institute is operated by the Association of Universities for Research in Astronomy, Inc. under NASA contract NAS 5-26555.

\clearpage
\appendix

\section{Integrated Quantized Intensity-Difference Criterion}
\label{A}

The following appendices provide further details about the arcfinder algorithm. Specifically, we provide short summaries of the key steps performed to go from the initial science image to the final arc catalog.
We begin by convolving our HST images with a square 
Top-hat kernel with an edge dimension of $0.065''$ to modestly enhance the 
contrast of the faint and thin arcs. Most  source detection algorithms work in 
intensity space, which means the performance of these algorithms largely depends
on how the detection threshold is chosen. A higher threshold will yield a 
catalog with lower completeness for faint objects while a lower threshold will 
lead to less precise segmentation and a higher false positive rate. To avoid the
non-trivial determination of an optimal detection threshold, we focus on three 
very general properties of giant arcs:

\begin{enumerate}
\item  Giant arcs, like all real astronomical sources, have a net positive 
amount of flux on average after subtracting off a suitable background level.
\item  Giant arcs have substantial angular lengths.
\item  Giant arcs are highly elongated objects.
\end{enumerate}

The above general properties imply that, on average, the intensity difference 
between the pixels belonging to the arc should be positive and the elongated and
distorted morphologies of arcs should also be reflected in the angular 
distribution of these intensity differences. Use of the non-parametric intensity
differences has a genuine advantage in the arc detection game: we can, in 
principle, detect faint structures almost as easily as bright structures. For 
this key reason, we perform the primary arc detection process in 
intensity-difference space. To do this, we first lay down a grid of points on 
the smoothed image, at spatial scale $n$, that is somewhat larger than the arc 
widths we wish to find. At each grid point we then determine whether each of its
8 adjacent grid points (up, down, left, right, upper-left, upper-right, 
lower-left, lower-right) is brighter or fainter than this pixel. We quantify 
this local set of flux differences by assigning a value of  $+1$ for positive 
difference (the central pixel at grid position $(i,j)$ is brighter than a given 
surrounding pixel) and a value of $-1$ for a negative difference (the central 
pixel at grid position $(i,j)$ is fainter than a given surrounding pixel). We 
sum up these values for all 8 directions. A grid point that was brighter than 
all of its surrounding grid points would thus have a final value of $+8$. A grid
point that was brighter than 6 of its surrounding grid points would have a final
value of $6 - 2 = +4$. And so on. As arcs are highly elongated, pixels lying 
along the ridge line of an elongated arc will tend to have at least 4 or 5 
adjacent pixels that are fainter than those at a given grid position. The value 
assigned to these pixels will thus be at least 2 or higher ($5 - 3 = 2$). 
In general, the brighter pixels in an arc will tend to have higher integrated quantized
intensity-difference values than the fainter pixels. Given
that some giant arcs may have complex intensity profiles we set the threshold for the 
integrated quantized intensity difference to be the lowest positive value, which is $+2$.
If we adopt a higher positive threshold, we find that some complex arcs are segmented
into several smaller arc detections. The threshold of $+2$ is the most conservative in
maintaining the overall structural shape of the arc candidates. We note that the exact choice
of threshold value, however, does not significantly impact the contents of the final sample of large ($l \ge 6"$) and
highly elongated ($l/w \ge 7$) arc candidates. The effect of the quantized intensity difference threshold is 
primarily on the number of small and less elongated sources in the initial detection process.

\begin{figure}
\centering
\includegraphics[width = 0.7\linewidth]{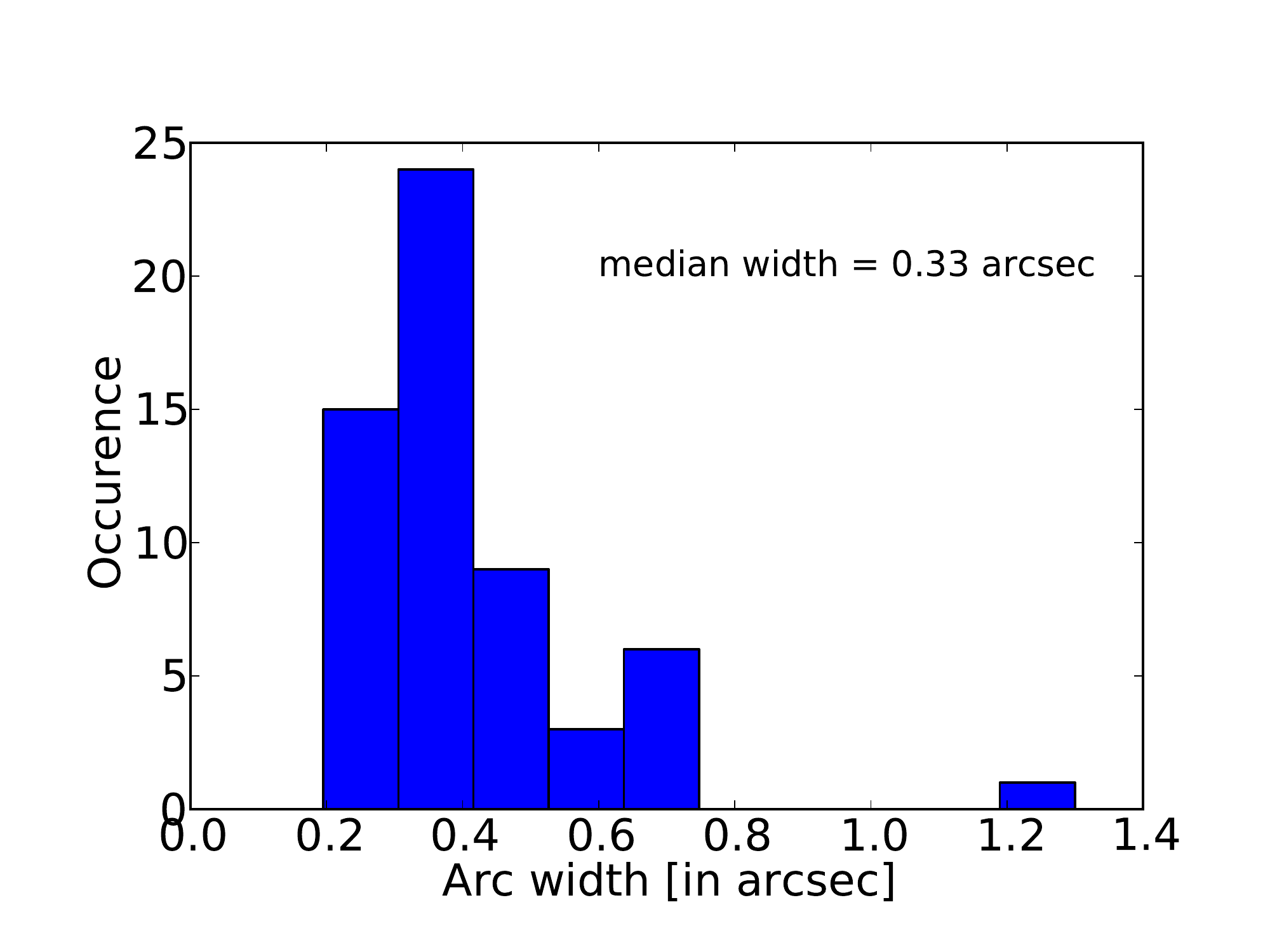}
\caption{The FWHM distribution of the pre-selected 58 giant arcs from ClASH 
F814W images. The median FWHM is around $0.33''$, and most of arc
widths are less than $0.72''$. The exception amongst this sample is from the 
cluster MACS1206, which includes a giant arc with width $\sim 1.3''$.}
\label{fig1a}
\end{figure}

Choosing a proper grid spacing scale, $n$, is important. 
Generally, the spacing scale $n$ should be larger than the typical arc's width, and it should neither be too large nor too small, to avoid extending the grid 
points to nearby bright structures or limiting the grid points around the arc 
rigid lines. To determine the scale, we visually select 58 giant arcs from our 
CLASH F814W images, and manually measure the arcs' full width at half maximum 
(FWHM) in the direction  perpendicular to their ridge lines 
\footnote{To measure the FWHM, we first draw a line crossing the intensity 
maxima which is perpendicular to the arc's ridge line, then use Gaussian profile
to fit the intensity of pixels that fall on the line. We approximate the 
Gaussian FWHM as the FWHM of the arc. }. Figure~\ref{fig1a} shows the 
distribution of the pre-selected arc's FWHM. Note that the median value of these
58 arcs is $0.33''$ and most of the arcs widths are less than $0.72''$. In 
principle we should traverse as many grid scales as possible to optimize the 
detection of the arcs, which is computationally expensive. We adopt two 
different scales: $0.39''$ and $0.78''$, to make sure that both narrow and wider
arcs can be effectively detected in a relatively short computational time. The 
results based on each scale are combined as the input to the next step.

As noise pixels may have regions with zero-valued or negative integrated 
quantized intensity-difference \footnote{For noise pixels, if their 
distributions are independent, the integrated quantized intensity-difference 
should be equal to 0.}, another obvious advantage is that we are able 
to effectively clip out noise pixels and make the arc detection task 
significantly easier, even in the presence of a bright diffuse background, as 
might be encountered in the halo of a bright foreground cluster galaxy.

\section{The Local Intensity Difference Criterion}
\label{B}

In certain regions (especially in the inner cores of bright galaxies), 
applying the integrated quantized intensity-difference criteria only will leave 
the segments with the diffraction pattern (see Figure~\ref{fig2:subfig:a}). To 
suppress these effects, we apply another criterion by comparing the intensity of
the central pixel with the mean value of all 8 adjacent pixels over the image. 
The selected pixels should satisfy two criteria below:

\begin{eqnarray}
 \sum_{l,m\in(-n,0,n)} SIGN(I(i,j) - I(i+l,j+m)) \ge 2 \\
 \sum_{l,m\in(-n,0,n)} (I(i,j) - I(i+l,j+m)) > 0    
\end{eqnarray}

Figure~\ref{fig2:subfig:b},~\ref{fig2:subfig:c} show the integrated quantized 
intensity-difference maps of MACS0717 before and after applying the above 
criteria. We can see that number of diffraction artifacts in the image is 
significantly reduced. 

\section{Initial Image Segmentation}
\label{C}

\begin{figure}
 \centering
  \subfigure[Segment with the diffraction pattern]{
  \label{fig2:subfig:a}
  \includegraphics[width = 0.4\linewidth]{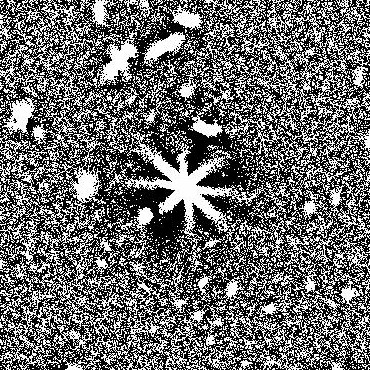}
}
  \centering
  \subfigure[Image before applying the local average criteria]{
  \label{fig2:subfig:b}
  \includegraphics[width = 0.4\linewidth]{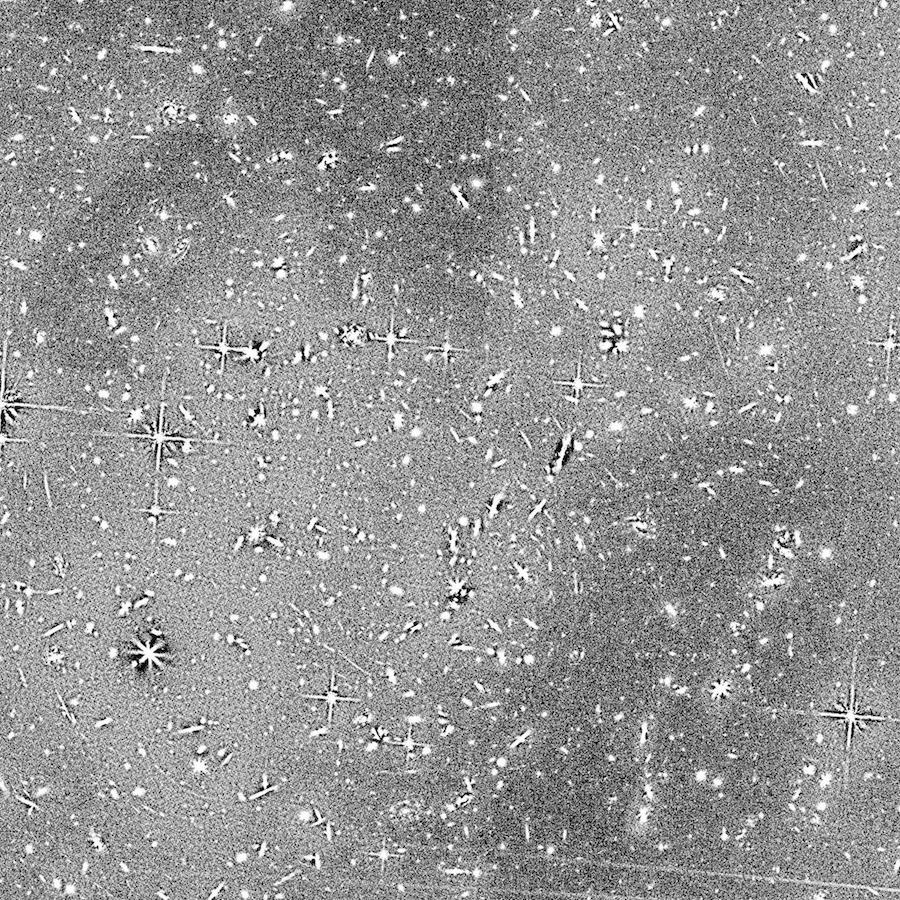}
}
  \centering
  \subfigure[Image after applying the local average criteria]{
  \label{fig2:subfig:c}
  \includegraphics[width = 0.4\linewidth]{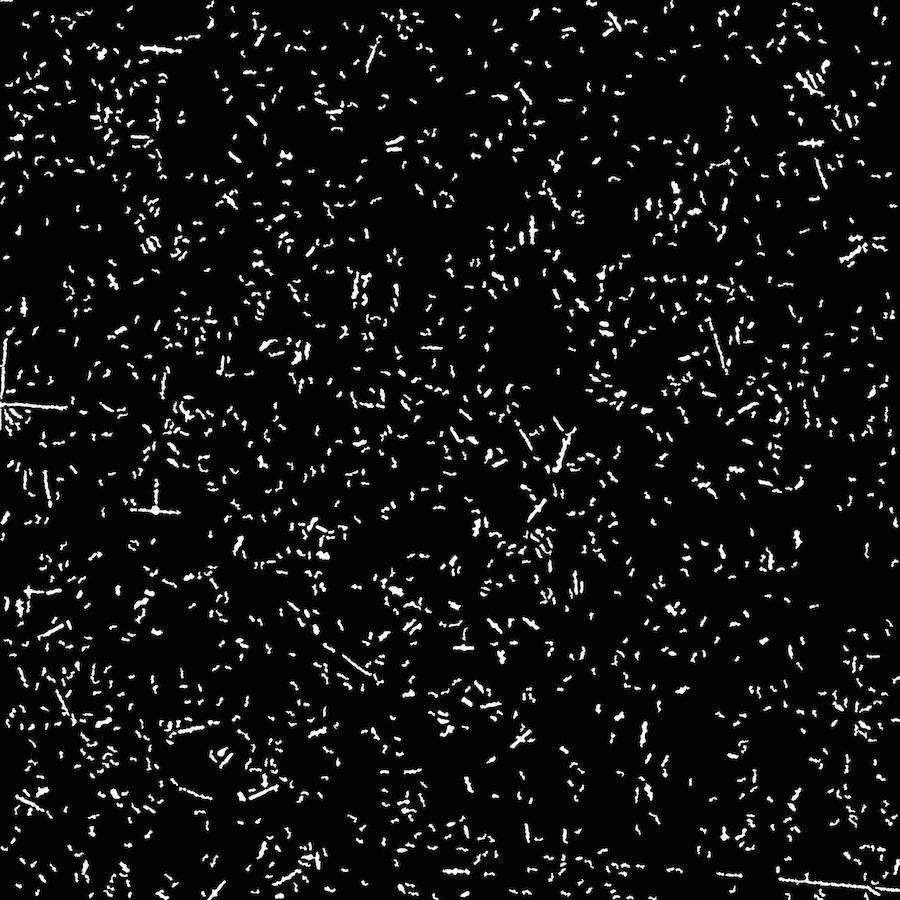}
}
  \centering
  \subfigure[Local contiguous regions selection]{
  \label{fig2:subfig:d}
  \includegraphics[width = 0.4\linewidth]{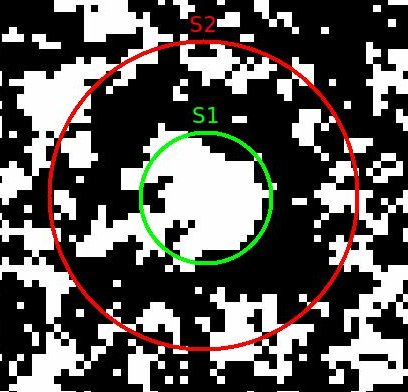}
}

\caption{(a) shows the segment with the diffraction pattern which needs to be 
suppressed by the local average criteria; (b), (c) are the integrated quantized 
intensity-difference map of MACS0717 F814W image, before and after applying the
local average criteria; (d) illustrates the idea of local selection of 
contiguous regions: to draw circles with different size on each pixel and 
calculate the average number density within the circles, and select those grid
points which have higher average number density within smaller circles. }
\label{fig2:subfig}
\end{figure}

To identify specific arcs, we need to locate regions of contiguous grid points 
in the integrated quantized intensity-difference map with sums in excess of 
$+2$. We have now replaced the challenge of finding objects in intensity space 
with the task of finding contiguous regions in this quantized 
intensity-difference space. We avoid using any global selection criterion on 
number density since the number density varies largely across the whole image. 
So the contiguous regions are selected by their local number density of the 
grid points in the quantized intensity-difference space. Based on the simple 
fact as Figure~\ref{fig2:subfig:d} shows: if the contiguous region is enclosed 
by circle S1, the local averaged number density inside S1 must be larger than 
that inside circle S2 which has the same center with S1 but larger radius. The 
details of contiguous regions selection are as follows: (1) we make three 
convolved images using three spherical uniform kernels (k1, k2 and k3) with 
increasing size ($0.52''$, $1.04''$ and $1.56''$); (2) we subtract an image 
convolved with a broader kernel from one convolved with a narrower kernel, to 
obtain two residual images (k2 - k1, k3 - k2); (3) we then select all the pixels
which have positive values in both residual images.

The selected contiguous regions include a few small and less elongated blobs 
that are not real sources. We set an area threshold $A > $ 100 pixls and an 
eccentricity \footnote{The eccentricity here is equal to the eccentricity
of the ellipse that has the same second-moments as the measured object} 
threshold $e > 0.85$ to remove these artifacts. As shown in 
Figure~\ref{fig2:subfig:b}. the noise has been suppressed and most giant arcs 
have been retained.

\clearpage

\section{Suppression of Diffraction Spikes}
\label{D}

Diffraction spikes from bright stars are the features  likely to account for
most of the false positive detections. The normal way to remove the star spikes 
is to locate the position of bright stars and then manually mask out the 
diffraction pattern. Here we adopt a different approach which eliminates the 
need to know the position of the bright stars or the direction of the spikes in 
advance.

\begin{figure}
\centering
\subfigure[The unsharp masked image]{
 \label{fig3:subfig:a}
 \includegraphics[width = 0.4\linewidth]{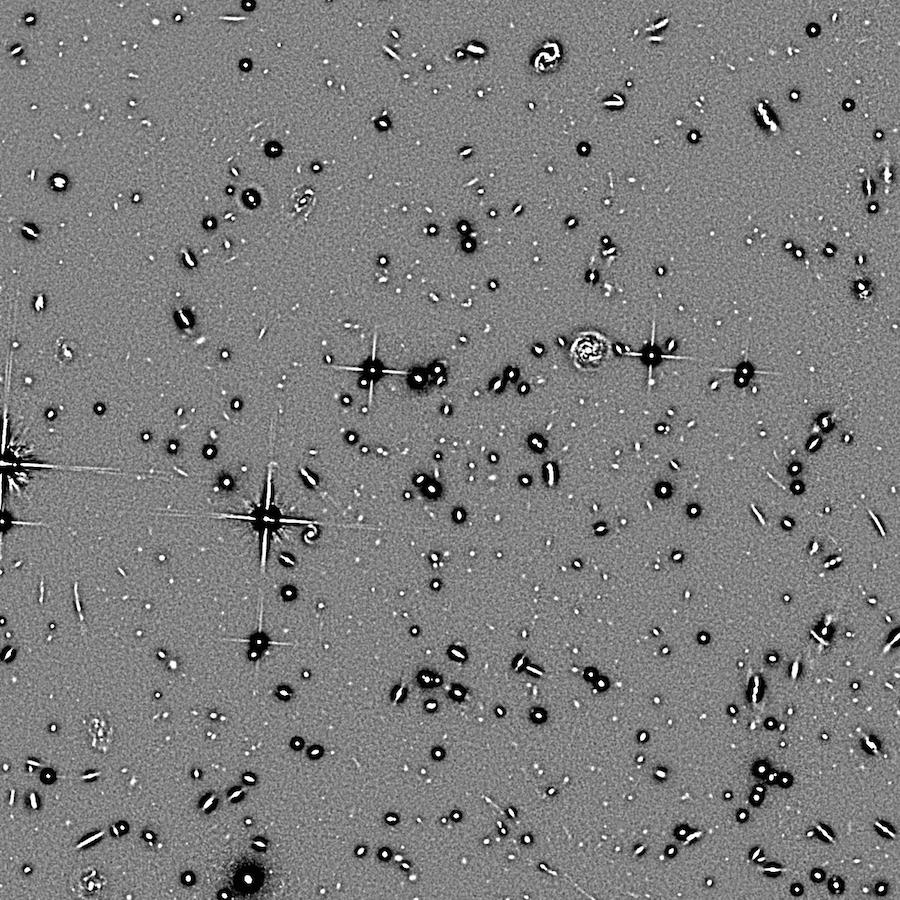}
}
\subfigure[The "dilated" halo image]{
 \label{fig3:subfig:b}
 \includegraphics[width = 0.4\linewidth]{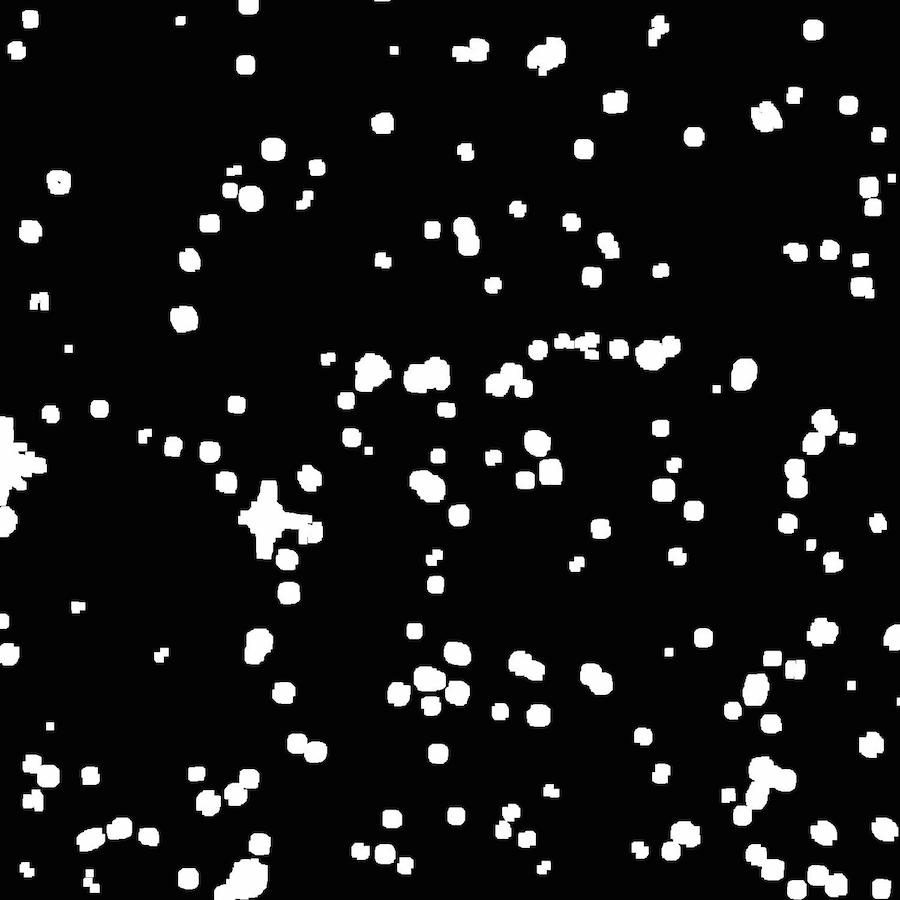}
}
\centering
\subfigure[The halo image combined with the detection image]{
 \label{fig3:subfig:c}
 \includegraphics[width = 0.4\linewidth]{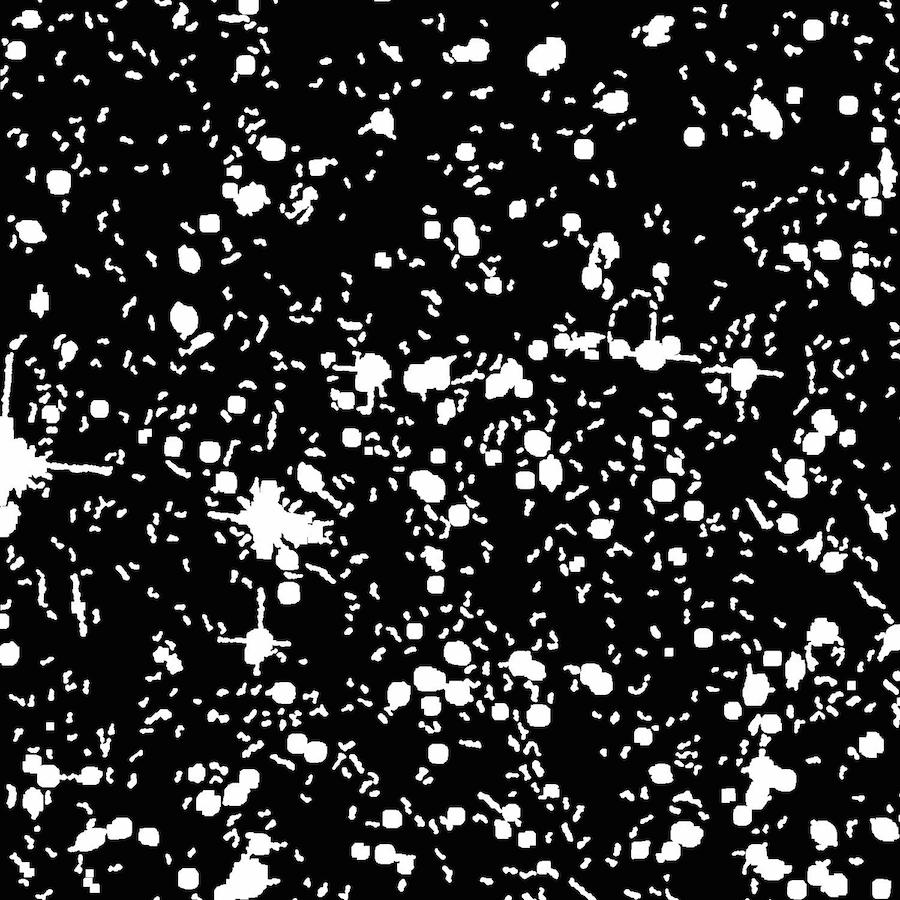}
}
\centering
\subfigure[The image with the star spikes removed ]{
 \label{fig3:subfig:d}
 \includegraphics[width = 0.4\linewidth]{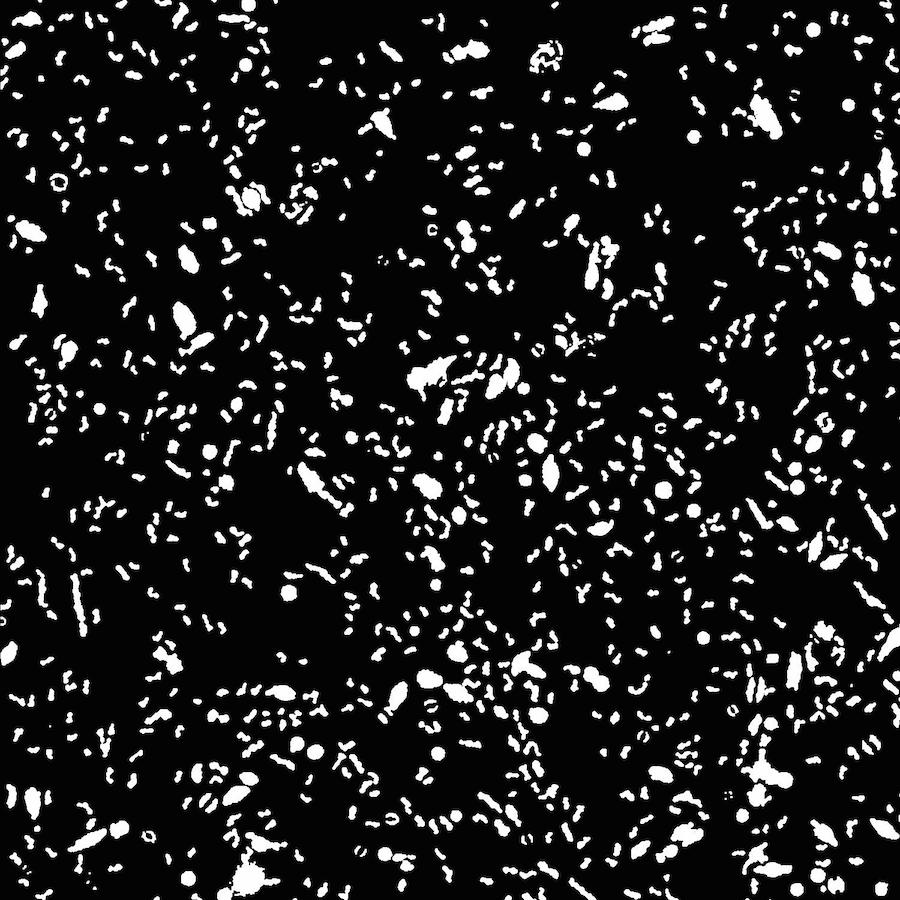}
}
\caption{The black halo regions are identified from the unsharp masked image 
by setting the threshold to $-0.01 e^{-} s^{-1}$. The halo segments are 
``dilated" (expanded) and combined with the normal detection image so that the 
stars start to merge with the diffraction spikes. Most of the diffraction spikes
can be removed by setting the maximum intensity value of the labeled segments 
less than $10 e^{-} s^{-1}$.}
\label{fig3:subfig}
\end{figure}

Our approach is to merge the diffraction spikes with each associated star and 
eliminate the combined source as a whole. To do this, we enhance the strong 
intensity gradients near bright stars and their diffraction spikes by applying 
unsharp masking. The unsharp masking enhances the peak and dampens the wings of 
the intensity distribution. As a result, some dark halos can be observed around 
the stars or bright elliptical galaxies, which are shown in 
Figure~\ref{fig3:subfig:a}. Empirically, we note that most of the pixels 
belonging to the dark halo regions in CLASH data tend to have a intensity value 
lower than $-0.01 e^{-} s^{-1}$, and we use this intensity as a threshold to 
identify these halos. We then dilate the segmentation boundaries around a dark 
halo in all directions to fill the gaps between the segments (
Figure~\ref{fig3:subfig:b}), and combine the ``dilated" images with the initial 
segmentation image obtained in Appendix~\ref{C} (Figure~\ref{fig3:subfig:c}).  
Most of the diffraction spikes merge with the segments of their mother stars as 
a result of performing this combination. We then label all the connected 
components \footnote{Whether a pixel connects to its neighbors or not is 
characterized by the pixel connectivity. Usually there are two types of 
connectivity: 4-connected and 8-connected. 4-connected pixels are connected 
horizontally and vertically, or diagonally; 8-connected pixels are connected 
horizontally and vertically, AND diagonally. In terms of pixel coordinates, in 
4-connected case, every pixel that has the coordinates ($x\pm1,y$) or 
($x,y\pm1$) is connected to the pixel at ($x,y$); in 8-connected case, in 
additional to 4-connected pixels, each pixel with coordinates ($x\pm1,y\pm1$) 
or ($x\pm1,y\pm1$) is connected to the pixel at ($x,y$). In this paper, all the 
adjacent 8-connected pixels are considered to belong to the same connected 
component.} 

in the combined image and calculate the maximum pixel intensity of each labeled 
connected component. Stars typically have maximum intensity values greater than 
$10 e^{-} s^{-1}$, while other objects barely have the maximum intensity value 
larger than $2 e^{-} s^{-1}$, therefore we can conservatively set 
$10 e^{-} s^{-1}$ as threshold to remove those bright stars along with the 
diffraction spikes (Figure~\ref{fig3:subfig:d}).
 
\section{Final Image Segmentation} 
\label{E}

\begin{figure*}
\centering
\begin{minipage}[t]{0.3\linewidth}
  \centering
  \includegraphics[width = \linewidth]{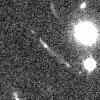}
 \end{minipage}
 \begin{minipage}[t]{0.3\linewidth}
  \centering
  \includegraphics[width = \linewidth]{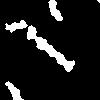}
 \end{minipage}
 \begin{minipage}[t]{0.3\linewidth}
  \centering
  \includegraphics[width = \linewidth]{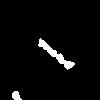}
 \end{minipage}

\begin{minipage}[t]{0.3\linewidth}
  \centering
  \includegraphics[width = \linewidth]{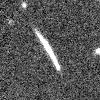}
 \end{minipage}
 \begin{minipage}[t]{0.3\linewidth}
  \centering
  \includegraphics[width = \linewidth]{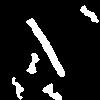}
 \end{minipage}
 \begin{minipage}[t]{0.3\linewidth}
  \centering
  \includegraphics[width = \linewidth]{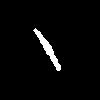}
\end{minipage}

\caption{The $left$ panel shows the original images of arc; the $middle$ panel
shows the primary segmentation; The $right$ panel shows the images
after the segmentation re-determination. The local background and noise level
within the box are estimated and the detection threshold is set to be 
proportional to its signal-to-noise level.}
\label{fig4}
\end{figure*}

The initial segmentation boundaries for objects detected in intensity-difference
space tend to have systematically larger surface area than the corresponding 
segmentation boundaries in pixel intensity space. So we refine the initial 
segmentation map to correct this small effect. We first define, for each 
detected segment, a ``bounding box" that spans the region from the minimum $x,y$
coordinates to the maximum $x,y$ coordinates. We then iteratively clip out 
pixels with very high (low) intensity within this box until the pixel intensity 
reaches convergence at $\pm 3\sigma$ around its median value. We then estimate 
the local background and noise level within the box. Since faint arcs are most 
likely missed or broken into small arclets at a high detection threshold, we set
the threshold for the re-segmentation to be proportional to the object's 
estimated local signal-to-noise ratio. Hence, objects with low surface 
brightness will be remapped using a lower detection threshold than that used for
brighter objects, allowing all sources to achieve their best segmentation (see 
Figure~\ref{fig4}).

\section{Utilization of the Length, Length-to-Width Ratio and 
Perimeter-to-Length Ratio}
\label{F}

\begin{figure}
\centering
\includegraphics[width = 0.7\linewidth]{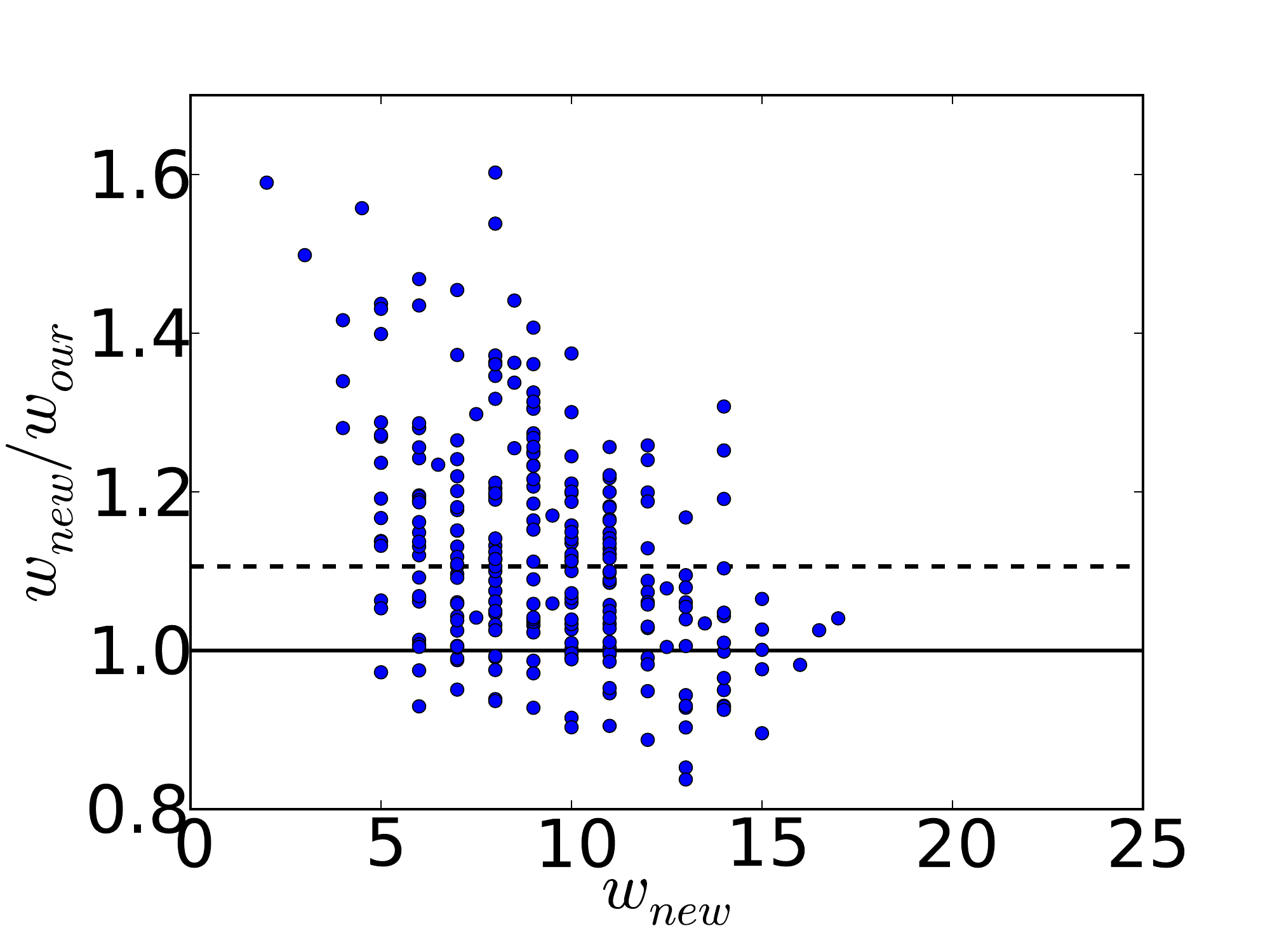}
\caption{The comparison of two definition of the width of arc: the y-axis is
the ratio of the newly defined width to our width; the x-axis is the newly defined
width. The solid line represents the $w_{new} = w_{our}$, while the dashed line 
denotes the median value of the ratio of the two widths.}
\label{fig4a}
\end{figure}

Once all images are processed through the preceding steps, we can begin the arc 
identification process. We identify giant arcs from among all detected sources 
primarily by their large ellipticity. For each source, we calculate the total 
area, perimeter length, and position of the peak intensity from the distribution
of all the connected pixels\footnote{We utilize {\tt ndimage} (a Python image 
processing module) to quickly calculate the mentioned parameters of the detected
objects}. Using the coordinates of the pixel with the peak intensity value in a 
given source, we locate the furthest point away from that maximum that is still 
within the boundaries of the source. We also locate the furthest point away from
that point, then calculate the sum of the distances from these two points to the
peak pixel position, and take this distance as the length of the segment. There 
are many ways to define the width of the segment: the image segments can be 
fitted by simple geometrical figures such as ellipse, circles, rectangles and 
rings \citep{mir93,bar94}; and therefore the width of the segment is 
approximated by the minor axis of the ellipse, the radius of the circle, the 
smaller side of the rectangle, or the width of the ring; \citet{dal04,hor05,
hen07} approximated the width by dividing the area by its length; \citet{men08}
proposed a more robust way to measure the width, by traversing the width profile 
of the arc and approximating the arc width as the median value of the profile. 
In this study, considering the computational efficiency, we adopt the former method: 
i. e. all the giant arcs are treated as rectangles and width = area / length, to 
determine the width of the segment in this paper. To test whether this definition 
of width will introduces bias in the measurement of $l/w$, we use the approach in 
\citet{men08} to re-calculate the width of all the detected arcs and compare with 
those in former definition. Figure~\ref{fig4a} shows the comparison of the ratio of 
two widths with the newly defined width. The dashed line denotes the median value 
of the ratio which is about 10\% higher than that in our definiton. Therefore, our 
$l/w$ (width) measure may be slightly biasing high (low).

The final step is to remove those detected segments that are not very likely to
be large lensed galaxies by requiring objects to satisfy three additional 
criteria \footnote{ In this study, we do not need to specify the orientation of the giant
tangential arcs relative to the cluster center. This allows us to apply our 
algorithm to less relaxed clusters that may not have a well-defined center.}: 
(1) their perimeter-to-length ratio must be $\ge 3$; (2) their minimal length
must be greater than a fixed value which is discussed in Section~\ref{s3.2}; 
(3) their minimal length-to-width ratio must be greater than a fixed value which
is determined in Section~\ref{s3.2}. The criterion (1) eliminates elongated 
objects with irregular morphology and criterion (2) both maintains the 
consistency with the concept of the ``giant'' arcs and prevents from the 
domination of the spurious detection as we will discuss later. We include all 
objects that satisfy these three constraints into our final arc candidate 
catalog.

\end{document}